\documentclass[useAMS,usenatbib]{mn2e}
\usepackage{pdfsync}
\usepackage{amsmath,amssymb}
\usepackage{delarray}
\usepackage{graphicx}

\newcommand{\Zmin}{{Z_\mathrm{min}}}
\newcommand{\Zmax}{{Z_\mathrm{max}}}
\newcommand{\Tmin}{{t_\mathrm{min}}}
\newcommand{\Tmax}{{t_\mathrm{max}}}
\newcommand{\Mmin}{{M_\mathrm{min}}}
\newcommand{\Mmax}{{M_\mathrm{max}}}

\newcommand{\sfr}{\operatorname{SFR}} 
\newcommand{\imf}{\operatorname{IMF}} 
\newcommand{\Bmass}{B^0}
\newcommand{\Bflux}{B}
\newcommand{\Frest}{F_\mathrm{rest}} 

\newcommand{\fext}{f_\mathrm{ext}}
\newcommand{\Fext}{\mathbf{f}_\mathrm{ext}}

\newcommand{\Eq}[1]{Eq.~(\ref{e:#1})}   

\newcommand{\Sec}[1]{Sect.~\ref{s:#1}}
\newcommand{\Fig}[1]{Fig.~\ref{f:#1}}

\newcommand{\diag}{\operatorname{diag}} 
\newcommand{\tr}{\operatorname{tr}}     
\newcommand{\mdot}{{\cdot}}             
\newcommand{\R}[1]{{\mathrm{#1}}}       
\newcommand{\M}[1]{\mathbf{#1}}         
\newcommand{\T}[1]{#1^{\top}}           

\newcommand{\mathd}{\mathrm{d}}

\newcommand{\pdrv}[2]{\frac{\partial #1}{\partial #2}} 

\newcommand{\Msun}{\ensuremath{\mathrm{M}_{\odot}}}

\newcommand{\AMR}{age-metallicity relation}
\newcommand{\CN}{conditioning number}
\newcommand{\GCV}{generalized cross validation}
\newcommand{\GSOd}{Gram-Schmidt othonormalized}
\newcommand{\GSO}{Gram-Schmidt othonormalization}
\newcommand{\GSVD}{generalized {\SVD}}

\newcommand{\LWSAD}{luminosity weighted stellar age distribution}

\newcommand{\PHR}{\textsc{P\'egase-HR}}
\newcommand{\SAD}{stellar age distribution}

\newcommand{\SED}{spectral energy distribution}
\newcommand{\SFHs}{star formation histories}
\newcommand{\SFH}{star formation history}
\newcommand{\SFR}{star formation rate}
\newcommand{\SSPs}{single stellar populations}
\newcommand{\SSP}{single stellar population}
\newcommand{\SVD}{singular value decomposition}
\newcommand{\TLS}{total least squares}

\newcommand{\STECMAP}{STECMAP}

\newcommand{\nicefrac}[2]{\leavevmode\kern.1em
            \raise.5ex\hbox{\the\scriptfont0 #1}\kern-.1em
      /\kern-.15em\lower.25ex\hbox{\the\scriptfont0 #2}}

\title[STEllar Content via Maximum A Posteriori]{STEllar Content from high resolution galactic spectra\\  via Maximum A Posteriori}

\author[P.\ Ocvirk,  C.\ Pichon, A.\ Lan\c{c}on,
E.\ Thi\'ebaut]{P. Ocvirk$^{{1}}$ C.\ Pichon$^{{2}}$ A.\ Lan\c{c}on$^{{1}}$
  \& E.\ Thi\'ebaut$^{{3}}$\\
$^1$Observatoire de Strasbourg (UMR 7550), 11 rue de l'Universit\'e,
67000 Strasbourg, France. \\
$^2$Institut d'Astrophysique de Paris, 98 bis boulevard Arago, 75014 Paris, France. \\
$^3$Observatoire de Lyon, 9 avenue Charles Andr\'e F-69561 Saint Genis Laval
 Cedex, France
}

\begin{document}
\date{Typeset \today; Received / Accepted}

\pagerange{\pageref{firstpage}--\pageref{lastpage}} \pubyear{2005}

\maketitle
\label{firstpage}


\begin{abstract}
This paper describes \STECMAP\ (STEllar Content via Maximum A Posteriori), a flexible,
non-parametric inversion method  for the interpretation of the
integrated light spectra of galaxies, based on synthetic spectra of {\SSPs} (SSPs). We
focus on the recovery of a galaxy's {\SFH} and stellar age-metallicity
relation.  We use the high resolution {SSPs} produced by {\PHR} to
quantify the informational content of the wavelength range
$\lambda\lambda = 4000 - 6800$\,\AA. Regularization of the inversion is
achieved by requiring that the solutions are relatively smooth functions
of age. The smoothness parameter is set automatically via generalized
cross validation.

A detailed investigation of the properties of the corresponding simplified
linear problem is performed using {\SVD}. It turns out to be a powerful tool
for explaining and
predicting the behaviour of the inversion, and may help designing {SSP}
models in the future. We provide means of quantifying the fundamental limitations of the problem
considering the intrinsic properties of the {\SSPs} in the spectral range of
interest, as well as the noise in these models and in the data.  We
demonstrate that the information relative to the stellar content is
relatively evenly distributed within the optical spectrum.  
{We show that one should
not attempt to recover more than about 8 characteristic episodes in the
{\SFH} from the wavelength domain we consider.}
STECMAP preserves optimal (in the cross validation sense) freedom in the
characterization of these episodes for each spectrum.

We performed a systematic simulation campaign and found that, when the time
elapsed between two bursts of star formation is larger than 0.8\,dex, the
properties of each episode can be constrained with a precision of 0.04\,dex
in age and 0.02\,dex in metallicity from high quality data ($R=10\,000$,
signal-to-noise ratio $\mathrm{SNR}=100$ per pixel), not taking model errors into account.
We also found that the spectral resolution has little effect on population
separation provided low and high resolution experiments are performed with the
same SNR per \AA. However, higher spectral resolution does improve the
accuracy of metallicity and age estimates in double burst separation
experiments. When the fluxes of the data are properly calibrated,
extinction can be estimated; otherwise the continuum can be discarded or
used to estimate flux correction factors.

The described methods and error estimates will be useful in the design and
in the analysis of extragalactic spectroscopic surveys.
\end{abstract}


%




\section{Introduction}
%
%
%
The diversity of shapes and colors of galaxies illustrates the wealth of
physical mechanisms acting in these complex objects.  Their formation
history, including the building of their halos, bulges, disks and disk
patterns, is still controversial. Empirical constraints on the formation
scenarii are engraved in the distribution of stellar ages, metallicities,
and kinematics.  Unless the galaxies can be resolved into stars, this
crucial information must be extracted from integrated spectra.  This
spectral energy distribution is a recording of the whole life of a galaxy:
the condition of its birth, the formation and assembly of its first blocks,
its passive evolution and the recycling of its material, or its active
evolution through merging, all these determine the current stellar content.
Yet, this information is embedded in a non-trivial manner in the light we
receive.

While a wealth of such data is currently being gathered from spectroscopic
surveys (for example the Sloan Digital Sky Survey or the 2DF Galaxy
Redshift Survey), using them to probe the general properties of the stellar
populations on a cosmological timescale is an exciting perspective.

In the literature, the stellar content of a galaxy is often characterized
by a luminosity weighted age, a luminosity weighted metallicity, a global
velocity dispersion, and a parameter characterizing extinction.  Since
\citet{worthey94}, the Lick indices have been readily used in order to
describe the nature of the stellar populations.  Spectral indices are
convenient because they are robust to a number of observational
perturbations, but they exploit only small wavelength domains.  The use of
a larger fraction, and eventually of all the information in a spectrum
must, at least in principle, help separate, age-date and characterize
coexisting stellar components, steps required to access the actual
evolution of the galaxies under study. Individual spectral features with
specific sensitivities to age or metallicity may add information to the
Lick data points, and the redundancy provided by many lines spread over a
wide spectral range reduces the sensitivity to noise. Recently, methods
have emerged that use the whole available spectral range, relying on
compression \citep{moped01} or on non-negative least squares
\citep{mateu01,CF04-1}.

The introduction of these methods gave birth to a field of research, whose
goal is to measure the cosmic {\SFH} by summing the individual {\SFHs} of a
large number of galaxies. This results in an estimate of the mean history of
star formation (a so called ``Madau plot'') in principle free from the
uncertainties related to pure emission-line diagnostics
\citep{dopita-sed2004}. Moreover, the distribution of individual {\SFHs} is
even more constraining than a Madau plot alone. If feasible, this approach
indeed constitutes a very powerful test for the current cosmological
models.  In fact, such techniques have been used recently to support the
idea of galactic downsizing, \textit{i.e.} to argue the stellar activity
has shifted in the recent past towards less massive galaxies, something
that some authors have presented as a problem for hierarchical clustering.
As more results of this kind are published, it becomes clear that different
authors have very different conceptions of what is a reasonable
interpretation of a galactic spectrum \citep{CF04-1,heavensnature}.
Indeed, the problem of characterizing star formation histories based on a
spectrum is strongly ill-conditioned as we will demonstrate extensively
below (see also \citealt{moultaka00,moultaka04}). This remains true in the
restrictive framework of evolutionary population synthesis, although this
approach incorporates the simplifying assumption that the intrinsic spectra
of mono-metallic, single-aged \emph{single stellar populations} (SSPs) are
known. Over-interpretation of the data is a common pitfall when
ill-conditioning is misjudged or overlooked.  A useful approach to
ill-conditioned inverse problems is the maximum penalized likelihood, which
is formally equivalent to a maximum a posteriori likelihood (MAP). It has
been applied in the past in a variety of fields in astronomy such as light
deprojection \citep{kochanek96}, stellar kinematics \citep{saha94,merritt97,pi-th98},
image deblurring \citep{thiebautp02,ThiebautCargese2005} or the
interpretation of low resolution energy distributions of galaxies
\citep{vergely02}.

This paper discusses the interpretation of high resolution optical spectra
of galaxies. A maximum resolving power $R=10\,000$ is considered, which is
adequate in particular for the studies of low mass galaxies or of massive
star clusters in galaxy cores.  We focus on the object's stellar
content. The simultaneous extraction of the kinematical
information with a direct extension of the adopted method is the subject of
a companion paper.  
Our work is positioned at the interface between single
stellar population models and observations. Its purpose is not to question
the particular ingredients and assumptions of a specific population
synthesis code, although some of the discussion will be specific to the
model package {\PHR} of \citet{PEG-HR}, since it is the first package to have provided a similar spectral
resolution (see \citet{GD05} for a medium resolution package).
Rather, we intend to clarify how the intrinsic properties of a basis of
single stellar population spectra can be used to infer consequences for the
study of composite stellar populations.

The general problem, where additional
constraints such as positivity of the {\SFH} are included is a non-linear
problem. Nevertheless, we give special importance to the linear problem
because it provides firm footing to explain the processes that
determine the reliability of a recovered \SFH. It also clearly displays
many of the features found in the more realistic inversions as well.

We also study the feasibility of the inversion in different observational
regimes (in terms of spectral resolution and noise), and give simple
scaling laws and error estimates to predict the accuracy and relevance of
the results. The main characteristics of our approach are:
\begin{itemize}
\item It is non parametric, and thus provides properties such as the
  stellar age distribution with minimal constraints on their shape.
\item The ill-conditioning of the problem is taken into account through
  explicit regularization.
\item Optimal interpretation of the data is achieved by the proper setting
  of the smoothing parameter.
\end{itemize}

The organization of the paper is as follows. We start in \Sec{models} by
describing the inversion problems that will be tackled.  In \Sec{age}, we
provide a comprehensive investigation of the idealized linear problem of
finding the stellar age distribution of a mono-metallic, reddening-free
stellar population.
\Sec{valid} investigates the performance of these inversions in a
set of simulations in terms of resolution and separability of bursts.
\Sec{aze} addresses the problem of the simultaneous study of
stellar ages and metallicities, while allowing for extinction (or other
transformations of the continuum).  Conclusions are drawn in
\Sec{2ndc}, while the paper closes with a discussion for prospects.

\section{Non parametric models of spectra}

\label{s:models}

The {\SED} (SED) that we measure for each spatial pixel of an observed
galaxy results from light emitted by coexisting stellar populations of
various ages, metallicities and kinematics, and from the interactions of
the stellar light with the interstellar medium (reddening, nebular
emission). The example of the Milky Way tells us that any given stellar
population of a galaxy may consist of stars with non-trivial distributions
in age, metallicity, or even relative abundances
\citep{feltzing01,prochaska2000,gratton2000}. In principle, age, abundances
and velocity distributions should thus be treated as independent parameters
in a galaxy model meant for an exploration without preconceptions.

In the following, we will restrict ourselves to simplified models that
balance, in our view, technical feasibility (in view of current models and
data) and scientific interest.  We assume that metallicity describes the
stellar abundances, mainly because our population synthesis model does not
allow for abundance variations (\citet{thomas2003} specifically address this
issue).  Except for the discussion of a more general case in
\Sec{aze}, we restrict ourselves to the assumption of a one to one
relationship between stellar ages and metallicities.  This allows us to
search for significant trends, as predicted by simple evolutionary scenarii
for galaxies. We adopt a simple parameterized formulation for extinction.
Finally, we will deal with stellar populations at rest (or with known
velocity distributions).

Emission lines are out of the aim of this study.  They may be used in the
future, in particular to obtain further constraints on the youngest stars
and on obscuration by dust, or to constrain properties of the interstellar
medium.

\subsection{The spectral basis}

\label{s:basis}

The basic building block to model the spectrum of an observed galaxy is the
spectral energy distribution $S(\lambda,m,t,Z)$ of a star of initial mass
$m$, age $t$ and metallicity $Z$ (mass fraction of metals at the formation of
the star).  Integrating over stellar masses yields
the intrinsic spectrum $\Bmass(\lambda,t,Z)$ of the {\SSP} of age $t$,
metallicity $Z$ and unit mass:
\begin{equation}
  \label{e:Bmass}
  \Bmass(\lambda,t,Z)
  \triangleq \int_{\Mmin}^{\Mmax} \imf(m)\,S(\lambda,m,t,Z)\,\mathd m \, ,
\end{equation}
where $\imf(M)$ is the Initial Mass Function and $\Mmin$ and $\Mmax$ are
the lower and upper mass cut-offs of this distribution.
Assuming that the metallicities of the stars can be described by a
single-valued Age-Metallicity Relation (AMR) $Z(t)$, it is possible
to derive the unobscured spectral energy
distribution of the galaxy at rest:
\begin{equation}
  \Frest(\lambda) =
  \int_{\Tmin}^{\Tmax} \sfr(t)\,\Bmass(\lambda,t,Z(t))\,\mathd t \, ,
\end{equation}
where $\sfr(t)$ is the Star Formation Rate (i.e.\ mass of new stars born
per unit of time, with the convention that $t=0$ is today) and $\Tmax$ is an upper integration limit, for
instance the Hubble time. Similarly, $\Tmin$ is a lower integration limit,
ideally 0. Both $\Tmin$ and $\Tmax$ must in practice be set
according to the validity domain of the SSP basis $\Bmass(\lambda,t,Z(t))$.

The Luminosity Weighted Stellar Age Distribution (LWSAD) $\Lambda(t)$ gives
the contribution to the total emitted light of stars of age $[t,t+ \mathd
t]$. It is related to the SFR by:
\begin{equation}
  \label{e:lwsad}
  \Lambda(t) \triangleq \frac{\operatorname{SFR}(t)}{\Delta\lambda}\,
  \int_{\lambda_{\min}}^{\lambda_{\max}}
  \Bmass\bigl(\lambda,t,Z(t)\bigr)\,\mathd\lambda \, ,
\end{equation}
where $\Delta\lambda=\lambda_{\max}-\lambda_{\min}$ is the width of the
available wavelength domain. In order to use the \LWSAD, we define the
flux-normalized {\SSP} basis $\Bflux(\lambda,t,Z)$ where each spectrum
is normalized to a unitary flux:
\begin{equation}
  \label{e:favgb}
  \Bflux(\lambda,t,Z)
  = \frac{\Bmass(\lambda,t,Z)}
         {\displaystyle\frac{1}{\Delta\lambda}\,
          \int_{\lambda_{\min}}^{\lambda_{\max}}\Bmass(\lambda,t,Z)\,\mathd\lambda} \, .
\end{equation}
Using $\Lambda(t)$, $\Bflux(\lambda,t,Z)$ and $Z(t)$, the unobscured {\SED}
of any composite population at rest reads:
\begin{equation}
  \label{e:v2}
  \Frest(\lambda) = \int_{\Tmin}^{\Tmax} \Lambda(t) \,
  \Bflux(\lambda,t,Z(t)) \, \mathd t  \, .
\end{equation}

For a given {\SSP} basis, dealing with the {\SFR} or the {\LWSAD} is
apparently equivalent.  Yet, because of the strong dependence of the
mass-to-light ratio of {\SSP} fluxes on time, $\Lambda(t)$ is more directly
related to observable quantities than $\sfr(t)$.  We therefore prefer the
formulation based on $\Lambda$ (see also \Sec{fab}).

\begin{figure*}
\begin{center}
\rotatebox{-90}{\resizebox{10cm}{16cm}{\includegraphics{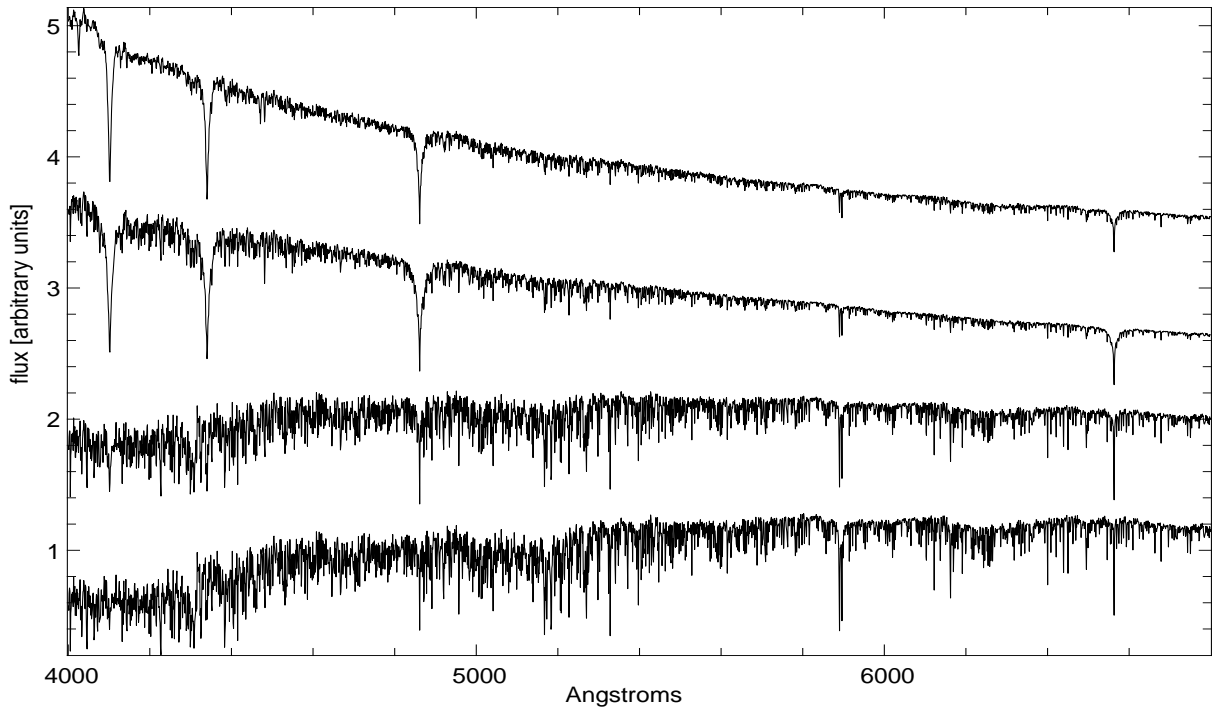}}}
\rotatebox{-90}{\resizebox{10cm}{16cm}{\includegraphics{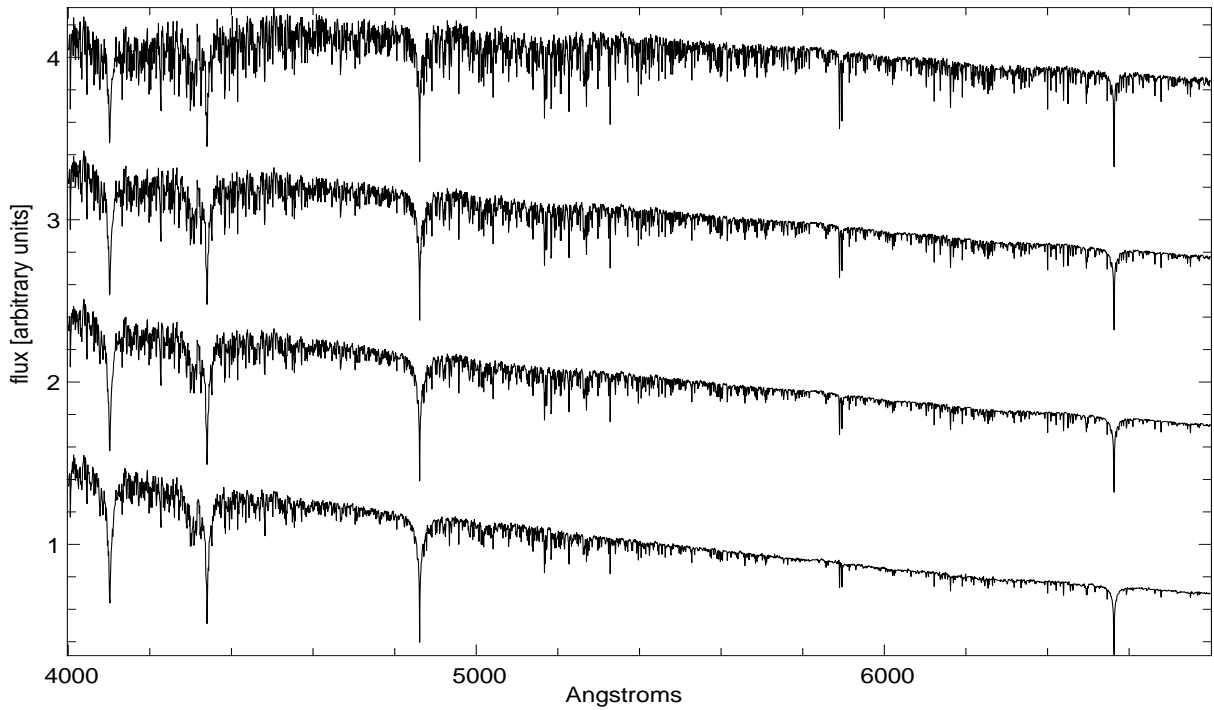}}}
\caption{Example of high resolution {\SSPs} produced by {\PHR}. Panel a:
solar metallicity {\SSPs} of age $50$, $400$, $2\,500$, and $15000$ Myr (from top to
bottom). Panel b: $1$ Gyr {\SSP} for several metallicities, Z$=0.05$,
$0.02$, $0.004$ and $0.0004$ (from top to bottom). The spectra are normalized
to a common mean flux and offset for clarity.}
\label{f:base}
\end{center}
\end{figure*}

Many codes are available to construct $\Bflux(\lambda,t,Z)$.
The {\SSP} library adopted here is computed with {\sc P\'egase-HR}
\citep{PEG-HR}, a version of
{\sc P\'egase}\footnote{Projet d'Etude des GAlaxies par
Synth\`ese Evolutive. {\tt http://www.iap.fr/pegase}} that provides
optical spectra at high resolution ($R=10\,000$), based
on the ELODIE stellar library \citep{elodie01}. It consists of
{\SSPs}  generated by single instantaneous starbursts with a set of
metallicities $ZZ= [0.0001,0.1]$.
The wavelength range of the spectra is
$\lambda \lambda=[4000 $\AA$ ,6800  $\AA$]$, sampled in $\delta \lambda=0.2$
{\AA} steps. \Fig{base} shows example spectra of such {\SSP}s, at fixed
metallicity (panel a) and fixed age (panel b). The large number
of lines is supposed to improve the accuracy of stellar content
analysis. The IMF used is described in
\cite{kimf} and the stellar masses range from $0.1$\,\Msun\ to
$120$\,\Msun. The IMF is an input of {\PHR}, that we do not attempt to
constrain. On the contrary, we assume it is universal and known a priori.
The generated spectra are considered most reliable from
$\Tmin=10\,\R{Myr}$ to $\Tmax=20\,\R{Gyr}$ \citep{PEG-HR}.  The spectra of
the different {\SSPs} are computed for a set $S_{t}$ of logarithmically
spaced ages between $\Tmin$ and $\Tmax$. The set of mono-metallic {\SSPs}
obtained is referred to as the \emph{basis} or \emph{kernel} in the rest of
the paper.
\subsection{Extinction models}
In most cases, the intrinsic emission of the stars
of a galaxy is affected by dust. Both the composition and the spatial
distribution of the dust determine the extinction. The ISM of galaxies
is rarely homogeneous, and the stars may be
seen through different amounts of dust. One could therefore
envisage an age-dependent extinction law or extinction parameter.
Indeed, there is evidence that the
obscuration of an ensemble of stars varies systematically with age over
the first $\sim 10^7$ years of their evolution, while these young stars
leave or destroy their parent molecular clouds (Charlot \& Fall, 2000,
and references therein). However, the early epochs relevant to starbursts
are currently slightly out of reach with {\sc P\'egase-HR}, although they will become
accessible with improvements of the stellar library.
\citet{vergely02} suggest that recovering such a trend with age is
possible with high quality data ranging from the ultraviolet to the
infrared. In this paper, we deliberately chose not to search for
an age-dependence of extinction. The main reason is that we are
considering only a limited section of the electromagnetic spectrum.
We postpone a systematic study to
future work. In the following, we adopt a unique extinction law
$\fext(E,\lambda)$ parameterized by the color excess $E \equiv
E(B-V)$ and normalized to have a unit mean. Accounting for extinction, the model {\SED} then reads:
\begin{equation}
  \Frest(\lambda) = \fext(E,\lambda)\,\int_\Tmin^\Tmax\Lambda(t) 
  \Bflux\bigl(\lambda,t,Z(t)\bigr)\,\mathd t \, .
\label{e:evergely}
\end{equation}

Note that $\fext$ can be a function of more than one time-independent parameter, and may for
example be a more complex attenuation law, function of the
distribution of dust in the galaxy and its mixing with the stars, or a low
order polynomial accounting for the instrumental spectrophotometric
calibration error.

\subsection{General properties and problems with SSPs}

\label{s:problemsSSPs}

Synthetic spectra of single stellar populations are the building blocks
involved in the interpretation of galaxy spectra.  Their properties have a
strong effect on the behaviour of the inversion problem.


Both the theory of stellar evolution and observations tell us that
{\SSP} evolution with time is fundamentally smooth in the optical except for a number of
specific evolutionary transitions (e.g.\ helium flash, carbon flash,
supernova explosion, envelope expulsion at the end of the Asymptotic Giant
Branch), and that it shows some linearity.  This means, for instance, that
a 500\,Myr old population looks very similar to the average between a
600\,Myr and 400\,Myr old one.  Our ability to identify the differences
depends strongly on the signal-to-noise ratio (hereafter SNR) of the models
and data.  Section~\ref{s:age} shows how to quantify this quasi-linearity
and its consequences.


The synthetic spectra of {\SSPs} are affected by uncertainties in the
stellar evolutionary tracks and in the stellar library used to
construct them. Despite permanent progress, some aspects of stellar
evolution remain difficult to model (e.g.\ the Horizontal Branch, the
Asymptotic Giant Branch, the Red Supergiant phase; effects of convection,
of rotation, of a binary companion). The errors propagate to the SSPs,
resulting in unknown systematic errors in age and metallicity estimates.
Some insight to the amplitude of these errors is given by the direct
comparison between results obtained using different sets of tracks.
Nevertheless, it is beyond the scope of this paper to discuss the pros and
cons of the different set of tracks and the reader is refered to
\cite{charlot96} and \cite{lej&fer2002} for an extensive discussion.

The input library of stellar spectra can be either empirical or
theoretical. The latter situation has the advantage of providing spectra
for any parameter set $(T, g, Z)$ with no observational noise. However,
these are not free of intrinsic uncertainties, due for instance to
shortcomings of atomic and molecular data, to assumptions on partial
thermodynamical equilibrium, or to inappropriate abundance ratios.
Empirical spectra, on the other hand, are hampered by a number of issues:


\begin{enumerate}
\item The library is discrete. Therefore interpolation between existing
  stars is needed. This can be a tricky issue, especially on the borders of
  the grid and in underpopulated regions of $(T, g, Z)$ space.  Moreover,
  when stars are interpolated, the noise patterns are also carried along.
We will see in \Sec{ill}
  that this has noticeable effects on the behaviour of the
  inverse problem.
\item The library generally consists only of Milky Way or even Solar
  Neighbourhood stars. Thus, the solar metallicity is the best populated
  region of parameter space, while other regions may be depleted,
  especially for extreme cases as young metal poor or old metal rich stars.
  We also know that outer galaxies may involve abundance ratios that are
  not found within the Milky Way.  One example is found in the metal-rich
  and $\alpha$-enhanced populations of large elliptical galaxies. This
  difficulty is known as \emph{template mismatch} and results in biases that
  would be best studied using simulations based on theoretical spectra with
  various sets of abundances. The library used in {\sc P\'egase-HR} is
  known to be deficient in high metallicity, high $\alpha$-element
  abundance red giants \citep{PEG-HR}, which may lead to an
  over-estimate of age or metallicity in observed galaxies\footnote{Work is
    being done to improve the underlying library.}.
\item Empirical stellar spectra have a finite SNR, and so do the
  averaged or interpolated spectra involved in the synthesis of a galaxy
  spectrum.  It should then be considered useless to observe stellar
  populations at SNR's larger than the library's.
\item The fundamental parameters of each star in the library are estimates,
  in the case of {\sc P\'egase-HR} based on a subset of standards and the
  automated code TGMET \citep{TGMET}.  Even though error bars on these
  parameters are provided, some glitches and outliers happen.  The final
  error resulting from interpolating between correct and ill-parametered
  stars and summing is unknown.
\end{enumerate}
Notwithstanding the above limitations of spectral synthesis, our purpose
here is to investigate the behaviour of the inverse method for a given
model.  Hence, in this paper we will be restricted to one given SSP model.

\section{A simplified inverse problem: the age distribution recovery}

\label{s:age}
This section discusses the inverse problem of recovering the age
distribution of a purely mono-metallic unobscured population at rest. This
simplification is deliberate and yields a linear relationship between the observed
spectral energy distribution $F_{\rm{rest}}(\lambda)$ and the stellar age distribution
$\Lambda(t)$. It allows
us to address its fundamental properties and behaviour, characterized by
simple quantities and criteria. These turn out to be precious tools in the
process of understanding and diagnosing the ill-conditioning and
pathological behaviour of such a problem and their non linear
generalization. It also allows us to
introduce the automated regularizing method required to solve the problem
in practice.

\subsection{The linear inverse problem}

Our idealized mono-metallic unobscured model stellar population is characterized by its
{\LWSAD} $\Lambda(t)$ and its constant age-metallicity relation $Z(t)=Z_0$, the {\SED}
of the emitted light $\Frest(\lambda)$ then reads:
\begin{equation}
  \Frest(\lambda) =
  \int_\Tmin^\Tmax \Lambda(t)\,\Bflux(\lambda,t,Z(t))\,\mathd t \, ,
  \label{e:fred}
\end{equation}
where $\Bflux(\lambda,t,Z(t))$ is the flux-normalized {\SSP} basis (cf.\
\Eq{favgb}) which is just a function of the wavelength and time as the AMR
$Z(t)$ is supposed to be known. Solving \Eq{fred} where
$\Bflux(\lambda,t,Z(t))$, and $\Frest(\lambda)$ are given and $\Lambda(t)$
is the unknown, is as we will demonstrate, a classical example of a
potentially 
ill-posed problem \citep{hansen}, i.e.\ it can be shown that small
perturbations of the data can cause large perturbations of the solution.
Hence any noise in the data, $\Frest(\lambda)$, or in the kernel,
$\Bflux(\lambda,t,Z(t))$, can lead to a solution very far from the true
solution.

\subsection{Discretization: the matrix form}
\label{s:disc}

Intuitively, after discretization of the wavelength and age ranges, the
linear integral equation~(\ref{e:fred}) can be approximated by:
\begin{equation}
  \label{e:dgfred}
  s_i \approxeq \sum_{j=1}^{n} B_{i,j}\,x_j
  \,,\quad i \in \{1,..,m\} \, ,
\end{equation}
with:
\begin{equation}
  \label{e:lincoefs}
  \begin{array}{rcl}
  s_i &=&
  \left\langle\Frest(\lambda)\right\rangle_{\lambda\in\Delta\lambda_i} \, ,\\[1ex]
  B_{i,j} &=&
  \left\langle\Bflux\bigl(\lambda,t,Z(t)\bigr)\right\rangle%
  _{\lambda\in\Delta\lambda_i,t\in\Delta{}t_j} \, , \\[1ex]
  x_j &=& \left\langle\Lambda(t)\right\rangle%
  _{t\in\Delta{}t_j} \, , \\
  \end{array}
\end{equation}
where the notation, e.g.,
$\langle\Frest(\lambda)\rangle_{\lambda\in\Delta\lambda_i}$ indicates some
kind of weighted averaging or sampling of the argument $\Frest(\lambda)$
over the \hbox{$i$-th} wavelength interval $\Delta\lambda_i$ and similarly
for the age interval.

More rigorously, let
$\{g_i:[\lambda_{\min},\lambda_{\max}]\mapsto\mathbb{R};i=1,\ldots,m\}$ and
$\{h_j:[\Tmin,\Tmax]\mapsto\mathbb{R};j=1,\ldots,n\}$ be two
ortho-normalized bases of functions spanning the wavelength and age
intervals respectively.  Then the \emph{best approximation}\footnote{In the
  sense of the $\ell_2$ norm defined by the ortho-normalized basis
  of functions.} of $\Lambda(t)$ writes:
\begin{equation}
  \Lambda(t) \approxeq \sum_{j=1}^{n} x_j\,h_j(t)\,,
  \mbox{ with }
  x_j = \!\!\int\!\! \Lambda(t)\,h_j(t)\,\mathd t\,,
\end{equation}
similarly, the best approximation of $\Frest(\lambda)$ writes:
\begin{equation}
  \Frest(\lambda) \approxeq \sum_{i=1}^{m} s_i\,g_i(\lambda)\,,
  \mbox{ with }
  s_i = \!\!\int\!\! \Frest(\lambda)\,g_i(\lambda)\,\mathd\lambda\,.
\end{equation}
It is straightforward to obtain the coefficients of the matrix $\M{B}$ in
\Eq{dgfred} by inserting these approximations in \Eq{fred}:
\begin{equation}
  B_{i,j} = \iint \Bflux\bigl(\lambda,t,Z(t)\bigr)\,
  g_i(\lambda)\,h_j(t)\,\mathd t\,\mathd\lambda\,.
\end{equation}

In practice, we adopt equally spaced $\lambda_i$ and equally spaced
$\log(t_j)$ to sample the wavelength range and the evolutionary timescales
of {\SSPs}. Then we simply use gate functions for $g_i$ and $h_j$.  In
other words, $s_i$ is the average flux received in
$\lambda_i\pm\delta\lambda$ and $x_j$ is the mean flux contribution of the
sub-population of age $[t_{j-1},t_{j}]$ -- hence the notation used in
Eqs.~\ref{e:lincoefs}.

Note that if $t_{j}-t_{j-1}$ is too large, significantly different
populations are already entangled in the sampled basis
$\Bflux_{j}(\lambda)=\big\langle\Bflux(\lambda,t,Z(t))\bigr\rangle_{t\in\Delta{}t_j}$.
For this reason the number $n$ of {\SSP} elements in the basis should not
be too small. The signatures of the populations of each age should be
expressed in the adopted basis. On the other hand (see \Sec{ill}), we will
sometimes want to use a small $n$, i.e.\ a basis that is coarser in time,
and we will see that the overall adopted value strongly depends on the
observational context (SNR, spectral resolution and range \dots).

Using matrix notation and accounting for data noise, the observed
SED reads:
\begin{equation}
  \M{y} = \M{B}\cdot\M{x} + \M{e} \, ,
\end{equation}
where $\M{y}=\T{(y_1,\ldots,y_m)}$ is the observed spectrum (including
errors), i.e.\ $y_i$ is the measured flux in the range
$\lambda_i\pm\delta\lambda$, and $\M{e}=\T{(e_1,\ldots,e_m)}$ accounts for
modelling errors and noise.  The vector of sought parameter $\M{x}$ is the
discretized \SAD, i.e.\ the $x_j$ is the luminosity contribution of the
stars of age $[t_{j-1},t_j]$ to the total luminosity, averaged over the
available wavelengths.  The vector $\M{s}=\M{B}\cdot\M{x}$ is the
\emph{model} of the observed spectrum and $\M{B}$ is the discrete model
matrix, sometimes also referred to as the kernel.

\subsection{Maximum a Posteriori}
\label{s:mlf1}

In a real astrophysical situation, the data $\M{y}$ is always contaminated
by errors and noise.  Following Bayes' theorem, the \emph{a
  posteriori} conditional probability density
$f_{\mathrm{post}}(\M{x}|\M{y})$ for the realization $\M{x}$ given the data
$\M{y}$ writes:
\begin{equation}
  f_{\mathrm{post}}(\M{x} | \M{y}) \propto f_\mathrm{data}(\M{y}|\M{x}) \,
  f_{\mathrm{prior}}(\M{x})  \, ,
  \label{e:bayest}
\end{equation}
where $f_{\mathrm{prior}}(\M{x})$ is the \emph{a priori} probability
density of the parameters, and $f_\mathrm{data}(\M{y}|\M{x})$, sometimes
referred as the \emph{likelihood}, is the probability density of the data
given the model.  For Gaussian noise,
$f_\mathrm{data}(\M{y}|\M{x})\propto\exp[-\frac{1}{2}\,\chi^2(\M{y}|\M{x})]$,
with:
\begin{equation}
  \chi^2(\M{y}|\M{x}) =
    \T{\bigl(\M{y} - \M{s}(\M{x})\bigr)}{\cdot}\M{W}{\cdot} 
       \bigl(\M{y} - \M{s}(\M{x})\bigr) \, ,
   \label{e:chi2}
\end{equation}
where the weight matrix is the inverse of the covariance matrix of the
noise: $\M{W}=\mathrm{Cov}(\M{e})^{-1}$.  Maximizing the posterior
probability (\ref{e:bayest}) is equivalent to minimizing the penalty:
\begin{equation}
   Q(\M{x}) = \chi^2(\M{y}|\M{x}) - 2\,\log\left(f_{\mathrm{prior}}(\M{x})\right) \,.
   \label{e:logMAP}
\end{equation}
Without \emph{a priori} information about the sought parameters, the
probability density $f_{\mathrm{prior}}$ is uniformly distributed and this
term can be dropped.  In this case, $Q(\M{x})$ simplifies to
$\chi^2(\M{y}|\M{x})$, the traditional goodness of fit estimator for
Gaussian noise.

When the errors are uncorrelated the matrix
$\M{W}$ formally assigns a weight $1/\mathrm{Var}(y_{i})$ to
each pixel $i$ of data. Practically, one may want to modify the
variance-covariance matrix in order to use it as a mask. For example, a
dead pixel can be assigned null weight. In the same way, we may also mask
emission lines. Because of this particular usage of the matrix $\M{W}$, it
will often be called the weight matrix. It need not be exactly a
variance-covariance matrix, even though it can be built upon one.


\subsection{Ill-conditioning and noise amplification}
\label{s:ill}
As mentioned earlier, the linear problem corresponding to the recovery of the
stellar age distribution $\M{x}$ by maximizing the likelihood term only,
qualifies as a discrete ill-conditioned problem, \textit{i.e.} it might
therefore be extremely sensitive to noise, both in the data and in the
kernel. It thus will require some form of regularization in order to obtain
physically meaningful solutions.

\subsubsection{Noisy data}

\begin{figure}
  \begin{center}
    \includegraphics[width=1.05\linewidth,clip]{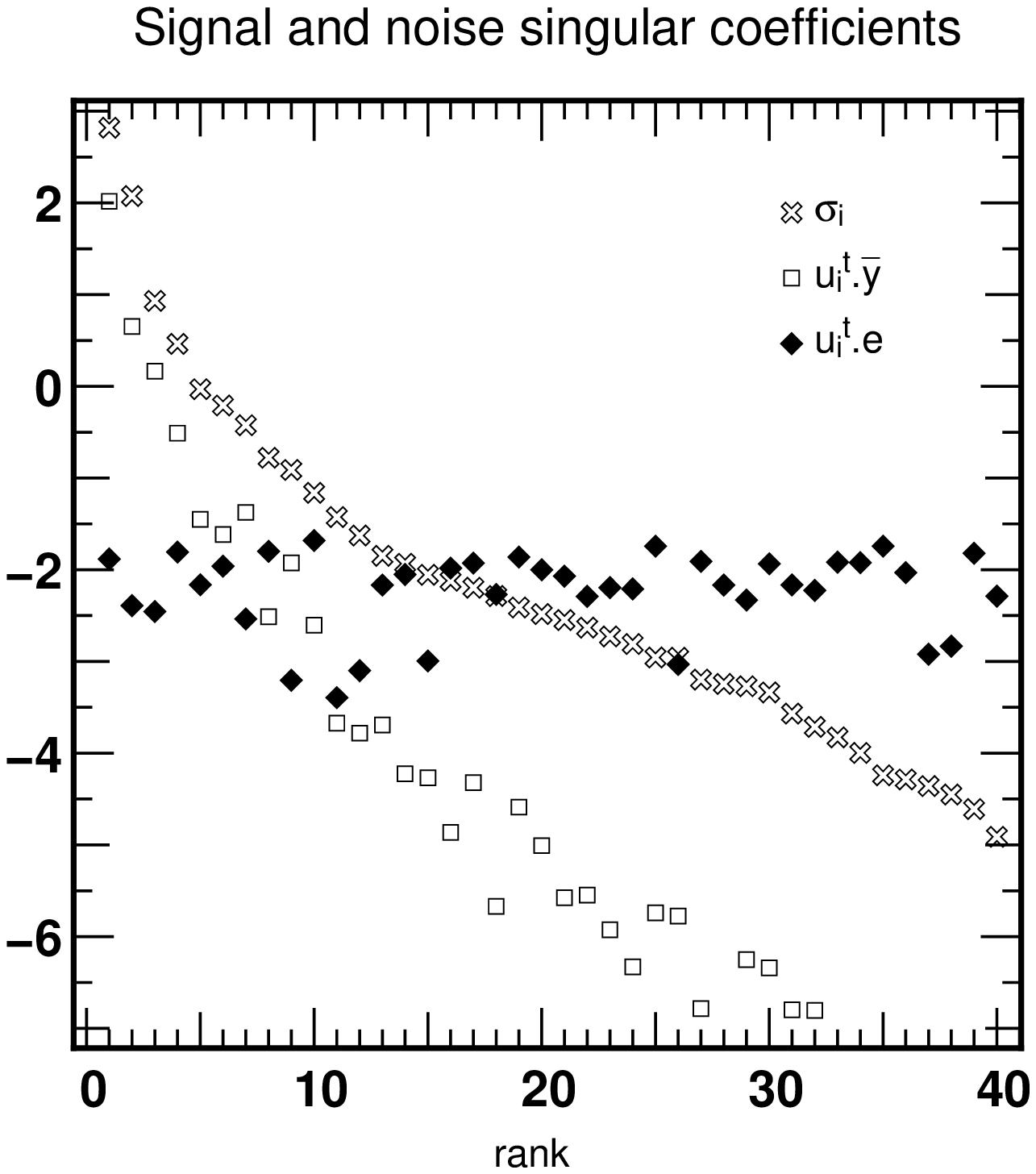}
  \end{center}
  \caption{The decay of the singular values of the kernel (crosses) is the
    origin of the bad behaviour of the problem, through the amplification
    of the last singular vectors.  In this example, the data $\M{y}$ is
    perturbed by Gaussian noise of constant $\mathrm{SNR_d}=100$ per pixel.  The
    unperturbed singular coefficients (white squares) decay, while the
    noise singular coefficients (black diamonds) remain spread around
    $1/\mathrm{SNR_d}$ for any $i$ (we chose
    $\langle\overline{\M{y}}\rangle=1$ in this example). The perturbed
    singular coefficients $\T{\M{u}_i}\cdot \M{y}$ are thus noise dominated
    as soon as $i \geq 7-9$, and so are the terms of the SVD solution
    (\Eq{svdsol}). The increasing difference between the true and noise
    singular coefficients is worsened by the division by smaller
    $\sigma_i$. The solution $\M{x}$ is dominated by the last few solution
    singular vectors, and its norm is purely noise dependent.  }
  \label{f:picard}
\end{figure}

First, let us see how ill-conditioning arises, in the case of a noiseless
kernel but with noisy data. We solve for $\M{x}$ by maximizing the
likelihood of the data $\M{y}$ given the model; this is the same as
minimizing:
\begin{equation}
  \chi^2(\M{y}|\M{x}) =
  \T{\bigl(\M{y} - \M{B}\cdot\M{x}\bigr)}\cdot\M{W}\cdot
     \bigl(\M{y} - \M{B}\cdot\M{x}\bigr) \, ,
\end{equation}
with respect to $\M{x}$.  The solution is the weighted least squares one:
\begin{equation}
  \M{x}_\mathrm{ML} =
  \bigl(\T{\M{B}}\cdot\M{W}\cdot\M{B}\bigr)^{-1}
  \cdot\T{\M{B}}\cdot\M{W}\cdot\M{y}\,.
\end{equation}
For sake of simplicity, we will consider stationary noise in this section.
The results of this section however apply for non-stationary noise by
replacing the model matrix $\M{B}$ by $\M{K}{\cdot}\M{B}$ and the data
vector $\M{y}$ by $\M{K}{\cdot}\M{y}$ where $\M{K}$ is the Choleski
decomposition of the weight matrix, i.e.\ $\M{W}=\T{\M{K}}{\cdot}\M{K}$.
For stationary noise, the weight matrix factorizes out:
\begin{equation}
  \chi^2(\M{y}|\M{x}) \propto
  \T{\bigl(\M{y} - \M{B}\cdot\M{x}\bigr)}
     \bigl(\M{y} - \M{B}\cdot\M{x}\bigr) \, ,
\end{equation}
and the maximum-likelihood solution becomes the ordinary least squares one:
\begin{equation}
  \label{e:xlsq}
  \M{x}_\mathrm{ML} =
  \bigl(\T{\M{B}}\cdot\M{B}\bigr)^{-1}
  \cdot\T{\M{B}}\cdot\M{y}\,.
\end{equation}
In order to clarify the process of noise amplification, we introduce the
singular value decomposition of $\M{B}$ as:
\begin{equation}
  \M{B} = \M{U} \cdot \M{\Sigma} \cdot \T{\M{V}} \, ,
\end{equation}
where $\M{\Sigma}=\diag(\sigma_{1}, \sigma_{2},\dots ,\sigma_{n})$ is a
diagonal matrix carrying the singular values, sorted in decreasing order,
of $\M{B}$ on its diagonal.  $\M{U}$ contains the orthonormal data singular
vectors $\M{u}_{i}$ (data-size vectors), and $\M{V}$ contains the
orthonormal solution singular vectors $\M{v}_{i}$ (solution-size vectors).
Replacing $\M{B}$ by its singular value decomposition in \Eq{xlsq} yields:
\begin{equation}
  \M{x}_\mathrm{ML}
  = \M{V}\cdot\M{\Sigma}^{-1}\cdot\T{\M{U}}\cdot\M{y}
  = \sum_{i=1}^{n} \frac{\T{\M{u}_{i}} \cdot \M{y}}{\sigma_{i}} \, \M{v}_{i} \, .
  \label{e:svdsol}
\end{equation}
The solution is obtained as the sum of $n$ solution singular
vectors $\M{v}_i$ times the scalar $\T{\M{u}_{i}}\cdot \M{y}/\sigma_i$.
For real data, we have $\M{y}=\overline{\M{y}} + \M{e}$, where the
noiseless data $\M{\overline{y}}$ is related to the true parameter vector
$\overline{\M{x}}$ via $\overline{\M{y}}={\M{B}} \cdot
\overline{\M{x}}$.  Instead of $\overline{\M{x}}$, the
solution recovered from the noisy data reads:
\begin{equation}
  \M{x}_\mathrm{ML} =
  \sum_{i=1}^{n} \frac{\T{\M{u}_{i}}\cdot\M{\overline{y}}}{\sigma_{i}}
  \, \M{v}_{i} +
  \sum_{i=1}^{n} \frac{\T{\M{u}_{i}}\cdot\M{e}}{\sigma_{i}}
  \, \M{v}_{i}
  \equiv \M{\overline{x}} + \M{x}_e \, .
  \label{e:svdsol2}
\end{equation}
Thus, we recover the true unperturbed solution $\M{\overline{x}}$ plus a perturbation,
$\M{x}_e$, related to the noise.  Comparing $\M{\overline{x}}$ and
$\M{x}_e$ is equivalent to comparing the unperturbed singular
coefficients ${\T{\M{u}_{i}} \cdot \M{\overline{y}}}$ and the noise
singular coefficients ${\T{\M{u}_{i}} \cdot \M{e}}$.
Figure~\ref{f:picard} shows an example with $40$ logarithmical age bins from
 $10$ Myr to $20$ Gyr, and where the data is perturbed by
Gaussian noise and has constant $\mathrm{SNR_d}=100$ per pixel (the
subscript d stands for data). The figure shows that the singular values
decay very fast and span a large range, giving a conditioning number,
defined by $\mathrm{CN}= \sigma_1/\sigma_n \approx 10^8$ characteristic of
an ill-conditioned problem. {Note that $\M{B}$ is the flux-normalized SSP 
basis defined by \Eq{favgb}, i.e. each spectrum of the basis has unitary flux,
and the $x_i$ are thus flux fractions and not mass fractions (see \Sec{fab}
for more details).} The noise singular coefficients remain rather constant for any rank $i$.  Indeed, $\T{\M{u}_{i}}\cdot \M{e}$
involves a normalized vector times noise, and has a constant statistical
{expected} value of $\langle \overline{\M{y}} \rangle /\mathrm{SNR_d}$.
On the contrary, the unperturbed singular coefficients decay. {In this
example, the model $\overline{\M{x}}$ is a Gaussian centered on
$1$ Gyr, and we find that changing the mean age of the model does not
significantly affect the decay of the ${\T{\M{u}_{i}} \cdot \M{\overline{y}}}$
{(see Appendix \ref{s:fid1})}.} We can
thus define two regimes, with a transition for $i_0 \approx 7-9$ in
this example:
\begin{itemize}
\item For $i\le i_0$ we have ${\T{\M{u}_{i}} \cdot \M{{y}}} \simeq
  {\T{\M{u}_{i}} \cdot \M{\overline{y}}}$ and the singular coefficients and
  modes are set by the unperturbed signal $\overline{\M{y}}$.
\item For $i > i_0$ we have ${\T{\M{u}_{i}} \cdot \M{{y}}} \simeq
  {\T{\M{u}_{i}} \cdot \M{{e}}}\simeq \langle\overline{\M{y}}
  \rangle/\mathrm{SNR_d}$. The singular coefficients are set by the noise
  in the data and saturate.
\end{itemize}
The division by  decreasing $\sigma_i$ makes the high rank terms in $\M{x}_e$
become very large. The solution $\M{x}$ is thus dominated by the last few
$\M{v}_i$. Its norm is several orders of magnitude larger than the true
solution.
We see that, for such ill-conditioned problems, pure
maximum-likelihood estimation results in huge noise amplification and useless
solutions.

\begin{figure}
{\includegraphics[width=0.9\linewidth,clip]{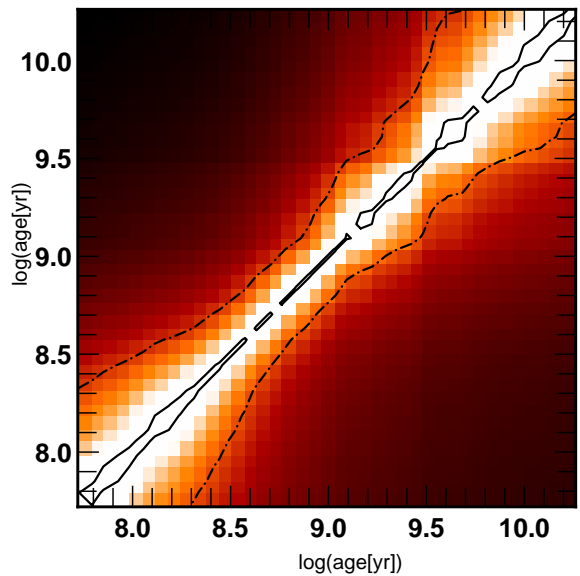}}
\caption{Distance map of the {\SED s} involved in the flux-normalized kernel $\M{B}$. The contours enclose a domain where
the $i$th spectrum can not be distinguised against the
$j$th at a $ 90 \% $ confidence level. The solid contour is for $\mathrm{SNR}=100$ per
pixel and the
dash-dotted one is for $\mathrm{SNR}=10$ per pixel. It is not possible to unambiguously
disentangle two spectra in such regions, i.e.\ the resolution in age of any
inversion method can not be finer than the width of these regions (which is
read on the axis), and it is not constant all along the age
range. This resolution in age in data space has a counter part in the
resolution defined in \Sec{agesep}
}
\label{f:chi2}
\end{figure}

The origin of ill-conditioning is, in most part, physical: it lies in the
evolution of the single stellar populations, which is dictated by stellar
physics and the relevant stellar evolution models. One aspect of the
situation is illustrated in \Fig{chi2}. It shows a map of the $\chi^2$
distances between the spectra (i.e.\ columns) of the kernel $\M{B}$, for
different SNRs.  In this figure, the time interval $[50 $Myr$, 15 $Gyr$]$
was arbitrarily divided in $40$ logarithmic age bins, and the SSP basis is flux
normalized as in \Eq{favgb}.  It shows that for low SNRs (of order 10), one
element of the basis can not be quantitatively distinguished from its
neighbours within a typical {log} age interval of $\sim$0.5 dex. It also
makes it clear that the logarithmic age-resolution of any inversion method
will not be constant all over the time range.

\subsubsection{Noisy correlated kernel}
\begin{figure}
\includegraphics[width=0.95\linewidth,clip]{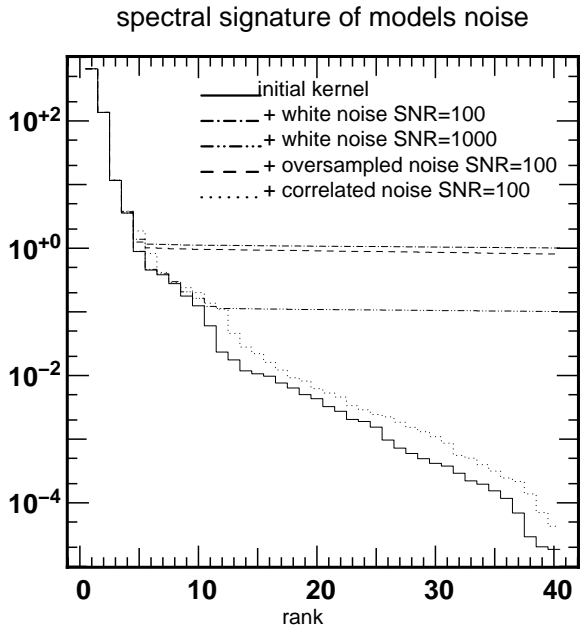}
\caption{Investigation of the noise signatures of the kernel.
For comparison, the kernel was noised in several different ways: with white
noise, oversampled noise and finally noise correlated in the age direction of
the kernel, each type of noise producing characteristic features in the
singular values. The expected spectral signature of the
noise in the initial basis (saturation of the singular values) does not
occur. This is likely to be caused by the interpolation between the stars of the stellar
library: the noise patterns are carried along in the interpolation, giving
rise to noise patterns correlated in the direction of ages.}
\label{f:SVstuff}
\end{figure}
\label{s:noisykernel}

As discussed in \Sec{problemsSSPs}, the models which are constructed from
observed spectra, are also contaminated by observational noise. Let us investigate the expected signature and basic properties of a noisy kernel.

PEGASE-HR {\SSPs} have a noise component estimated to
${\mathrm{SNR_b}}\approx 200$ per $0.2$ {\AA} pixel (the subscript b stands
for basis). From theoretical studies of random matrices \citep{hansen88}, it is known that a hypothetical noiseless {\SSP}
basis perturbated by adding white noise of root mean square $\sigma_0 $
should have its singular values settle around $\sqrt{m}\,\sigma_0$, where
$m$ is the number of samples in the observed {\SED}.  If the spectra are
normalized to unitary flux we have $\sigma_0\simeq1/\mathrm{SNR_b}$.
Figure~\ref{f:SVstuff} shows the singular values of the flux-normalized
kernel $\M{B}$ (thick line).  The singular values clearly do not settle
around the value expected for $m \simeq 10^4$, i.e.\ $\approx 1$ for
$\mathrm{SNR_b}=100$ (dash-dotted line) and $\approx 0.1$ for
$\mathrm{SNR_b}=1\,000$ (dash-double-dotted line). On the contrary, their
decay is typical of an ill-conditioned noiseless kernel, as if the {\SSP s}
involved had infinite SNR. Let us investigate some details of the synthesis
process, in an attempt to explain this unexpected property.

As every {\SSP} is actually the weighted sum of $p$ single stars from the
library, the noise level of the synthetic {\SED} should be lower (typically
divided by $\sqrt{p}$). However, the singular values of the kernel plus
white noise at a level $\mathrm{SNR}=1\,000$ (corresponding to summing
$p=100$ stars having $\mathrm{SNR}=100$) are still much larger than the
initial kernel's singular values. Having more stars available would lower
the saturation level, but one would need $10^{10}$ stars with
$\mathrm{SNR}=100$ to make the saturation vanish.

In order to test for the effect of wavelength resampling of the individual
stellar spectra, we
added $\mathrm{SNR}=100$ per pixel smoothed noise (i.e. noise with a
correlation between neighboring wavelengths) to the kernel. The corresponding singular values
are very similar to the former white noise case, except that they settle to
a slightly smaller value. They still saturate high above the singular
values of the initial kernel.

In contrast, when the added noise pattern is correlated in the direction of
ages instead of wavelength, one obtains a non-saturated singular value
spectrum very similar to the initial kernel, even with SNR as low as $100$
(a larger SNR would make it look even more similar).

Indeed, such correlated noise arises in part in the kernel because
individual stellar spectra are interpolated in $(T,g,Z)$ space.  

A single spectrum from the input stellar
library can thus significantly contribute to several ages. For instance,
the same limited number of red giants will be used (with slightly different
weights) to represent the red giant branch stars over a range of ages and
metallicities.  Their noise patterns will show up in several consecutive
synthetic {\SSP s}, and can therefore not be properly discriminated against
true physical signal. The expected saturation is washed out by the
interpolation between spectra, resulting in a degraded signature. This
correlation affects us in two ways: it prevents us from determining the
precise SNR of the basis, and then from computing the {\CN} of the real
problem (where $\mathrm{SNR_b} \rightarrow\infty$). Only a lower limit on
the {\CN} is obtained, meaning the real problem could actually be worse.

Whatever process is responsible for degrading the noise signature, the
properties of the problem in very high quality data regimes can not be
inferred from the apparently noiseless initial kernel $\M{B}$.
Let's return to the case of white noise, with a noisy
kernel $\M{B+E}$. Its singular values saturate at some rank $i_B$. The
singular vectors of lower rank are identical to those of $\M{B}$ but for higher
rank, they differ strongly.  Thus, the number of free parameters we can recover
can not be larger than $i_B$.  For {\PHR} we estimate $i_{B}=6$ for
$\mathrm{SNR_b} \approx 200$. This means that high frequency variations of
the {\SAD} are unreachable, no matter what is the SNR of the data. This is
a fundamental limitation of the problem, related specifically to the SNR of
the {\SSP} models.
When $\mathrm{SNR_{d}} \geq \mathrm{SNR_b}$, a pure maximum likelihood
estimation actually uses noise patterns inside the kernel as if it was true
physical signal, and simulations will give results with an illusory
accuracy. A useful technique, which explicitly accounts for modeling
errors, is then total least squares (hereafter TLS). The {\TLS} solution to
our linear problem (for simplicity we set $\M{W}$ to Identity here) is
defined by:
\begin{equation}
  \M{x}_\mathrm{TLS}=\arg\min_{\M{x}, \, \bar{\M{B}}}
  \, \left( \Vert \M{y}-\bar{\M{B}} \cdot \M{x} \Vert ^2 +
    \Vert \bar{\M{B}} - \M{B} \Vert ^2 \right) \, ,
  \label{e:TLS}
\end{equation}
where $\Vert\M{x}\Vert=\sqrt{\T{\M{x}}\cdot\M{x}}$ denotes the Euclidian
(or $\ell_2$) norm. More can be found in \cite{hansen96regularization} and
\cite{golub00tikhonov}.

However, in the rest of the paper, we will most frequently explore
regimes where the dominant error source is the data, so that the number of degrees
of freedom of the problem is dictated by $\mathrm{SNR_d}$ rather than
$\mathrm{SNR_b}$. It will also allow us to estimate what could be the
\emph{best} performance of the method, if the {\SSP} models were taken as
perfect. Thus, in the following sections, we will focus exclusively on
the treatment of noisy data, and will often drop the subscript ``d''.



\subsection{Regularization and MAP}
\label{s:reg}
This section explains how adequate regularization allows us to improve the
behaviour of the problem with respect to noise in the data. Perturbation of the solution arises from the noise-dominated higher rank
terms of \Eq{svdsol}. In order to ensure that $\M{x}_e$ remains small, one
could reduce the effective number of age bins. Several criteria are
applicable.
\begin{itemize}
\item The singular coefficients should always be dominated by the true
  signal. With plots such as \Fig{picard}, we find that $i_0$ is {between
  $7$ and $9$ }for $\mathrm{SNR_d}=100$ per pixel with {\PHR} {\SSP s}.
  Nevertheless, in a real situation only $\T{\M{u}_i}{\cdot}\M{y}$ is
  generally available, and $i_0$ is guessed from the rank for which the
  singular coefficients begin to saturate.
\item In the true signal dominated region, the singular coefficients
  decrease faster than the singular values. Inversely, singular
  coefficients decreasing faster than the singular values for any rank $i$
  guarantee the smallness of $\M{x}_e$.  This requirement is known as the
  discrete Picard condition. See \citet{hansen} for further details.
\item A useful criterion that does not require any plot involves choosing
  the number of age bins $n$ so that the conditioning number of the resulting
  kernel satisfies
  \begin{equation}
    \mathrm{CN}=\sigma_1/\sigma_n \lesssim \sqrt{m} \,\mathrm{SNR_d} \, ,
    \label{e:trunc}
  \end{equation}
  where ${m}$ is the number of pixels.
\end{itemize}
Note that this statement is SNR dependent.

Another way to prevent the noise component from being amplified into the
solution is to truncate the SVD expansion at some rank $i_{\mathrm{trunc}}$:
\begin{equation}
  \M{x}_\mathrm{TSVD} = \sum_{i=1}^{i_\mathrm{trunc}}
  \frac{\T{\M{u}_{i}} \cdot \M{y}}{\sigma_{i}}
  \,\M{v}_{i} \, .
\label{e:svdtrunc}
\end{equation}
This technique is known as truncated SVD (hereafter TSVD). The use of
this method dates back to \citet{TSVD1} and \citet{TSVD2}.  The truncation rank $i_{\mathrm{trunc}}$ can be chosen with the help
of plots such as \Fig{picard}

However, if the truncation is brutal, it will produce strong artifacts,
known as aliasing, which reflects the fact that higher frequencies are
projected onto a low frequency basis; the best fit leads to a non local
alternated expansion which rings.  Moreover, TSVD is best suited for
problems where a clear gap in the singular values is seen because in this
instance, the lower modes are well represented by the truncated basis.
Unfortunately, our kernel displays a smooth, continuously decreasing
spectrum of singular values.
This is very similar to the situation in image reconstruction.  When
deconvolution problems are addressed, the brutal truncation of the transfer
function (which corresponds to the singular coefficients of the point
spread function, hereafter PSF) results in the formation of strong
artifacts known as Gibbs rings.

Moreover, we here have an other degree of complexity
arising from the property that our problem is not shift-invariant. As a
consequence, the solution singular vectors are fairly unsmooth and
even more artifacts are expected as discussed in
\Sec{fab}.
In image deblurring, artifacts are reduced and reconstructions improved by apodizing
the Fourier transformed PSF (i.e.\ making it smoothly decrease to $0$), for example by Wiener
filtering.\footnote{Non quadratic penalty functions, such as $\ell_{1}$-$\ell_{2}$ penalties
which accomodate rare sharp jumps in the sought field, can also significantly reduce the effect of ringing.} In a similar manner, we wish to apodize the singular value spectrum
of the kernel
$\M{B}$.

We chose to regularize the problem by
imposing the smoothness of the solution through a penalizing
function. We define the objective function as
\begin{equation}
Q_{\mu}(\M{x}) \equiv -\frac{1}{2} \log( f_\mathrm{post})=
{\chi^2}({\M{s}}(\M{x})) + \mu  \, P(\M{x}) \, ,
\label{e:reg}
\end{equation}
which is a penalized $\chi^2$, where $P$ is the penalizing function: it has
large (small) values for unsmooth (smooth) $\M{x}$.  Adding the
penalization $P$ to the objective function is exactly equivalent to
injecting a priori information in the problem. We effectively proceed as if
we assumed a priori that a smooth solution was more likely than a rough
one. This is in part justified by the fact that any unregularized inversion
tends to produce rough solutions. If we identify $Q_{\mu}$ with the
expression of the logarithm of the maximum a posteriori likelihood
(\ref{e:logMAP}) we see that by building a penalization $P$ we have built a
prior distribution $f_{\mathrm{prior}}$
\begin{equation}
f_{\mathrm{prior}}({\M{x}})= \exp \left( - \mu P({\M{x}})  \right) \, ,
\end{equation}
omitting the normalization constant.  If $\mu=0$, the prior distribution is
uniform and contains no information. It is a pure maximum likelihood
estimation. If $\mu > 0 $ the prior probability density is larger for
smooth solutions, and we are performing a maximum a posteriori likelihood
estimation (MAP).

The smoothing parameter $\mu$ sets the smoothness requirement on the
solution. There are several examples of such regularizations in the
litterature (Tikhonov, least squares with quadratic constraint, maximum
entropy regularization \dots see \citet{pichsieb02} for a discussion). Here,
we define $P$ as a quadratic function of $\M{x}$, involving a kernel
$\M{L}$.
\begin{equation}
P(\M{x})=\T{\M{x}} \cdot \T{\M{L}} \cdot \M{L} \cdot \M{x} \, ,
\label{e:defP}
\end{equation}
If $\M{L}$ is the identity matrix $\M{I}_n$, then $P(\M{x})$ is just the
square of the Euclidian norm of $\M{x}$. To explicitly enforce a smoothness
constraint, we can use a finite difference operator
$\M{D}_2\equiv{\diag}_2 [-1,2,-1]$ 
that computes the Laplacian
of
$\M{x}$, defined in \citet{pichsieb02} by
\begin{equation}
\M{D}_2  \equiv    \left[
\begin{array}{rrrrrrrr}
 -1 & 2 & -1 &  0  & 0 & 0  & 0 & \cdots  \cr
0 & -1 & 2 & -1 & 0 & 0 & 0 & \cdots \\
0 & 0 & -1 & 2 & -1 & 0 & 0 & \cdots \\
0 & 0 & 0 & -1 & 2 & -1 & 0 &  \cdots \\
\cdots & \cdots & \cdots & \cdots & \cdots & \cdots & \cdots & \cdots
\end{array}
\right]\, \, \,
\,,
\label{e:l1l2}
\end{equation}
The objective function $Q_\mu$ is then quadratic and has an  explicit minimum:
\begin{equation}
  \M{x}_{\mu} \triangleq \widetilde{\M{B}} \cdot \M{y}
  =\Bigl(\T{\M{B}}\cdot\M{W}\cdot\M{B} + \mu\,\T{\M{L}}\cdot\M{L}\Bigr)^{-1}
   \cdot\T{\M{B}}\cdot\M{W}\cdot\M{y} \, ,
  \label{e:xtild}
\end{equation}
where $\widetilde{\M{B}}$ is defined here to be the regularized inverse
model matrix, whose properties we will investigate below.

We may now derive a more insightful expression for $\M{x}_{\mu}$ while
relying on the \GSVD (hereafter GSVD) of $(\M{B},\M{L})$
(assuming $\M{W}=\M{I}_m$ or using the Choleski square root of $\M{W}$).
According to Appendix~\ref{s:GSVD}, the regularized solution now writes:
\begin{eqnarray}
  \M{x}_\mu &=& \arg \min_{\M{x}} \left(  \Vert\M{B}\cdot\M{x}-\M{y}\Vert^2
  + \mu\,\Vert\M{L}\cdot\M{x}\Vert^2 \right)  \nonumber \, , \\
  &=& \bigl[\T{\M{B}}\cdot\M{B}+\mu\,\T{\M{L}}\cdot\M{L}\bigr]^{-1}\cdot
  \T{\M{B}}\cdot\M{y}\nonumber  \, , \\
  &=& \M{V}\cdot\bigl[\M{\Sigma}^2 + \mu\,\M{\Theta}^2\bigr]^{-1}\cdot
  \M{\Sigma}\cdot\T{\M{U}}\cdot\M{y}\nonumber \, , \\
  &=& \sum_{i=1}^{n} \eta_i\,\bigl(\T{\M{u}}_i\cdot\M{y}\bigr)\,
  \M{v}_{i} \, ,
  \label{e:ff}
\end{eqnarray}
where the filter factors $\eta_i$:
\begin{equation}
  \eta_i = \frac{\sigma_i}{\sigma_i^2 + \mu\,\theta_i^2}  \, ,
\end{equation}
depend on the type of penalization and the smoothness parameter $\mu$.  For
any quadratic penalization as in \Eq{defP}, the matrices $\M{U}$, $\M{V}$,
$\M{\Sigma}=\diag(\sigma_1,\sigma_2,\dots,\sigma_n)$ and
$\M{\Theta}=\diag(\theta_1,\theta_2,\dots,\theta_n)$ are given by the
{\GSVD} of the matrix pair $(\M{B},\M{L})$ (see Appendix~\ref{s:GSVD} for
details).  For the simple case of square Euclidian norm penalization, $\M{L}=\M{I}_n$,
the filter factors becomes:
\begin{equation}
  \eta_i=\frac{\sigma_i}{\sigma_i^2 + \mu} \, .
\end{equation}
We then have $\eta_i \approx 1/\sigma_i$ when $\sigma_i^2 \gg \mu$, and
$\eta_i \rightarrow 0$ for higher ranks (i.e.\ smaller singular values), so
that division by almost $0$ is avoided in high rank terms.  Thus, setting
$\mu$ actually sets the rank where the weights of the SVD solution
components begin to decrease. Note that the smooth cutoff (apodization) of
the singular values should allow us to recover models similar to relatively
high rank singular vectors provided that the weights associated to lower
rank vectors are small enough.  Small $\mu$ yield noise sensitive, possibly
unphysical solutions, whereas very large $\mu$ lead to flat solutions
whatever the data. The choice of $\mu$ thus appears as a critical step, and
should give a fair balance between smoothness of the solution and
sensitivity to the data.

\subsection{Setting  the weight for the penalty: $\mu$}
\label{s:mu}

The optimal weighing between prior and likelihood is a central issue in MAP
since it allows us to taylor the effective degree of freedom of each
inversion to the SNR of the data.  See, e.g., \cite{Titterington1985} for
an extensive comparison between various methods for choosing the value of
the hyper-parameter $\mu$.

\subsubsection{The automatic way: generalized cross validation}
\label{s:gcv}

\begin{figure*}
\includegraphics[width=0.3\linewidth,clip]{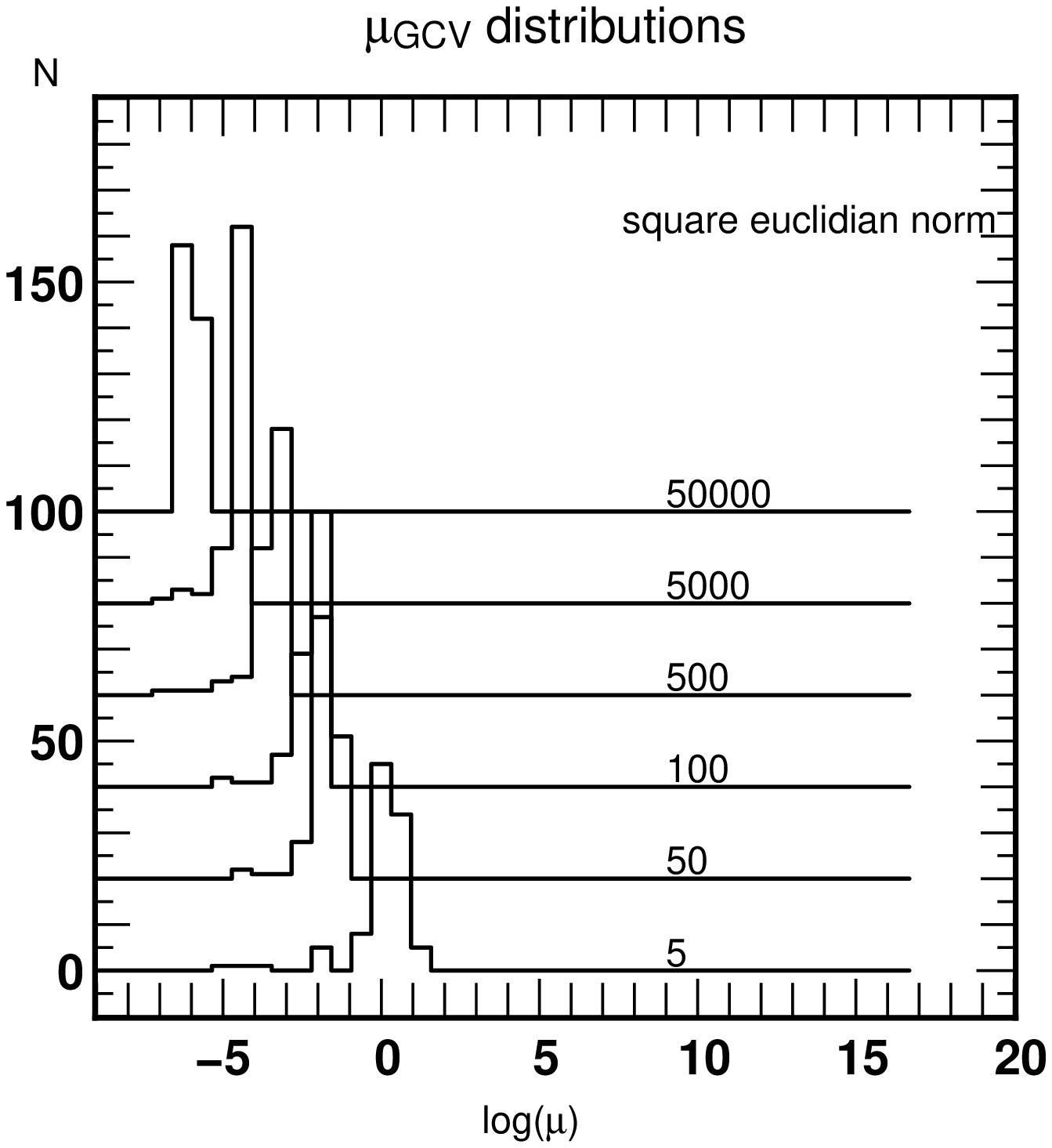}
\includegraphics[width=0.3\linewidth,clip]{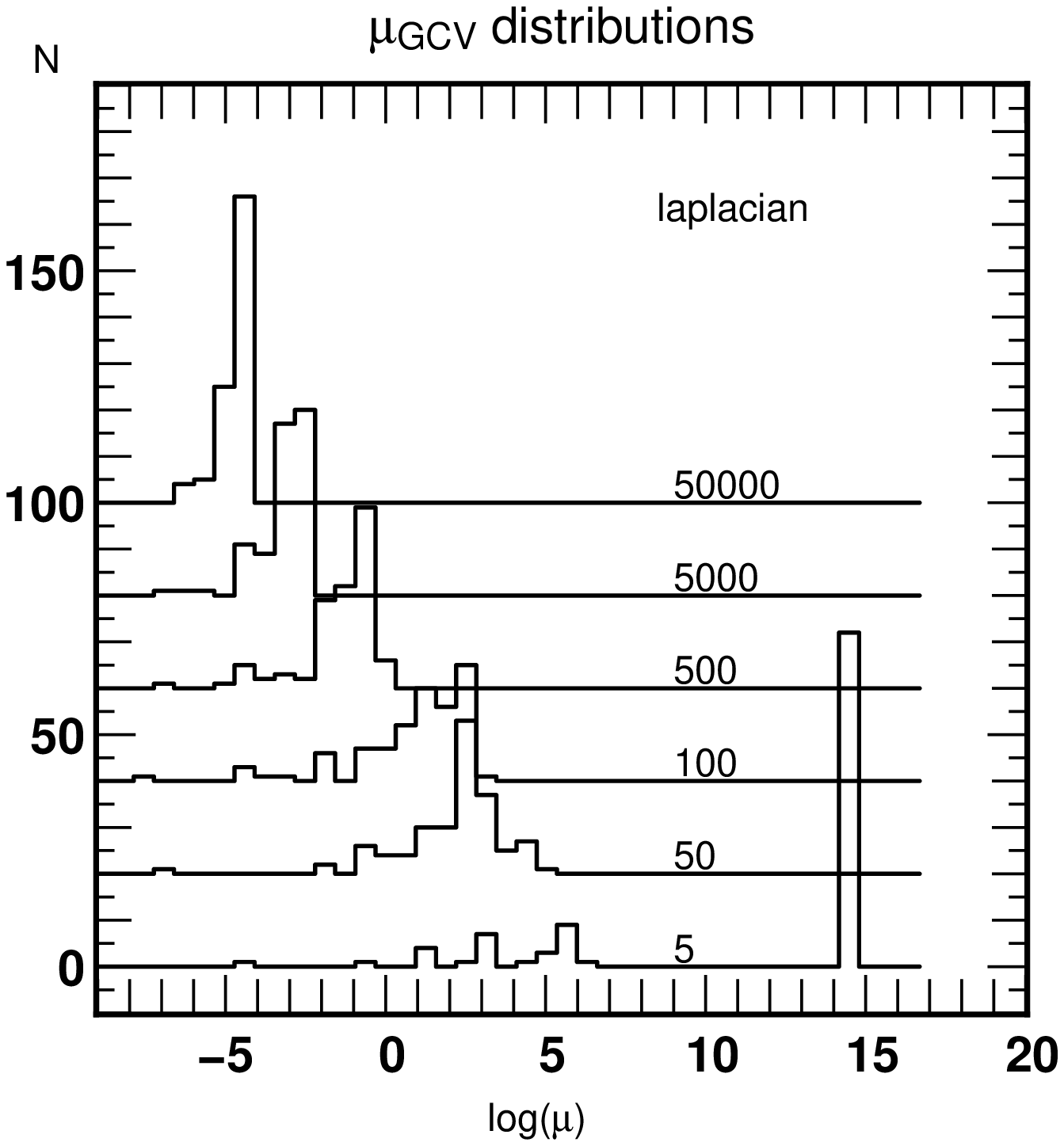}
\includegraphics[width=0.3\linewidth,clip]{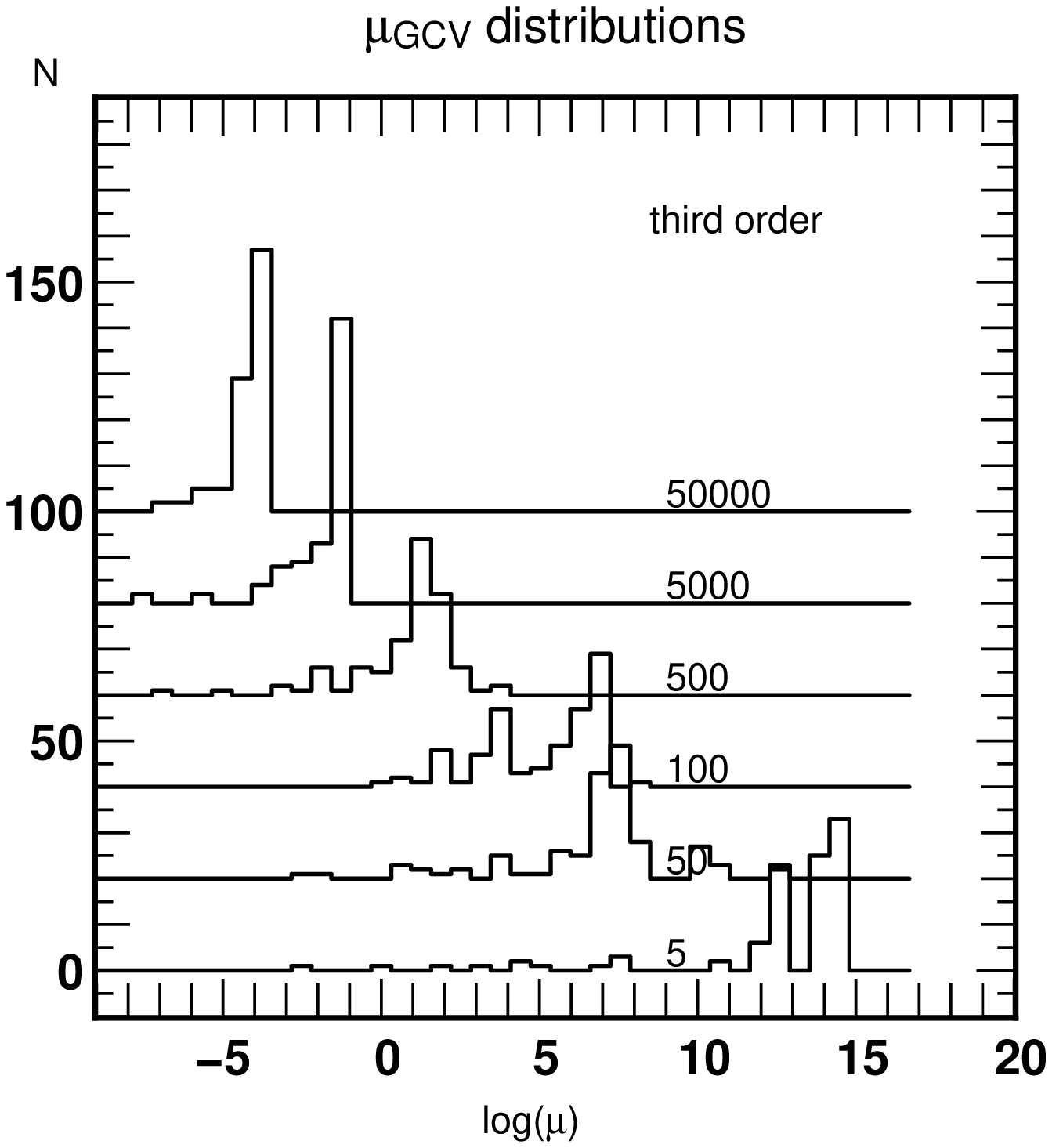}
\caption{Histograms of the distribution of $\mu_{\mathrm{GCV}}$ for a
  linear stellar age distribution inversion with 60 age bins, and several
  SNR per pixel and penalizations. From left to right: Euclidian
  norm, Laplacian and $\M{D}_3$ penalizations. The distributions are
  vertically offset for readility, and the SNR is given for each of them.
  The median of these distributions give the GCV-optimal smoothing
  parameter for each SNR and penalization. It is well defined in all cases
  except for very low $\mathrm{SNR}=5$ per pixel. The median parameter
  increases with the order of the penalization and decreasing SNR. Note the
  skewed distributions (this is quite generic in GCV).}
\label{f:gcv}
\end{figure*}

Generalized cross validation (GCV) is a function of the parameter $\mu$,
the data and the kernel $\M{B}$, defined as
\begin{equation}
  {\mathrm{GCV}}(\mu) =
  \frac{\bigl\Vert \bigl(\M{I} - \M{B}\cdot\tilde{\M{B}}\bigr)
    \cdot\M{y}\bigr\Vert^2
  }{%
    \tr^2\bigl(\M{I} - \M{B}\cdot\tilde{\M{B}}\bigr)
  } \, ,
\end{equation}
where $\tilde{\M{B}}$ is the regularized inverse model, defined by
\Eq{xtild} and $\tr(\cdot)$ is the trace of its argument. The minimum of GCV
optimizes the predictive power of the solution \citep{wahba}, in the sense
that if any pixel is left out of the data, this pixel's value should still
be well predicted by the corresponding regularized solution. For quadratic penalizations, one may
obtain very simple expressions for the GCV function, speeding up its
computation, and therefore the determination of $\mu$ by several orders of
magnitude.  Using the GSVD of $(\M{B},\M{L})$, we can derive:
\begin{equation}
  \mathrm{GCV}(\mu) =
  \frac{\sum_{i=1}^{n} \left(\rho_i \,\T{\M{u_i}} \cdot\M{y}\right)^2}
       {\left(\sum_{i=1}^{n} \rho_i \right)^2} \, ,
\label{e:gcvsvd}
\end{equation}
where
\begin{equation}
  \rho_i = 1 - \frac{\sigma_i^2}{\sigma_i^2 + \mu\,\theta_i^2}
   = \frac{\mu\,\theta_i^2}{\sigma_i^2 + \mu\,\theta_i^2}\,.
\end{equation}
where $\sigma_i$ and $\theta_i$ are the singular values obtained from the
{\GSVD} of the matrix pair $(\M{B},\M{L})$ (see Appendix~\ref{s:GSVD}).
Note that the $\mu$ in the denominator of $\rho_i$ factorizes out in the
expression of $\mathrm{GCV}(\mu)$.

When available, the minimum of GCV provides a good, data quality-motivated
value for $\mu$. Moreover, GCV has been extendedly tested and applied by a
number of authors, in several fields of physics.  Figure~\ref{f:gcv} shows
distributions of $\mu_{\mathrm{GCV}}$ for a mono-metallic inversion for
several SNR and penalizations. Each histogram results from $150$
experiments. The GCV determination of the smoothing parameter is successful
over a wide range of SNR, in the sense that the histogram shows a clear
maximum. This maximum is best defined for the Tikhonov penalization (square
of the Euclidian norm).  With laplacian and higher order penalizations,
especially for low SNR, the GCV values are more widely spread.
Nevertheless, we can still obtain a useful value by extrapolating the
higher SNR $\mu$ down to the desired SNR.

\subsubsection{Empirical approach: trial and error}

GCV and most of the automated smoothing parameter choice methods were
designed for linear problems. In the case of non-linear problems, it can
provide a useful value for $\mu$ to start with, but fine empirical tuning
is also required \citep{craig-brown}.  For instance, when positivity is
imposed through reparameterization or gradient clipping, $\mu$ should be
smaller than $\mu_{\mathrm{GCV}}$. Indeed, since the positive problem has a
better behaviour than the full linear one, it is expected that GCV
overestimates $\mu$. One can thus afford to lower it to some extent without
threatening the relevance of the solution. As a consequence, finer
structures can be recovered.  To set $\mu$ for the positive problem, we
used the simple following procedure. First, we set
$\mu=\mu_{\mathrm{GCV}}$.  We produce mock data, and perform successive
inversions, while decreasing $\mu$. As a consequence, finer structures are
recovered. At some point, we will enter a regime where the structures of
the solution can be identified as artifacts. This transition defines a
lower limit above which $\mu$ should remain.

\begin{figure*}
\begin{center}
\rotatebox{-90}{\resizebox{9cm}{18cm}{\includegraphics{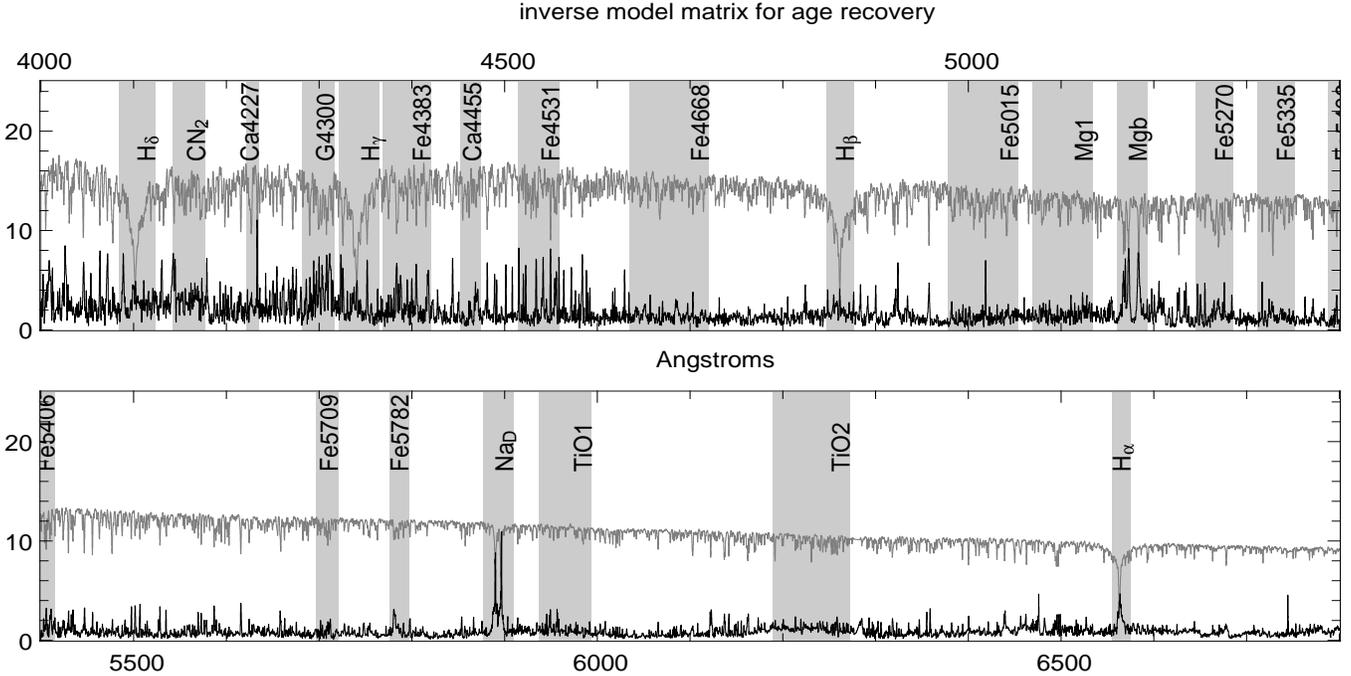}}}
\caption{Black solid line: peak to peak variations of the inverse model
matrix discussed in \Sec{invmodel}. In this example, we took $60$ age
bins and $\mu=10^2$ corresponding to $\mathrm{SNR}=100$ per pixel with Laplacian penalization. Large
values point at age-sensitive parts of the spectrum. A $500$ Myr {\SSP} with
half-solar metallicity is shown as reference (grey solid line).
The spectral domains corresponding to the Lick indices appear as grey-shaded areas.
Many of the spectral domains
involved in the Lick system seem to effectively
carry more information than the rest of the spectrum. However, the information
is still widely distributed along the whole optical range.}
\label{f:invmodel}
\end{center}
\end{figure*}

\subsection{Where is the age information?}
\label{s:invmodel}

Which spectral domains or lines are most discriminative in terms of
population age-dating? An answer to this can be given by inspecting the
properties of the regularized inverse model matrix $\tilde{\M{B}}(\mu)$
defined by \Eq{xtild}.  In effect, we expect the peak to peak amplitude of
a column of $\tilde{\M{B}}(\mu)$ to be largest for the most discriminatory
wavelengths for age-dating.
In \Fig{invmodel}, the inverse model matrix was computed for a Laplacian
penalization with $\mu_{\mathrm{GCV}}=10^2$ corresponding to
$\mathrm{SNR}=100$ per pixel with $60$ age bins from $10$ Myr to $20$ Gyr
and half-solar metallicity. It shows that the Balmer lines
${\mathrm{H}}_{\alpha,\beta,\gamma,\delta}$, along with the spectral
regions of the Lick index NaD, the magnesium indices ${\mathrm{Mg}}_1,
{\mathrm{Mg}}_2, {\mathrm{Mg}}_b$ and the calcium ${\mathrm{Ca}}_{4227}$
have strong weight in the age-dating process.  Note that the above analysis
is clearly noise dependent via $\tilde{\M{B}}(\mu_\mathrm{GCV})$.  The list
of relevant lines will change with the SNR. Many of the wiggles
and peaks of the inverse model remain so far uninterpreted, and many peaks
hit spectral domains where no referenced index is known, but still
contribute strongly to age separation. Another important feature of the
inverse model is that most of its norm is in the form of low value pixels.
If some of the peaks were $2$ or $3$ orders of magnitude larger than the
average value, we could conclude that most of the information is contained
exclusively in the corresponding lines. Yet, the \Fig{invmodel} does not
allow us to reach this conclusion. Even though the information seems denser
in the strongest, well known lines, most of it remains in the form of a
large number of weaker lines, more concentrated in the blue part of our
spectra. This supports the intuition that a lot of information is left
aside by looking exclusively at spectral indices, and that the constraints
obtained therefrom are not optimal.  Hence our effort to build a global
spectrum fitting tool.

\section{Validation: Behaviour of the linear inversion}
\label{s:valid}

Let us now apply {\STECMAP} to mock data, to study the biases and the
dispersion of the solutions, and to test for different penalizations. Producing mock data involves choosing a model age
distribution, $\M{x}_{M}$, and a noise model, $\M{e}$. A mock spectrum is
then obtained as $\M{y}=\M{B} \cdot \M{x}_{M} + \M{e}$.  The corresponding
astrophysical goal is the recovery of the star formation history of
mono-metallic stellar populations (for example superimposed clusters) seen
without extinction. The {\SAD} models for these objects are single
(\Sec{agebump}) or multiple (\Sec{agesep}) star formation episodes of
approximately Gaussian shape. Recall that no assumption on the shape of
the distribution is included in the inversion process. The only a priori is
the smoothness of the solution, while the smoothing parameter is set by
GCV.  {Here} we relate the results of our simulation to the properties of
the solution singular vectors, thereby explaining the generation of
artifacts.

\subsection{Single bump \SAD}
\label{s:agebump}

Let us discuss in turn the relationship between the artifacts of the reconstructions
and the shape of the solution vectors (\Sec{artefact}), the flux-averaging of
the basis and the behaviour of the problem regarding the fiducial model (\Sec{fab}), the choice of penalization  (\Sec{laplacian}), the
requirement to impose positivity  (\Sec{pos}), and the need for an extensive simulation campaign (\Sec{campaign})
\subsubsection{Artifacts and the shape of the solution vectors}
\label{s:artefact}

\begin{figure}
\begin{center}
\includegraphics[width=0.9\linewidth,clip]{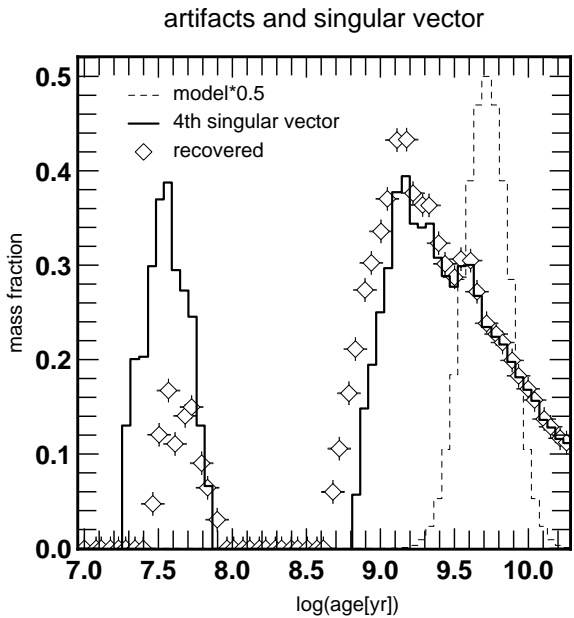}
\end{center}
\caption{Blow-up of the bottom right panel of \Fig{1bsim} showing only the
  mass reconstruction of the oldest bump. The dashed line is the model
  distribution, and the diamonds show the median of the recovered age
  distributions for $10$ realizations. The error bars showing the
  dispersion are smaller than the symbol itself. The details of the shape
  of the mass distribution reconstruction trace closely the $4$th singular
  vector of the kernel $\M{B}$, with very little dispersion, showing that
  the artifacts and the fine structures of the reconstructions are closely
  related to the properties of the {\SSP} models.}
\label{f:bu}
\end{figure}

Since any solution is a linear combination of the solution vectors
$\M{v}_i$ (see \Eq{ff}), their shapes impose what kind of shape for $\M{x}$
can or can not be reconstructed, depending on what feature in the observed
spectra is best matched by the corresponding data singular vectors.

\begin{figure*}
\begin{center}
\begin{tabular}{ll}
\includegraphics[width=0.4\linewidth,clip]{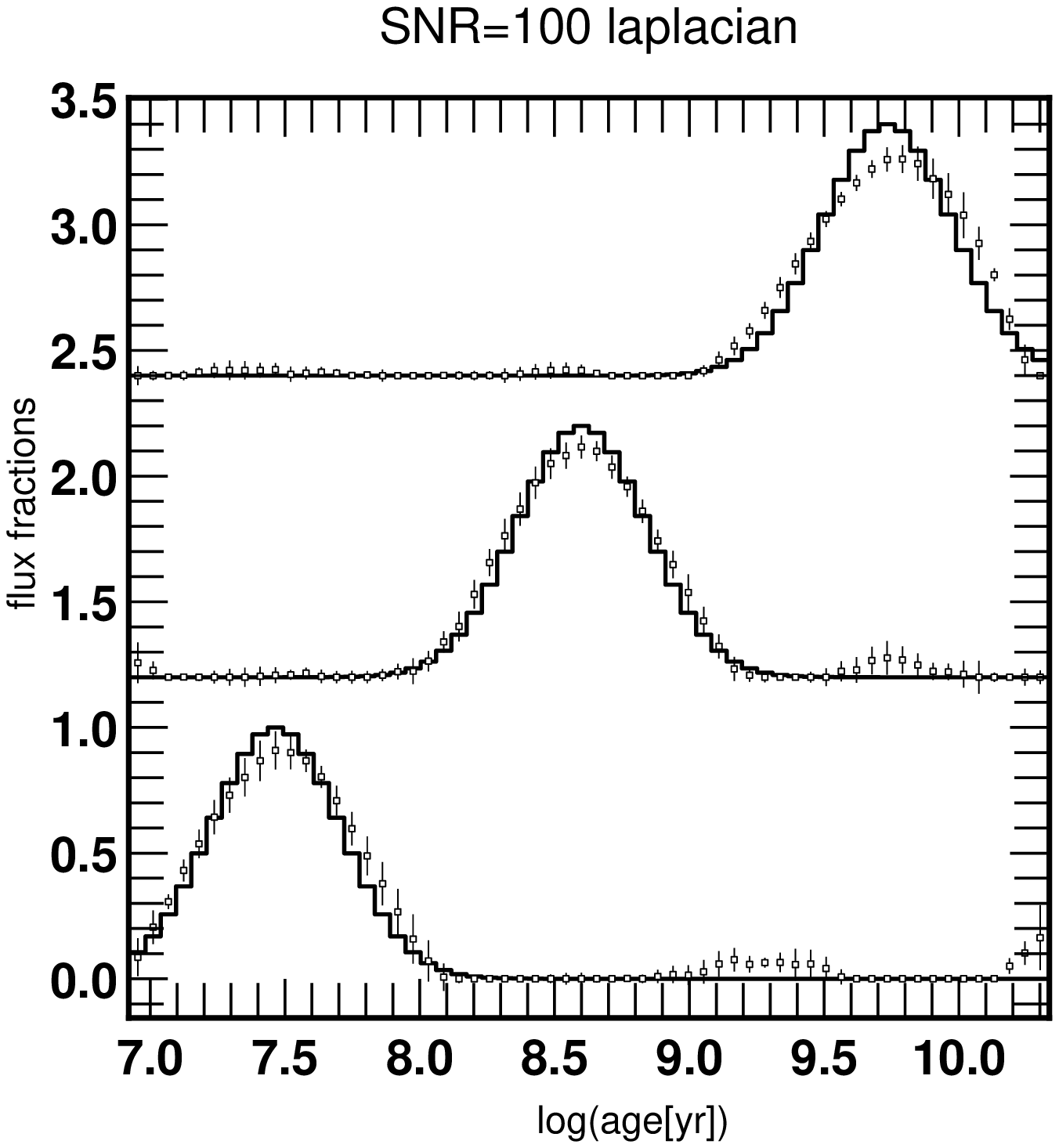} &
{\includegraphics[width=0.4\linewidth,clip]{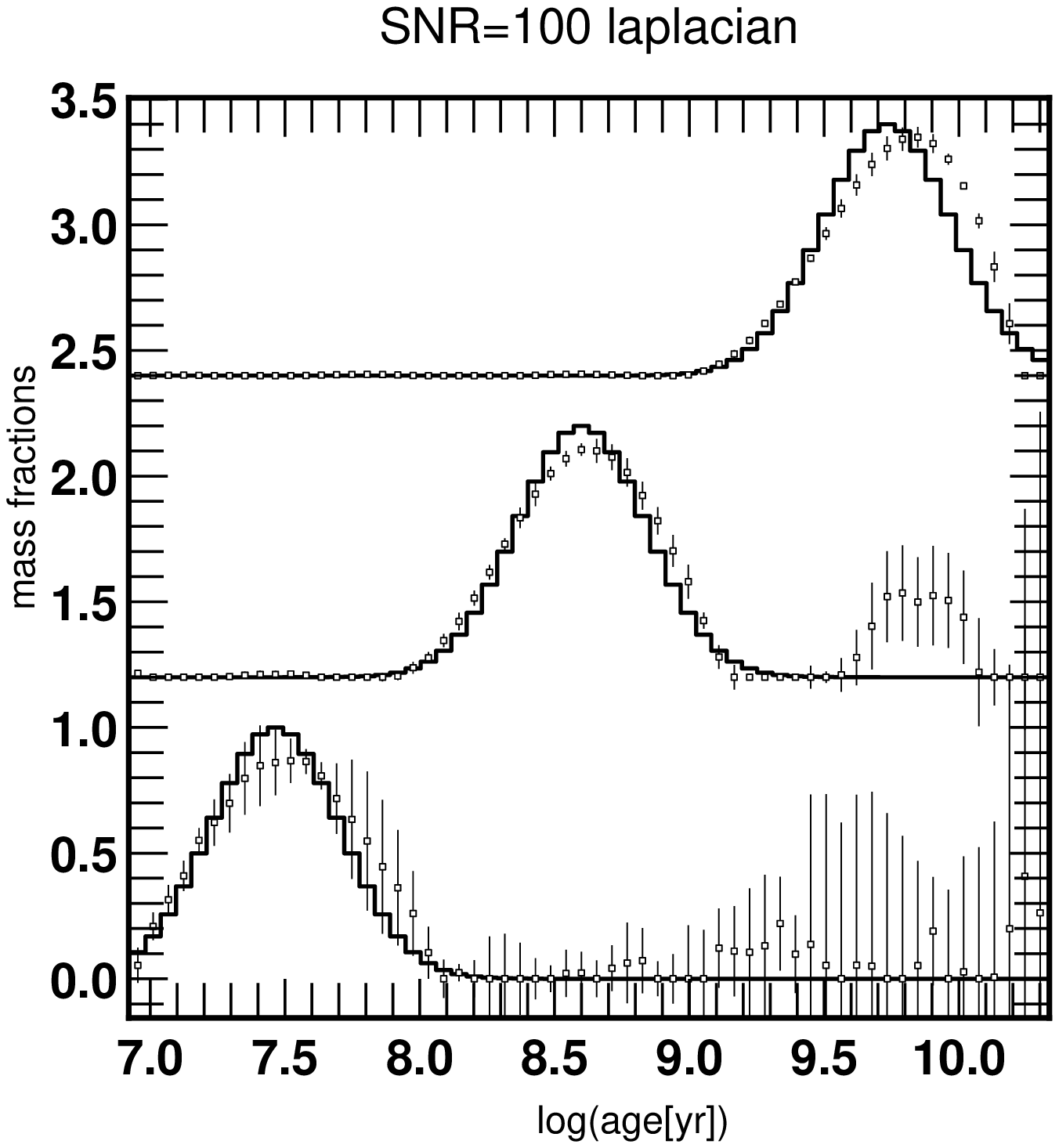}}
\\
{\includegraphics[width=0.4\linewidth,clip]{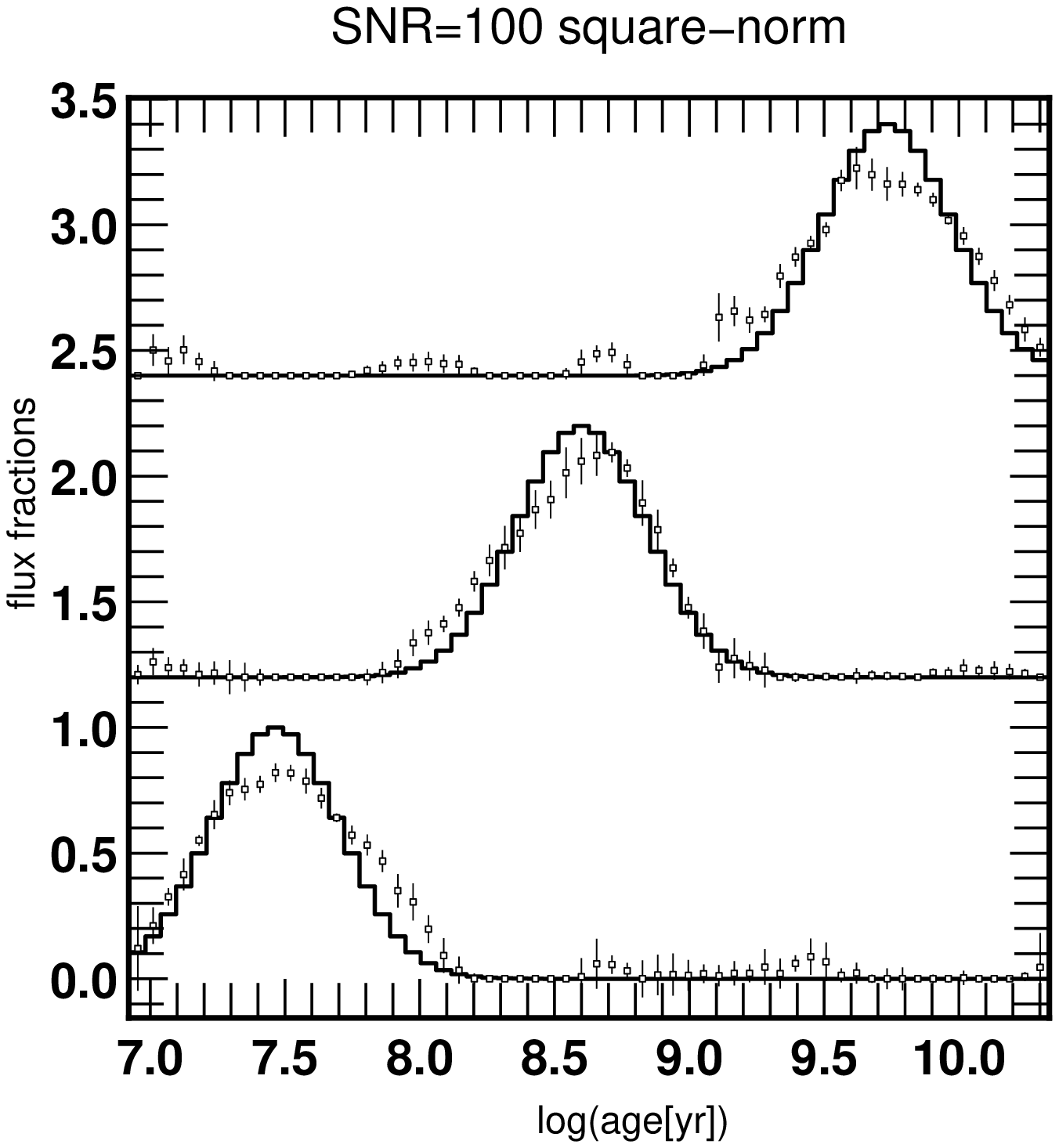}} &
{\includegraphics[width=0.4\linewidth,clip]{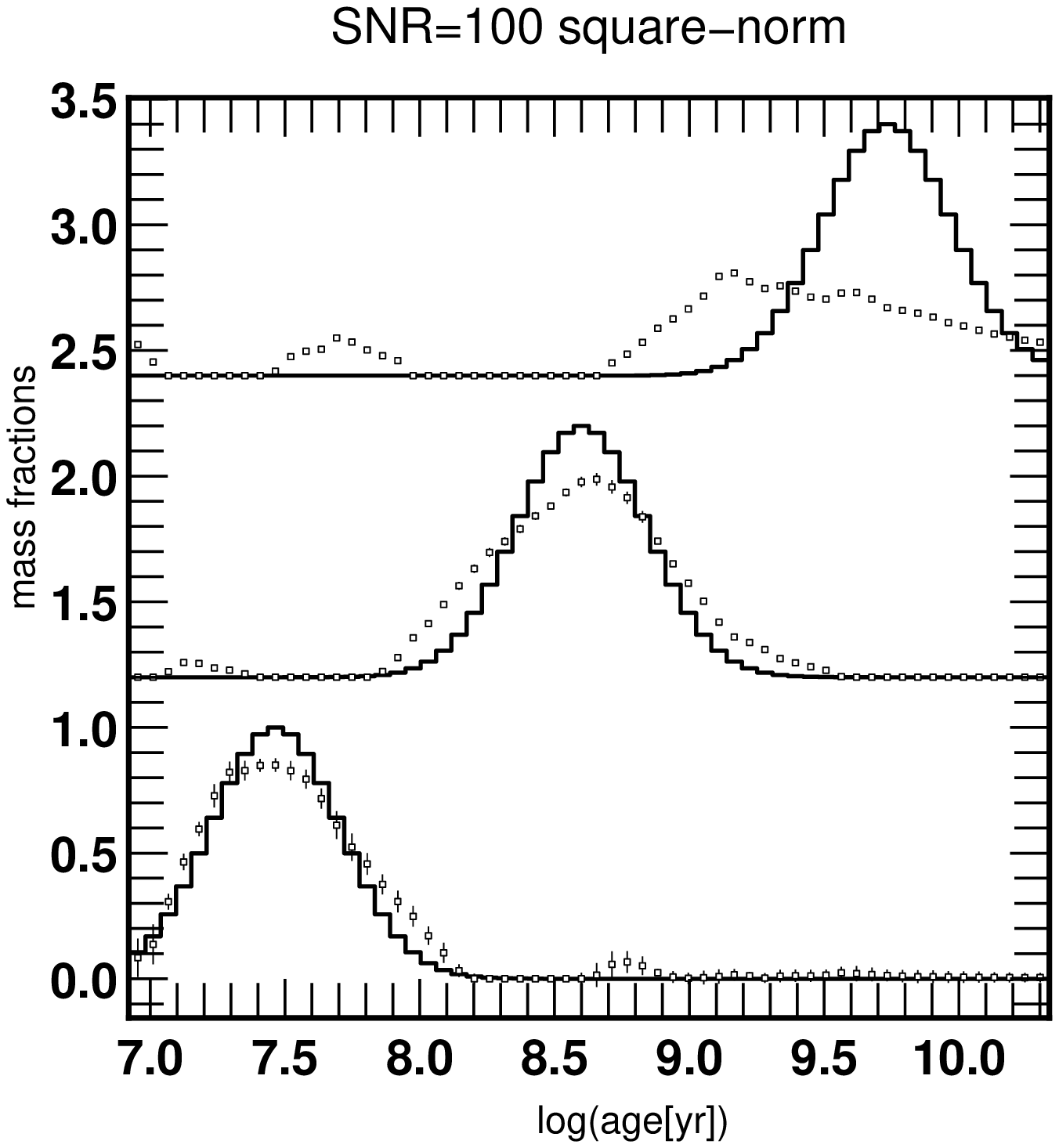}}
\end{tabular}
\end{center}
\caption{Simulations of the reconstruction of a young, intermediate and old
single burst populations. The thick
  histograms represent the models, while the symbols and vertical bars show the median and interquartiles of 10 inversions. Negative
values in these reconstructions have been set to zero for clarity.
\emph{Right}: Case of an absolute flux basis. The plots thus represent mass
fractions. \emph{Left}: Case of a flux normalized basis. Thus are represented
flux contributions. The SNR is fixed to $100$ per pixel with $R=10\,000$.
The penalizations are square Euclidian norm (bottom) and Laplacian (top).
In terms of distance to the model, the bumps are best reconstructed in flux
fractions, and the best penalization is Laplacian. We checked that
Laplacian penalization gave flux fraction reconstructions similar to the
third order penalization, showing that these do not strongly rely on the
details of the smoothness a priori.}
\label{f:1bsim}
\end{figure*}

Moreover, as regularizing the problem involves attenuating the high rank
terms of \Eq{ff}, the detailed shape of the solution is in general given by
the first few $\M{v}_i$. Figure~\ref{f:bu} shows the stellar mass
distribution reconstruction of an old population. It is actually a blow-up
of the recovery of the oldest burst in the bottom right panel of
\Fig{1bsim}. The penalization is square Euclidian norm, so that the relevant
singular vectors are given by the SVD of $\M{B}$.
The details of the solution are mostly those of the $4$th
solution singular vector, and appear as a systematic artifact (the diamonds
are the median of $10$ realizations, and the dispersion of the solutions is
smaller than the symbol itself). The spurious young component between
$10^{7.5}$ and $10^8$yr seems to be related to the 4th singular vector as
well, and also appears systematically even though it has no physical reality.
The fine structure and the artifacts of
any solution thus rely most on the properties of the {\SSP} basis rather
than on the data or even the realization of the noise.

It is generally impossible to reconstruct accurately the shape of the
distribution for ages where the singular vectors display no structure. The
right panel of \Fig{soleig}, shows that the $10$ first singular
vectors of the absolute flux kernel have very little structure for ages
larger than $\simeq 3$ Gyr. Correspondingly, the right panels of
\Fig{1bsim} show that indeed, in this range of ages, the shape of
the distribution is very poorly constrained.

For an inversion problem to be well behaved, the first solution singular
vectors, the $\M{v}_k$, should be rather smooth. They should display more
and more oscillations as the rank $k$ increases (typically $k-1$
oscillations), but remain smooth and regular. The unsmooth aspect of our
singular vectors arises from the temporal roughness in the spectral basis.
This could also be related to physical fast evolution of the {\SSP s} in
some specific stages of stellar evolution, producing variable distance
between the elements of the basis. It also reflects the non
shift-invariance of the problem, as is
also illustrated by \Fig{chi2}.


Some further artifacts can however not be trivially explained by the
solution singular vectors alone. For example many of the displayed
solutions, even with high SNR, show variations far away from the bulk of
the signal, seen as misleading spurious secondary bumps. This artifact is
the analog of Gibbs rings in imaging. It arises because the higher
frequency modes needed to suppress these secondary oscillations are
attenuated by regularization, and would be best identified by examining the
GSVD of $(\M{B},\M{L})$. It is the old age extension of the low
frequency mode involved in building the main bump. We will deal with this
by introducing positivity in \Sec{pos}.

\subsubsection{Flux-normalized basis and independence from the fiducial model}
\label{s:fab}

\begin{figure*}
\begin{tabular}{ll}
\includegraphics[width=0.45\linewidth,clip]{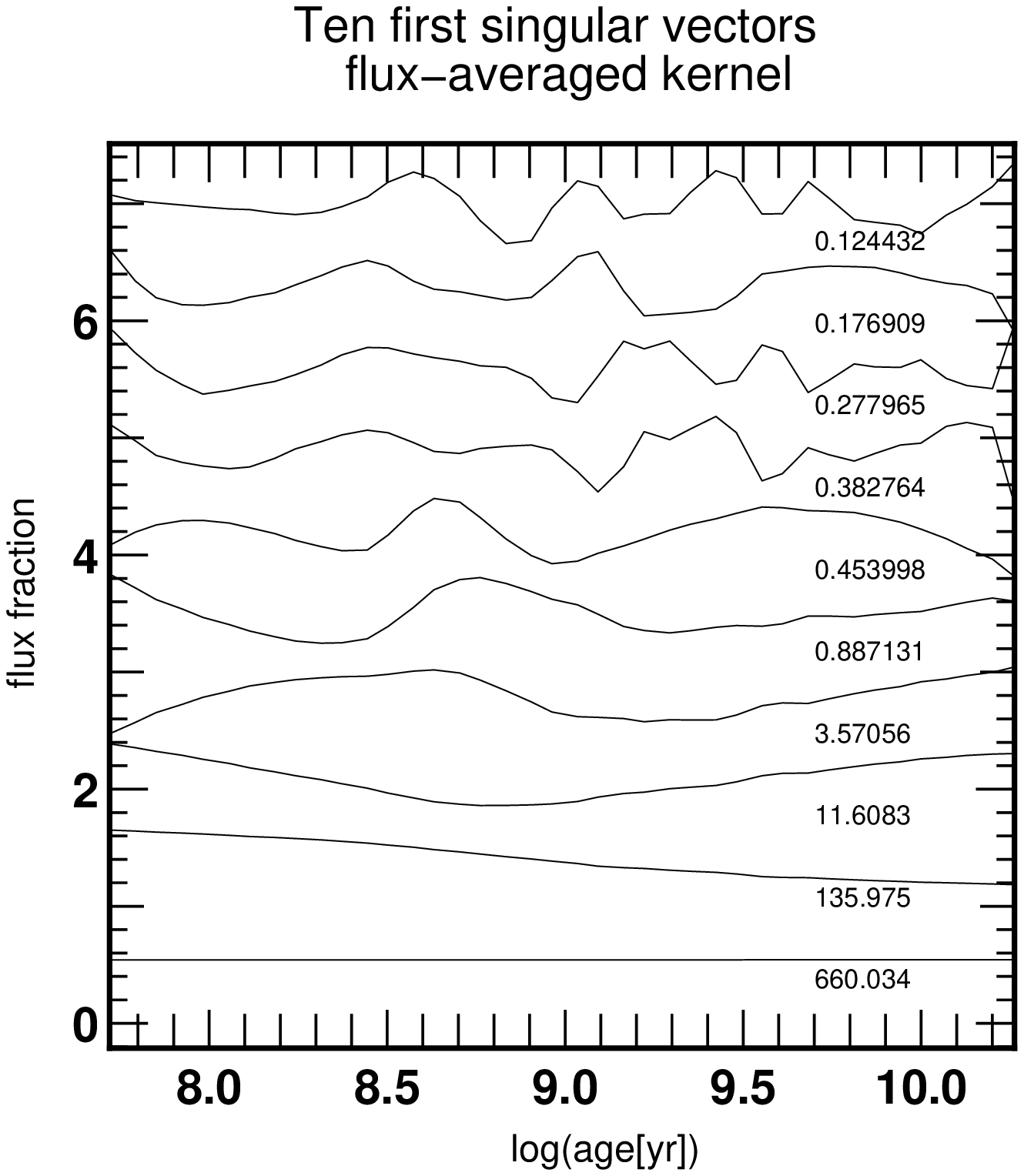}&
\includegraphics[width=0.45\linewidth,clip]{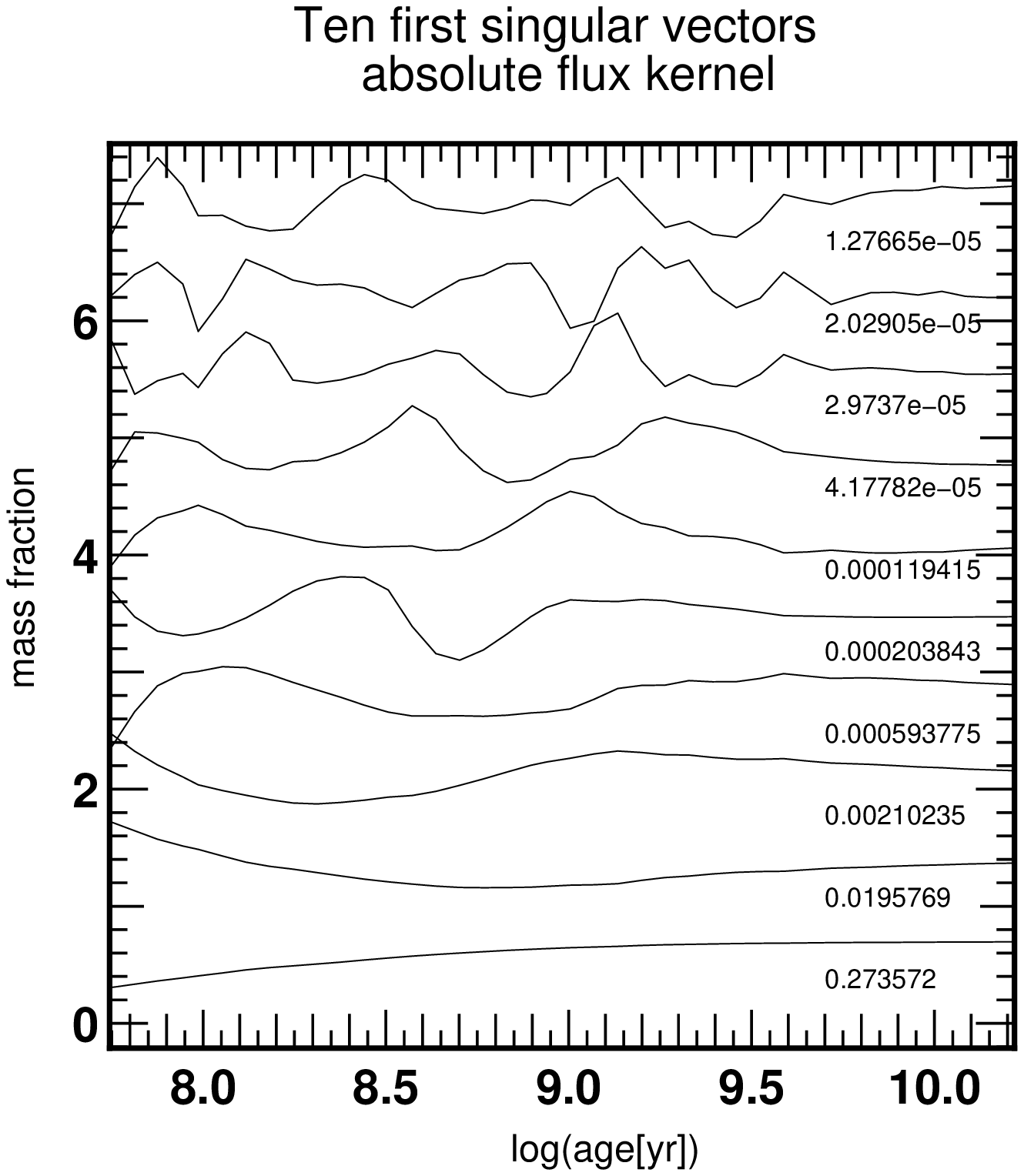}
\end{tabular}
\caption{Solution singular vectors of the flux-normalized kernel (left) and
  the absolute flux basis (right). The vectors are vertically offset for
  lisibility, and the associated singular values are given on the right.
  The low rank singular vectors of the absolute flux basis are very flat in
  the large ages, indicating that no information about these populations
  can be obtained unless we have very high SNR. On the contrary,
  fluctuations in large ages are already present in the low rank singular
  vectors of the flux-normalized basis, which indicates the better
  feasibility of reconstructing the age distribution in the older part.}
\label{f:soleig}
\end{figure*}
In practice, one can  choose between a basis where the flux of each
{\SSP} is given for $1 M_{\sun}$ (absolute flux basis or mass-normalized basis), and a basis
where the flux of each {\SSP} has been normalized to the same value (or
flux-normalized basis, cf.\ \Sec{basis}). This choice has a physical
meaning: in the first case, the unknown $\M{x}$ will contain mass
fractions, whereas in the latter case, it will contain flux fractions.



There are several reasons why we prefer to work with the
flux-normalized basis. 

It is more directly linked to the luminous properties of the observed
population (and thus less directly linked to the mass): a component of a given
flux can not ``hide'' behind another component of similar flux. This is not
true for components of similar  masses, due to the evolution of $M/L(t)$. 
For instance, in the upper right plot of \Fig{1bsim}, the mass of the older
components is poorly constrained when the model is a young burst. This is
expected, because when a young component is present, adding the
same mass of old stars will have very little effect on the integrated
optical light. 
This is predictable from the lack of
structure beyond $3$ Gyr in the singular vectors of the right panel of
\Fig{soleig} (see also the discussion in
\Sec{artefact}). Modulations in this range of ages are seen in the vectors
of the right panel for the higher rank vectors only.
On the other hand, the singular vectors of the flux-normalized basis (left
panel of \Fig{soleig}) display structure in the large ages even for low
ranks, indicating a better behaviour. And indeed, the upper left plot of \Fig{1bsim} shows that all the flux
fractions are satisfactorily constrained no matter if the model population is
young or old. 
In this respect, the 
``separability'' issues tackled later in the article for superimposed
populations (\Sec{agesep}) are more easily discussed in terms of flux
fractions.
Note that it is however not expected that the mass fractions obtained by multiplying
the flux fractions by $M/L(t)$ be accurate over the whole age range
(positivity will improve this particular aspect significantly; see \Sec{pos}).




The difference of behaviour between the mass and flux fractions reconstructions
is also reflected in the variation of the
transition rank $i_0$ (see \Sec{ill}) between the noise and signal dominated regimes, as shows \Fig{fid2}.
For a mass-normalized basis, the transition rank $i_0$
increases with the age of the fiducial model $\overline{\M{x}}$ (as
defined in \citet{panter1}), from 5 to 20. On the
other hand, for a flux-normalized basis, the transition rank remains
around 7-9 in this pseudo-observational setup, no matter the age of the
fiducial model. 
Ideally, we would like to come up with a problem whose behaviour is fixed only by the SNR. In this respect, independence of the
transition rank $i_0$ from the fiducial
model is a welcome property. We thus chose to carry on
with the flux-normalized basis for the rest of the paper.

\subsubsection{Laplacian or square Euclidian norm penalty}
\label{s:laplacian}
Figure~\ref{f:1bsim} allows us to check which penalization gives the
solutions with smallest distance to the model. First of all, it is quite
clear that the square Euclidian norm penalization is worst, because it
produces both flattened solutions and strong artifacts. Indeed, requiring
the norm of the solution to be small does not explicitly have an effect on
the smoothness of the solutions.

Laplacian penalizations give results very similar to the third order
penalization $\M{D}_{3}\equiv
{\mathrm{Diag}}_3 [-1,3,-3,1]$ defined as in \Eq{l1l2} . The latter are therefore not plotted, and perform equally
well at any SNR. Both produce moderately flattened solutions showing
increasing dispersion with decreasing SNR, without systematic bias in age. The width of these
bumps is a simple (but crude) measure of the time resolution of the reconstructions, because
any bump narrower than the models displayed would be broadened by the
inversion. The absence of
significant difference between the results of the Laplacian and third order
penalizations shows that the inversion does not rely strongly on the
details of regularization, as long as it involves a differential operator.
We chose to carry on with the Laplacian penalization for the rest of the
paper.

\subsubsection{Positivity and Gibbs apodization}
\label{s:pos}
\begin{figure*}
\begin{center}
\begin{tabular}{ll}
{\includegraphics[width=0.45\linewidth,clip]{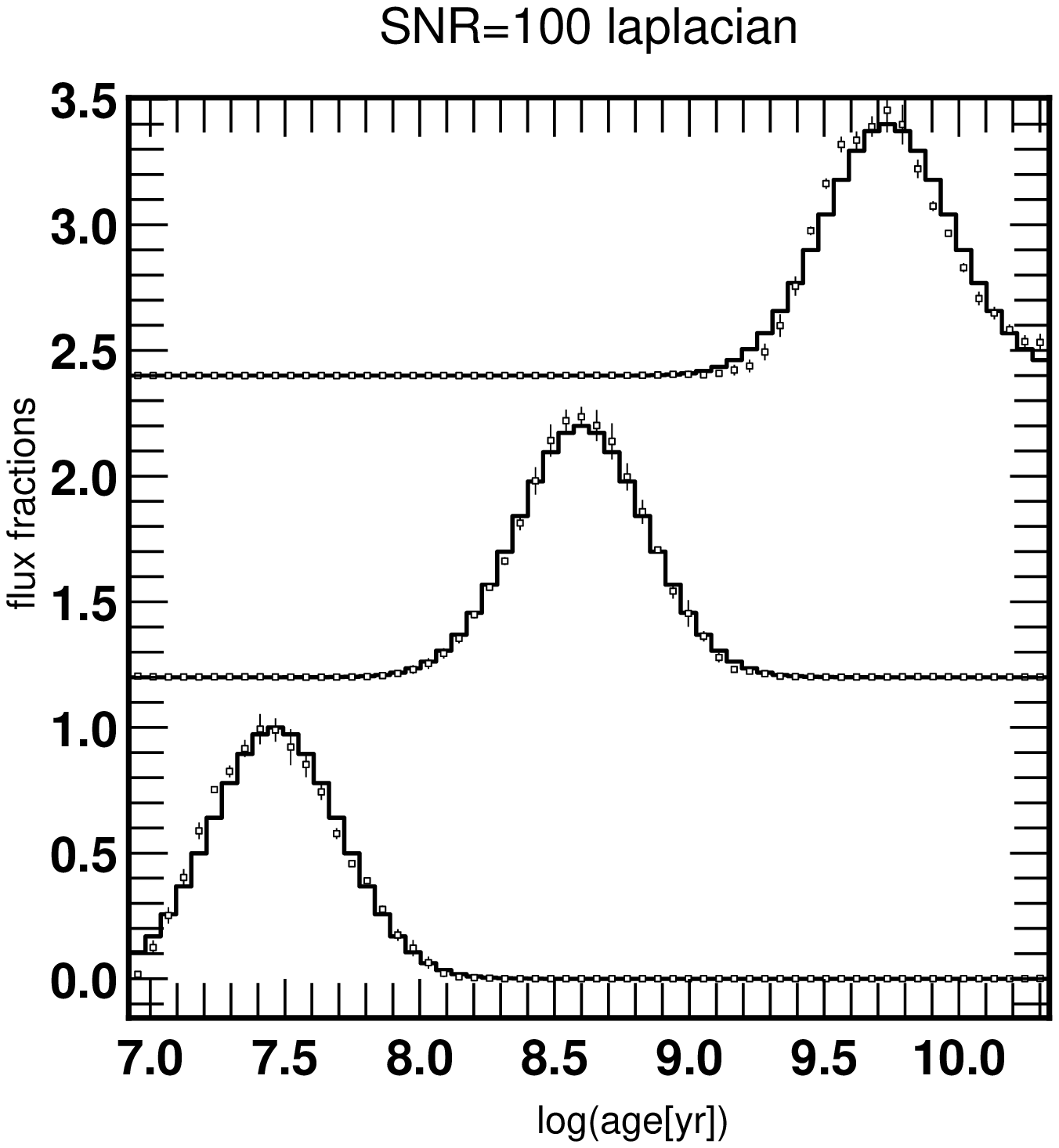}} &
{\includegraphics[width=0.45\linewidth,clip]{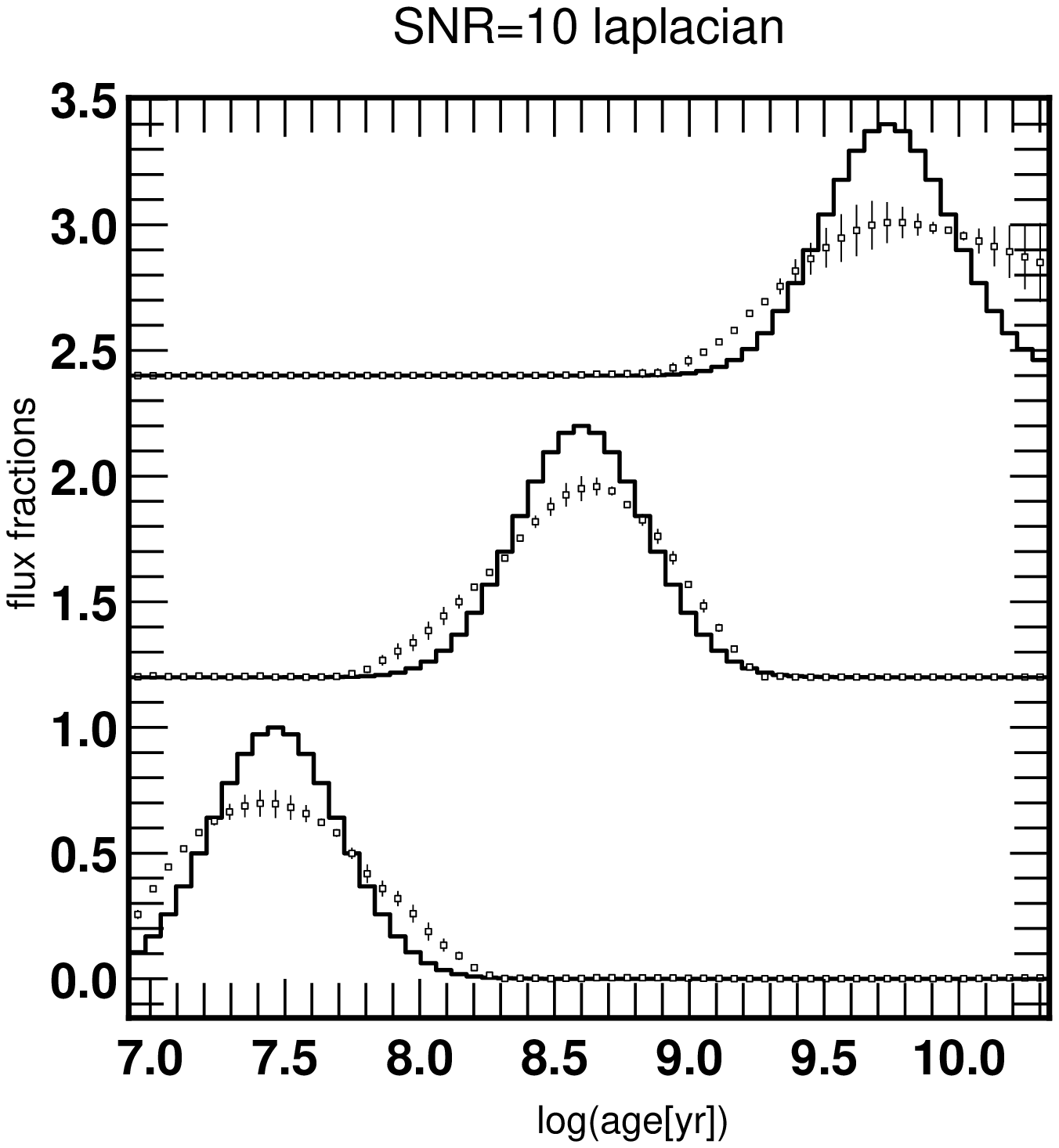}}
\end{tabular}
\end{center}
\caption{Same as \Fig{1bsim} with a flux-normalized basis, positivity
enforced by quadratic reparameterization and Laplacian
penalization. Results of simulations for $\mathrm{SNR}=100$ and $\mathrm{SNR}=10$ per pixel at
$R=10\,000$ are shown. Even though some residual remains, the solution sticks
to zero where it should, instead of displaying Gibbs rings.
}
\label{f:1bqsim}
\end{figure*}

Positivity of the solution is a physically motivated requirement, but it
also stabilizes the inversion by strongly reducing the explored parameter
space. The maximum frequency (or best resolution in age) that would be
obtained for infinite SNR is thus not only a matter of basis
ill-conditioning but also has a methodological component. This is
illustrated by the slightly better age-resolution (and thus higher
frequency) obtained while relying on positivity as shown in \Fig{1bqsim}.
Unfortunately there isn't any simple extension of the analytical
ill-conditioned problem diagnosis to the non linear problem.  Also the
minimization of $Q_{\mu}$ defined in \Eq{reg} requires efficient algorithms
as described in Appendix~\ref{s:agemin}.  As any regularization method,
positivity will also introduce some bias. Indeed, the solutions in
\Fig{1bqsim} seem to be slighlty asymmetrical compared to the linear
solutions.  However, one strong advantage of positivity is its ability to
reduce Gibbs ringing.  Linear solutions with any penalization exhibit
spurious oscillations even far from the main bump, which can be interpreted
as a superimposed component. These annoying artifacts do not appear in the
positive solutions as shows \Fig{1bqsim}. In many applications, this
property turns out to be more important than the possible bias it might
introduce in age estimation.

\subsubsection{Why carry out an extensive simulation campaign?}
\label{s:campaign}
\begin{figure}
\begin{center}
\includegraphics[width=0.9\linewidth,clip]{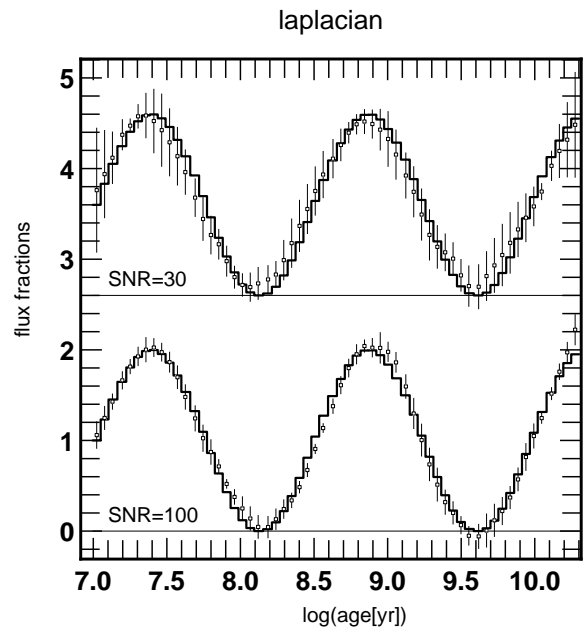}
\end{center}
\caption{Same as \Fig{1bqsim}, with a $1+\sin$ model for the
  {\SAD}. The SNR per pixel is given for each experiment ($10$ realizations), and the resolution
is $R=10\,000$. The
  smoothing parameter $\mu$ was adjusted by running several simulations and
  choosing the one providing the smallest distance to the model. The
  reconstruction is excellent, but there is a catch: it turns out that sine
  functions are intrinsically easier to recover than single bumps, given
  the shape of the solution vectors of the kernel. Hence, such reconstructions are
  very misguiding. More systematic simulations are required.}
\label{f:age}
\end{figure}

An inversion method can perform very well for some specially chosen cases
while performing poorly generally.  As an example we discuss the recovery
of the age distribution of a complex population consisting in a
superposition of young, intermediate, and old sub-populations. Each of
these 3 components contributes equally to the total observed spectrum
$\M{y}$. The noise is Gaussian. Figure~\ref{f:age} shows reconstructions of
the age distribution by the \Eq{xtild}, for 150 realizations, with a Laplacian
penalization. The
reconstruction seems to be satisfactory: it is unbiased and the
interquartile intervals of the solutions shrink with increasing SNR.  A naive
reading of \Fig{age} would suggest that we are able to recover nearly any
age distribution, without bias and with very small error for \emph{all} the
time bins, even with quite low SNR, but there is a trick. Why do the simulations in \Fig{age} look so good? First, the
temporal frequency of the solution is lower than in the single bump
simulations. Second, higher frequency sine fuctions are needed to represent
a single bump than to represent a sinusal curve (one is enough). Thus, as
the first singular vectors roughly form a basis of sine functions, one
needs fewer and lower order solution singular vectors to represent a sine
function than a bump, and lower SNR.

One simple (yet unadvisable!)  recipe to make good looking simulations even without regularization could involve the following steps:
\begin{enumerate}
\item choose as model $\M{x}$ one of the last few solution singular vectors
$\M{v}_k$ (or one of the first few if some penalization is implemented)
\item compute the corresponding pseudo-data $\M{y} = \M{B}\cdot\M{x}$
\item noise the data at chosen SNR
\item invert and show how close the recovered solution lies to the initial
  model
\end{enumerate}
By doing so, we managed to produce apparently  good looking simulations down to
$\mathrm{SNR}=0.1$ per pixel. Thus the requirement to assess and demonstrate the validity
and efficiency of the MAP method carried out in this section.

\subsection{Age separation versus $R$ and SNR}
\label{s:agesep}

\begin{figure*}
\begin{tabular}{ll}
\resizebox{6cm}{6cm}{\includegraphics{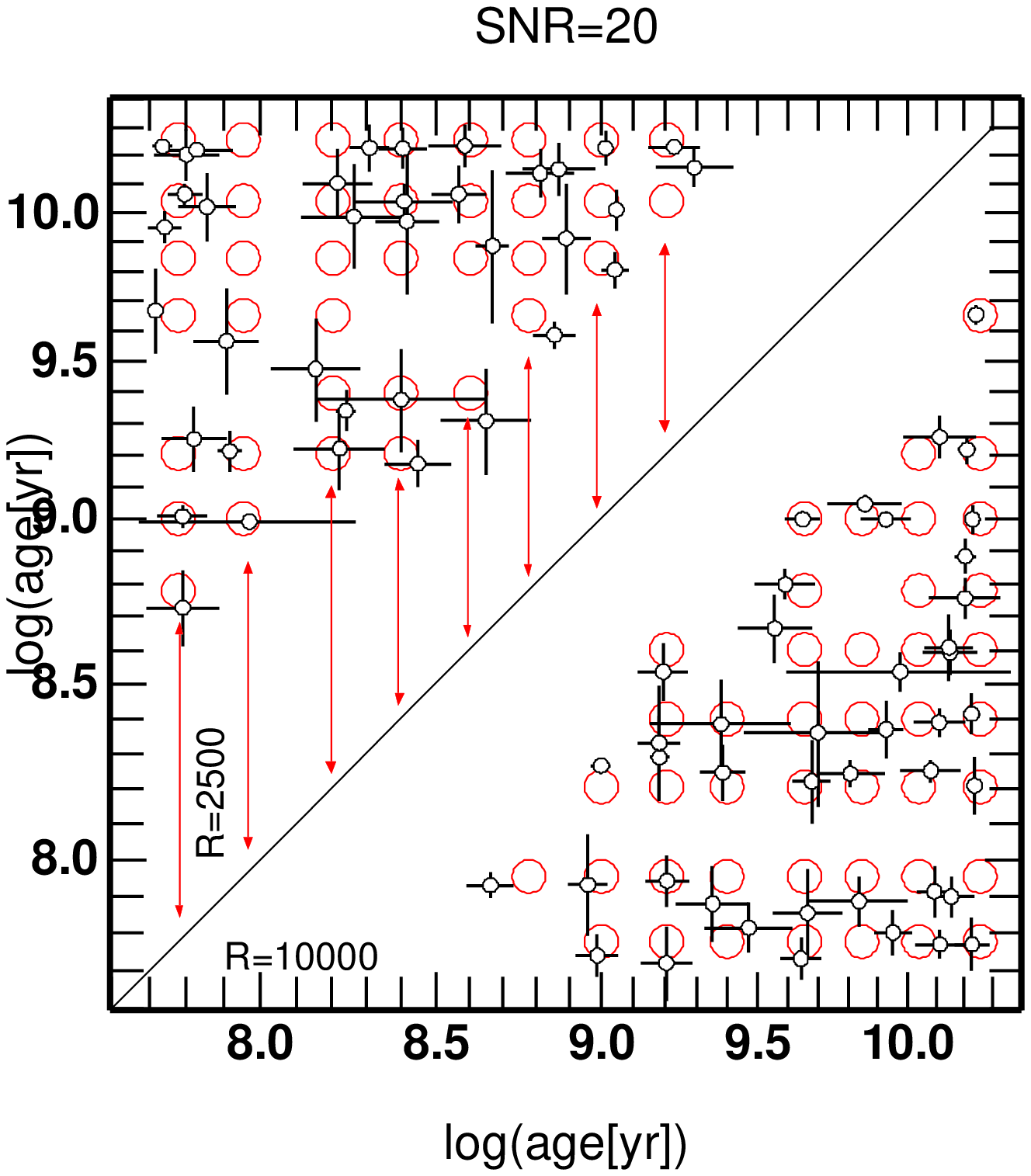}} &
\resizebox{6cm}{6cm}{\includegraphics{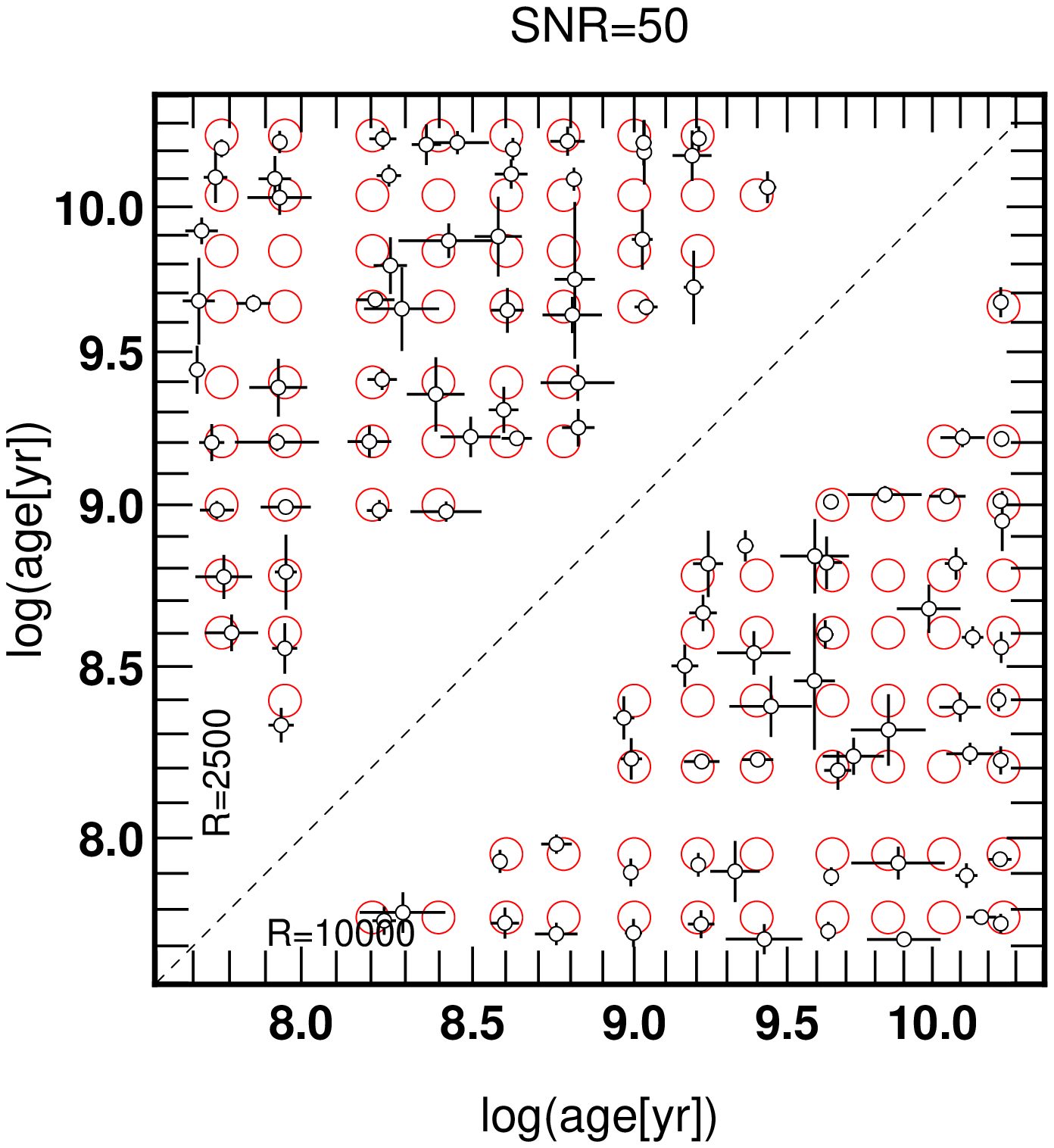}} \\
\resizebox{6cm}{6cm}{\includegraphics{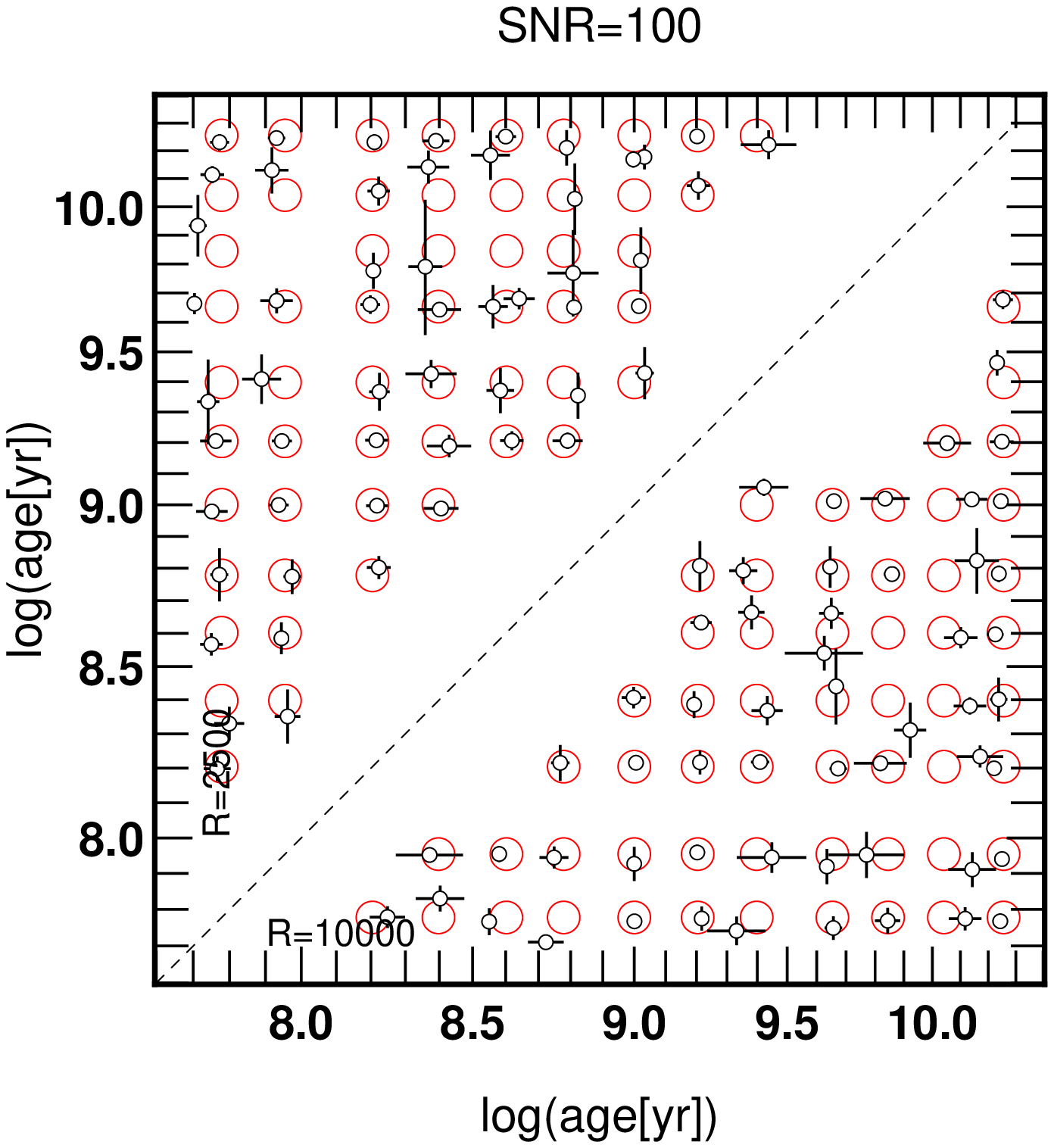}} &
\resizebox{6cm}{6cm}{\includegraphics{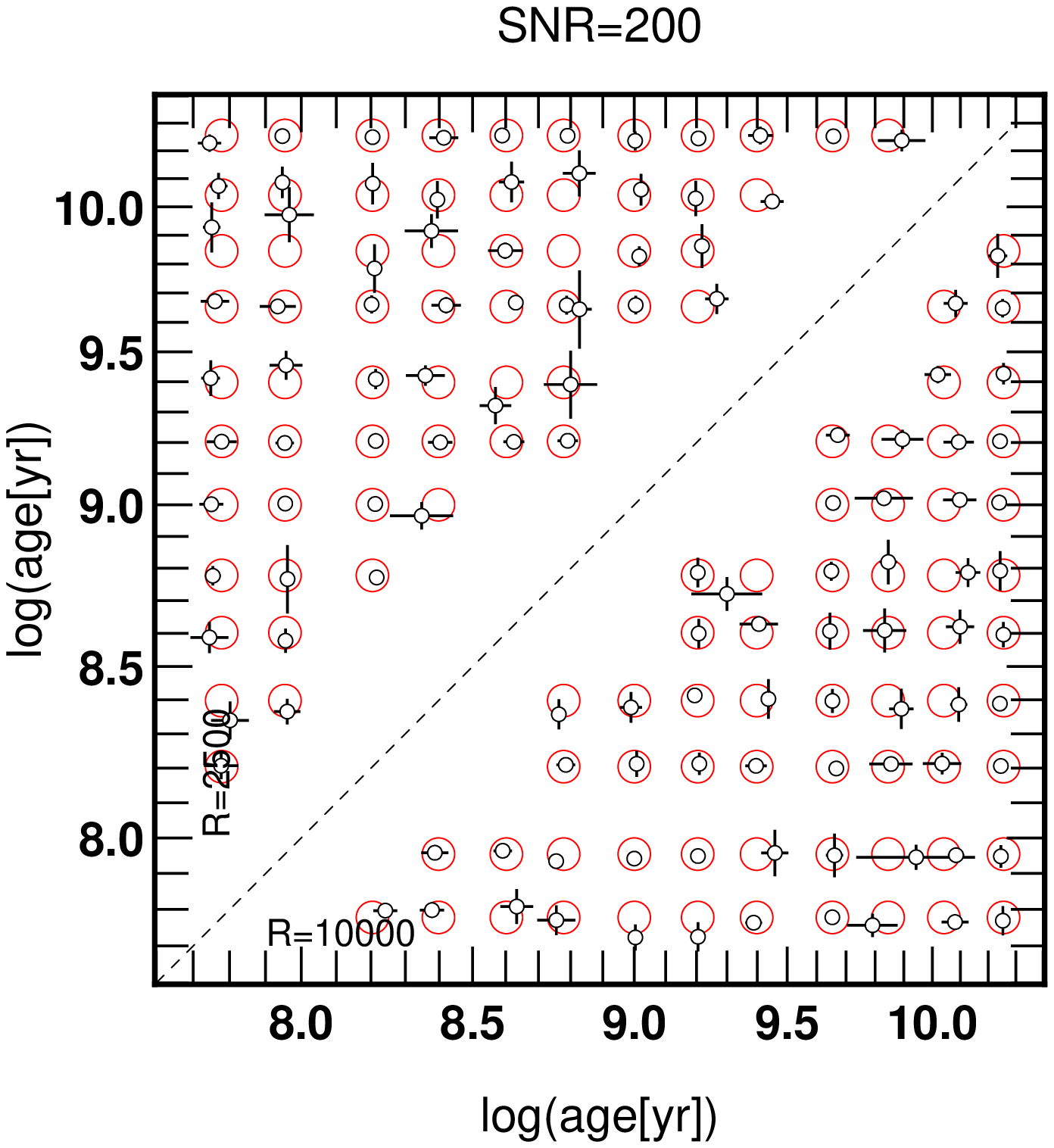}}
\end{tabular}
\caption{Recovery of double bursts for several SNR per {\AA}. The large circles are the models. Their coordinates $(a_1,a_2)$ are the ages of the
two bursts. The smaller circles with error
bars show the median and the interquartiles of the recovered ages in $10$ reconstructions each. The dotted line
represents the $a_{1}=a_{2}$ limit. Solutions that do not satisfy the quality
criteria illustrated in \Fig{dd} are rejected and not plotted. The
upper diagonal part of each panel shows $R=2\,500$ results while the lower
diagonal part shows $R=10\,000$ results. Results for the other spectral
resolutions down to $R \approx 1\,000$ are very similar and therefore not shown. Our
ability to separate close double bursts improves with increasing
SNR, but does not significantly change with spectral resolution. The top left panel illustrates the definition of the
resolution in age as the median length of the segments. Note that the shape
of the ``unseparable'' zone and its evolution with SNR are similar to that
shown in \Fig{chi2}.}
\label{f:r-snr-sim}
\end{figure*}

We have already made clear that we can not recover all the high frequency
oscillations of a given {\SAD} even with very high SNR, but rather
moderately slow variations, corresponding to smooth solutions. Let us
nonetheless consider the special case where a composite population consists
of two successive bursts, i.e.\ {\SAD}s with two bumps of same luminosity.
This is one order of complexity above the classical characterization of a
population through one unique age using Lick indices. And indeed, it
applies to many astrophysically interesting cases. The ability to separate
the two main populations would allow for example to age-date respectively
the disc and the bulge of unresolved spiral galaxies, or late stages of
accretion and star forming activity in ellipticals in surveys such as SDSS
and $2$DFGRS. It would also allow to better constrain the mass to light
ratio of such complex populations. We wish to investigate what
observational specifications (spectral resolution, SNR) are required to
reliably perform such a separation.  We thus ran extensive simulations of
reconstructions of double bursts populations. The spectral resolution, SNR
and the age separation $\Delta$age between the 2 bursts were varied, and
the recovered ages were studied as a function of $R$, SNR, and $\Delta$age.
Figure~\ref{f:r-snr-sim} shows the recovered and model age couples
$(a_1,a_2)$ in several experiments of double bursts superpositions, for
$\mathrm{SNR}=20$ to $200$ per {\AA}, at $R=10\,000$ and $R=2\,500$. The
model age grid takes $13$ values, separated by $0.2\;\mathrm{dex}$,
therefore defining $78$ age couples.

These systematic simulations allow us to estimate the resolution in age
achievable for a given $(R,\mathrm{SNR})$ and the corresponding errors. It
is a solid, systematic way for testing the method in different regimes. The
smoothing parameter was set for each $(R,\mathrm{SNR})$ by taking the GCV
value as a guess and fine tuning it in order to obtain stable
reconstructions of close bumps.
\begin{figure}
\begin{center}
{\includegraphics[width=0.9\linewidth,clip]{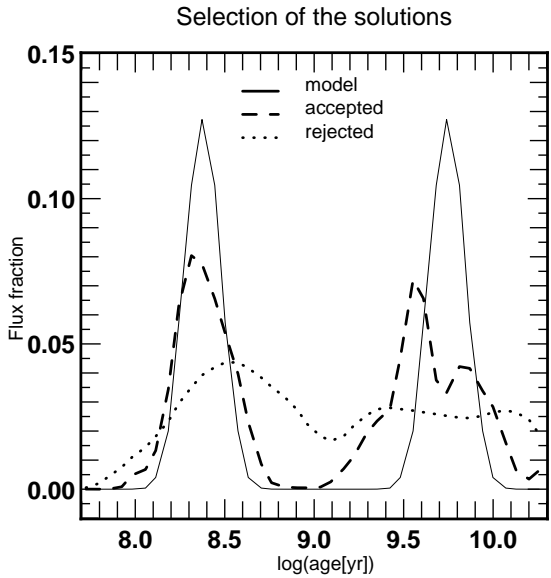}}
\caption{Selection criterion: the rejected solution shows no clear separation, while the accepted one has two clear bumps of similar area with a well defined minimum.}
\label{f:crit}
\end{center}
\end{figure}
\begin{figure*}
\begin{center}
\begin{tabular}{ll}
{\includegraphics[width=0.45\linewidth,clip]{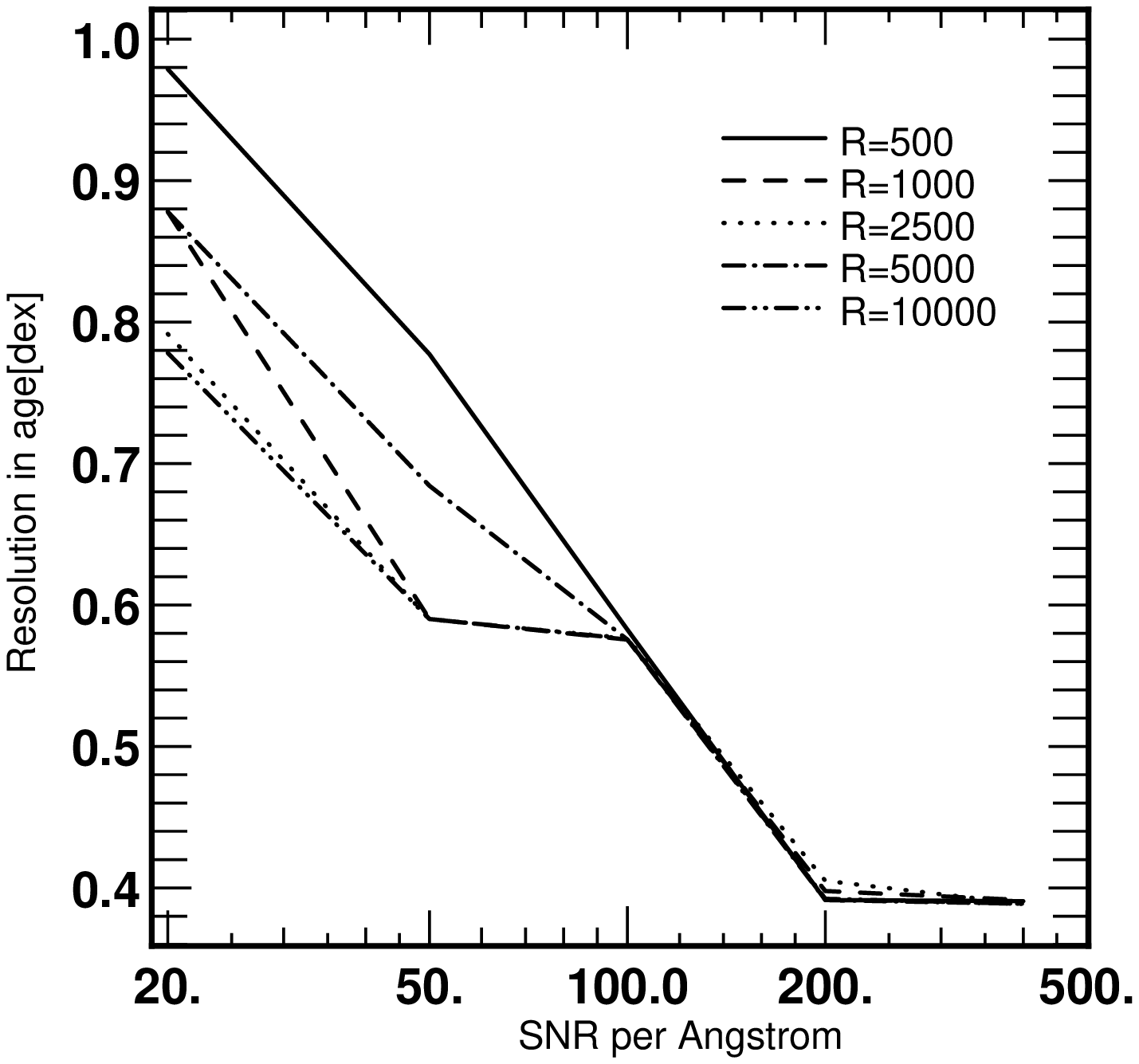}}&
{\includegraphics[width=0.45\linewidth,clip]{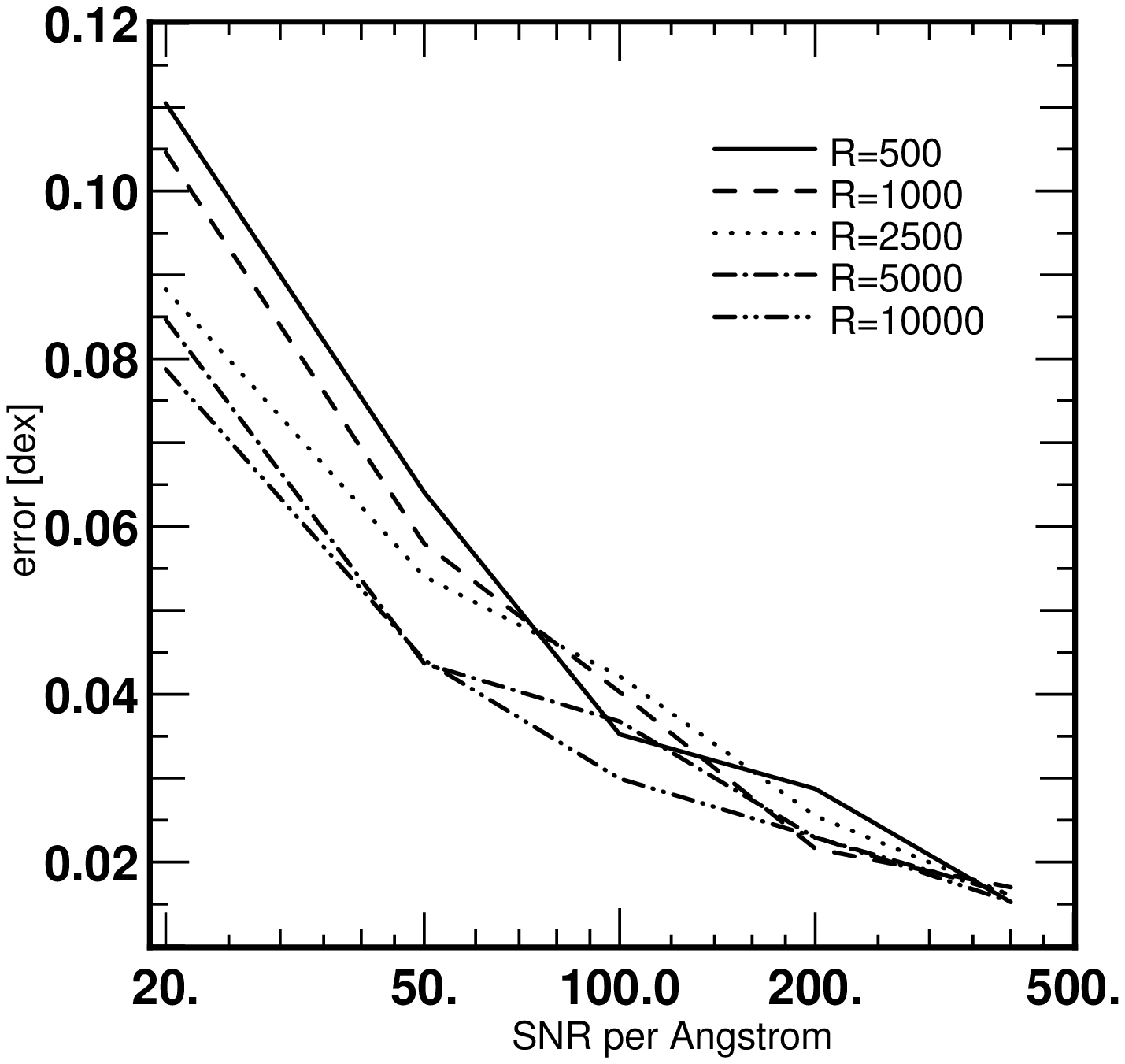}}
\end{tabular}
\caption{\emph{Left}: Resolution in age, in dex versus SNR per {\AA} for various spectral resolutions.
As expected, the age resolution improves with increasing SNR, and seems to
settle around $0.4\;\mathrm{dex}$ for the highest SNR. No significant trend is seen with
spectral resolution.\emph{Right}: Mean error of the age estimates for the successful
cases (according to our criteria). The mean error is approximately
one order of magnitude smaller than the resolution in age, and decreases with
increasing SNR.}
\label{f:dd}
\end{center}
\end{figure*}
The quality of the reconstructions is assessed using two criteria:
\begin{itemize}
\item since, in the model, the two bursts have exactly the same luminosity,
  we require that the areas of the two biggest bumps
  have a ratio smaller than $2$.
\item the minimum between the two main bumps of the solution should be
  fairly low, otherwise it is difficult to state whether the
  populations are truly distinct or part of an extended star formation
  episode. Here we required the minimum to be lower than $10$\% of the mean
  height of the biggest bumps.
\end{itemize}
The solutions are required to satisfy these two criteria to be considered
as ``good'' in terms of age separation.  Figure~\ref{f:crit} shows as an
example an acceptable (well-defined bumps, minimum at $0$), and a rejected
solution (bumps and minimum unclear). In \Fig{r-snr-sim}, we retained
exclusively the cases satisfying these criteria, i.e.\ for the other age
couples (not plotted), the recovered {\SAD s} failed one or both criteria.
A common failure is the recovery of one wide bump instead of two,
indicating that the sub-populations are not separated given the SNR and
spectral resolution.  Thus, the empty region between the successfully
separated couples and the bisector (dashed line) is a region of
``unseparable'' couples. The width of this region indicates the resolution
in age that we can achieve. This region shrinks with increasing SNR,
showing that we can separate two close sub-populations more accurately.  We
superimposed on the leftmost panel of \Fig{r-snr-sim} several vertical
segments spanning the ``unseparable'' region. We define the resolution in
age as the median length of these segments. The statistical error on this
quantity is of order $0.2\;\mathrm{dex}$ for $\mathrm{SNR}=20$ per {\AA}.

In a realistic observational context, a separation of two sub-populations
with an age difference lower than the computed resolution in age should not
be attempted, or at least not trusted.  The resolution in age achieved here
is a lower limit because no error source other than Gaussian noise is
considered. Other possible sources of noise are glitches, residual sky
lines, non-sky emission lines (when not masked in $\M{W}$),
spectrophotometric and wavelength calibration error, and models error,
along with effects of the age-Z-extinction degeneracy (in this section the
true metallicity of the observed system was known a priori).

Figure~\ref{f:r-snr-sim} also shows that the error on both ages of the
couple of sub-populations decrease on average with increasing SNR, as
expected. For small SNR, the figure is quite inconclusive, and the
recovered age couples are more or less randomly spread all over the age
domain, while for high SNR, every couple seems to be quite in place,
even though some couples remain slightly offset. For other resolutions, the
plots are quite similar, and therefore we do not reproduce them here.  The
left panel of \Fig{dd} gives a synthesis of all the experiments by showing
the resolution in age, computed according to the given definition, versus
the SNR per {\AA}, for several spectral resolutions. The resolution
in age improves with increasing SNR, from $0.9\;\mathrm{dex}$ at
$\mathrm{SNR}=20$ per {\AA} to $0.4\;\mathrm{dex}$ at $\mathrm{SNR}=200$
per \AA. Given the small number of measurements of the width of the
unseparable zone in each experiment, the variation of the resolution in age
with spectral resolution is not highly significant, and thus it seems that,
as long as the SNR per {\AA} is conserved, spectral resolution does not
significantly improve our ability to separate sub-populations. The right
panel of \Fig{dd} shows the error on recovered ages versus SNR for the
successful separations, for several spectral resolutions. The error
decreases with increasing SNR, as expected, and is about ten times smaller
than the resolution in age for the same SNR. Again, no strong trend is seen
with spectral resolution.

\subsection{Compressed versus uncompressed data}


In this section, we discuss the similarity between SVD and {\GSO} (GSO), the
decomposition scheme adopted by MOPED's authors \citep{moped01}. This
comparison is carried out in the
mono-metallic, extinction-less regime.
Data can be compressed by multiplying it by the $n$ singular vectors to
obtain $n$ numbers containing the same information as the whole
original spectrum. Appendix~\ref{s:CDvsSVD} shows that, the fact that the
singular vectors are provided by non-truncated SVD or GSO makes little difference in the
linear regime.
The compression can effectively be lossless, but the conditioning of the
problem is unchanged, as shown by the inspection of the singular
values in the left panel of \Fig{MOPED}. The right panel of Figure~\ref{f:MOPED} shows the result of a \GSO\
(\Eq{GSO-sol}) and a  
SVD (\Eq{svdsol}) inversion for a composite population in a 
moderately ill-conditioned example. They are equal down to machine
precision. 
Minimizing the $\chi^2$ of the compressed data involves the issues discussed
in \Sec{ill}, if the compression is provided via the SVD or GSO
singular vectors.

\subsection{Constraints  on metallicity?}

\begin{figure}
\begin{center}
\includegraphics[width=0.9\linewidth,clip]{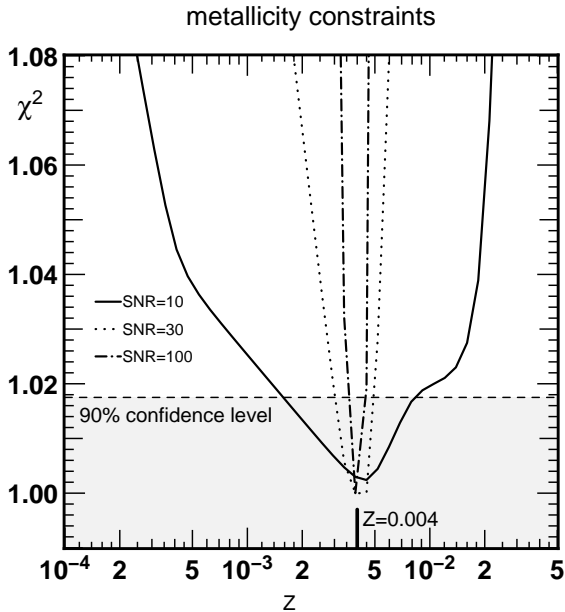}
\caption{The high resolution  {\SED}
of model extinctionless  mono-metallic population
with
$Z=0.004$ is inverted using spectral bases with different
metallicities for several SNR. For $\mathrm{SNR}=10$ per pixel, the
metallicity is moderately well
constrained ($\Delta Z \approx 1\;\mathrm{dex}$), while for $\mathrm{SNR}\geq 30$ per pixel all the
metallicities other than $0.004$ can be rejected at the $90 \%$ confidence level.}
\label{f:ztm}
\end{center}
\end{figure}

When attempting to reconstruct the stellar age distribution from real observations, one would still have
to guess the metallicity of the population. A classical parametric way to
proceed would be to perform a mono-metallic inversion for each of the
available metallicities in the basis. If the dominant observational error
is Gaussian, we expect the $\chi^2$ to be minimum when using the true
metallicity. However, because of the age-metallicity degeneracy, it might
not be so clear, and one could expect to reach a good $\chi^2$ even with an
erroneous metallicity guess, resulting in an error in age estimation.
Figure~\ref{f:ztm} shows a plot of the reduced $\chi^2$ when inverting a
population of metallicity $Z=0.004$ with a basis of different metallicity
for several SNR and $R=10\,000$. The smoothing parameter was chosen using
GCV with the $Z=0.004$ kernel. The best fit is always obtained when the
initial model metallicity is used. We computed the $90 \%$ confidence level
by taking as the number of degrees of freedom, the number of pixels in the
spectrum minus the number of age bins ($40$ in this example). This choice could be
discussed because the weights of adjacent time bins are correlated by the
penalization.
However, the number of time bins remains far smaller than the number of
pixels and thus plays no critical role.  For $\mathrm{SNR}=10$ per pixel
(i.e.\ $\mathrm{SNR}=20$ per \AA), we can not reject fits with wrong
metallicities $Z \in [0.002, 0.009]$. The error on metallicity can therefore
reach $0.35\;\mathrm{dex}$ for $\mathrm{SNR}=10$. The range of acceptable
metallicities however shrinks rapidly with increasing SNR, tightening the
constraints.
At $\mathrm{SNR}\ge30$, it is possible to break the
age-metallicity degeneracy, and thus to let metallicity be a free parameter of
the inversion problem.

This closes our detailed investigation of the idealized problem of recovering the stellar age distribution of a mono-metallic,
reddening-free  stellar population.

\section{Stellar content and reddening recovery}

\label{s:aze}
The previous section presented \STECMAP\ in an idealized regime, which
could only be applied to observations where both the metallicity and the
extinction are known a priori, which is rarely the case in reality.
We now present an extension of \STECMAP\ accounting for these
additional free parameters as well. In \Sec{linaz}, the full linear age-metallicity problem is examined, where
both metallicity mixing \emph{and} age mixing are allowed, and study its
behaviour.  Then, for simplicity, and given the extremely poor
conditioning of this problem, the unknown metallicity will be handled
specifically as an \AMR. The technique for reconstructing the \SAD, the
{\AMR} and the extinction will be presented in \Sec{age-metal}, along with a
few example simulations in \Sec{sims-aze}, and finally
its applicability and accuracy will be discussed while exploring several
observational regimes in \Sec{sep-aze}.

\subsection{2D Linear age-metallicity problem}
\label{s:linaz}

Here we consider a \emph{very} composite population where several
sub-populations with different ages and metallicities are superimposed. Let
us define a $2$D stellar age and metallicity distribution $\Lambda(t,Z)$
yielding the fraction of optical flux emitted by stars with age $t \in
[t,t+ \mathd t]$ and metallicity $Z \in[Z,Z+ \mathd Z]$. The model spectrum is the
integral of $\Lambda$ over age and metallicity space. Discretizing as in
\Sec{age}, we get the discrete model spectrum as the weighted sum of the
{\SSPs} for all the ages and all the metallicities in the basis. Here the
parameter vector is a $2$D map containing the weights $x_{ij}$ of the
{\SSP} of age $t_i$ and metallicity $Z_j$. The model matrix $\M{B}$ is the
concatenation of the mono-metallic bases described in \Sec{age}, i.e.\
sequences of {\SSPs} in age and metallicity.  Its conditioning number is
commonly of order $10^8$, telling us that thorough regularization is
required.

\subsubsection{Where is the information on $Z$?}

\begin{figure*}
\begin{center}
\rotatebox{-90}{\resizebox{9cm}{18cm}{\includegraphics{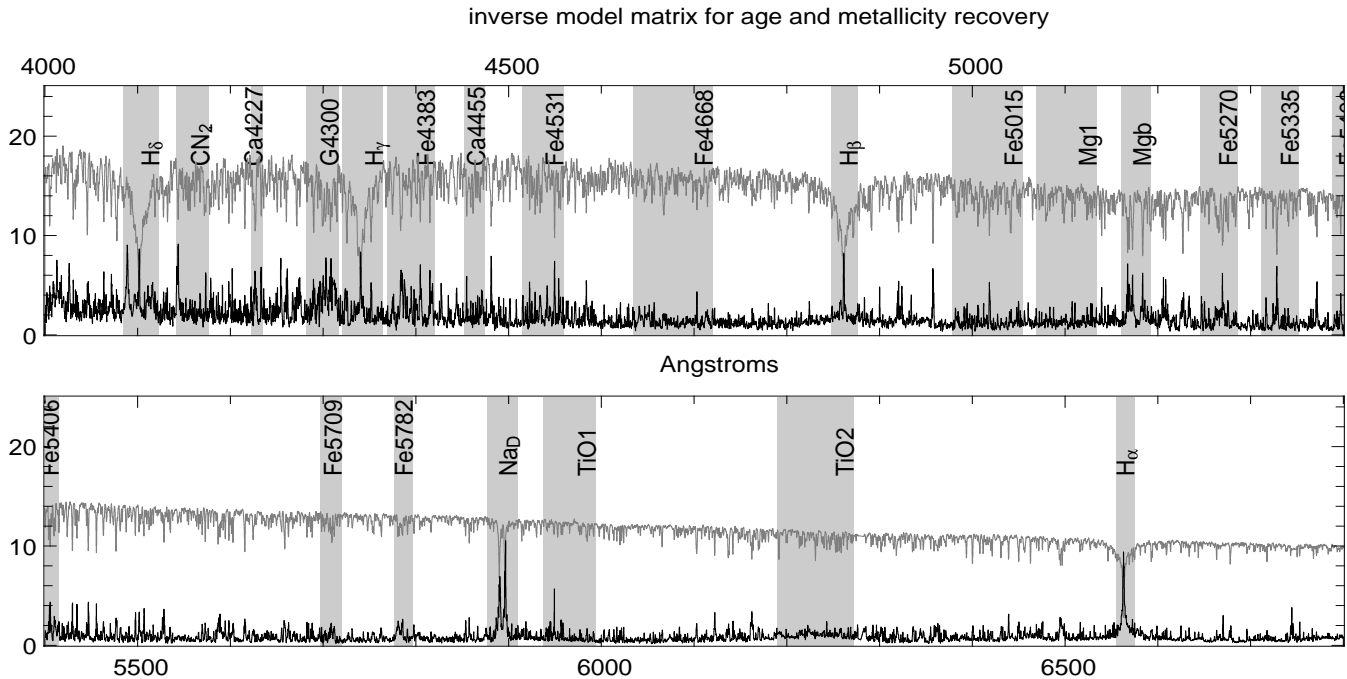}}}
\caption{Same as in \Fig{invmodel} for the linear age-metallicity
distribution recovery. The dimensions of the inverse problem are $60$ age bins
and $5$ metallicity bins. The smoothing parameter is set by GCV for
$\mathrm{SNR}=100$ per pixel.
The grey solid line is a $1$ Gyr half-solar metallicity
{\SSP} for reference.
Many of the spectral domains
involved in the definition of the Lick indices system seem to
carry more information than the rest of the spectrum. However, the information
is still widely spread along the whole optical
range in the form of medium depth lines, suggesting there is a large number of potential high resolution indices.}
\label{f:invmodelaz}
\end{center}
\end{figure*}

In a manner similar to \Sec{invmodel} we can determine which
spectral domains are important for age and metallicity determination.
We compute the inverse model matrix $\tilde{\M{B}}$ of the problem for a given
$\mathrm{SNR_d}$ and look for large peak to peak variations in this
matrix, indicating spectral features having strong discriminative
power, as shown in \Fig{invmodelaz}. Most of the bands involved in the Lick
indices carry a lot of information. However,
some of them, like TiO$_2$, seem to be unimportant, and a large number of medium and high resolution lines not involved
in Lick indices actually carry most of the information.The
comparison with \Fig{invmodel} shows that several metallic lines,
which were not important for a mono-metallic population age distribution
recovery, turn out to carry a substantial part of the information when the
metallicity is unknown. Again, the blue part of the spectrum seems to be more discriminative.

Since age sensitive and metallicity sensitive lines are spread along the
whole optical wavelength range, any small section of the spectrum has good
chances of containing such lines (see \cite{PEG-HR} for an example around
H$_\gamma$). Thus, if the available data does not allow reliable full
optical domain fitting, plots such as \Fig{invmodelaz} are a good
starting point for the search for new high resolution indices. The use of
the whole spectrum implies some redundancy, but considering the sensitivity
of the inversion problem to noise, this redundancy is highly
welcome\footnote{the redundancy is also useful in oder to address in part
  problems induced by the poor modeling of some spectral lines}.

\subsubsection{Age-metallicity degeneracy?}

\begin{figure*}
\begin{tabular}{ll}
\resizebox{6cm}{6cm}{\includegraphics{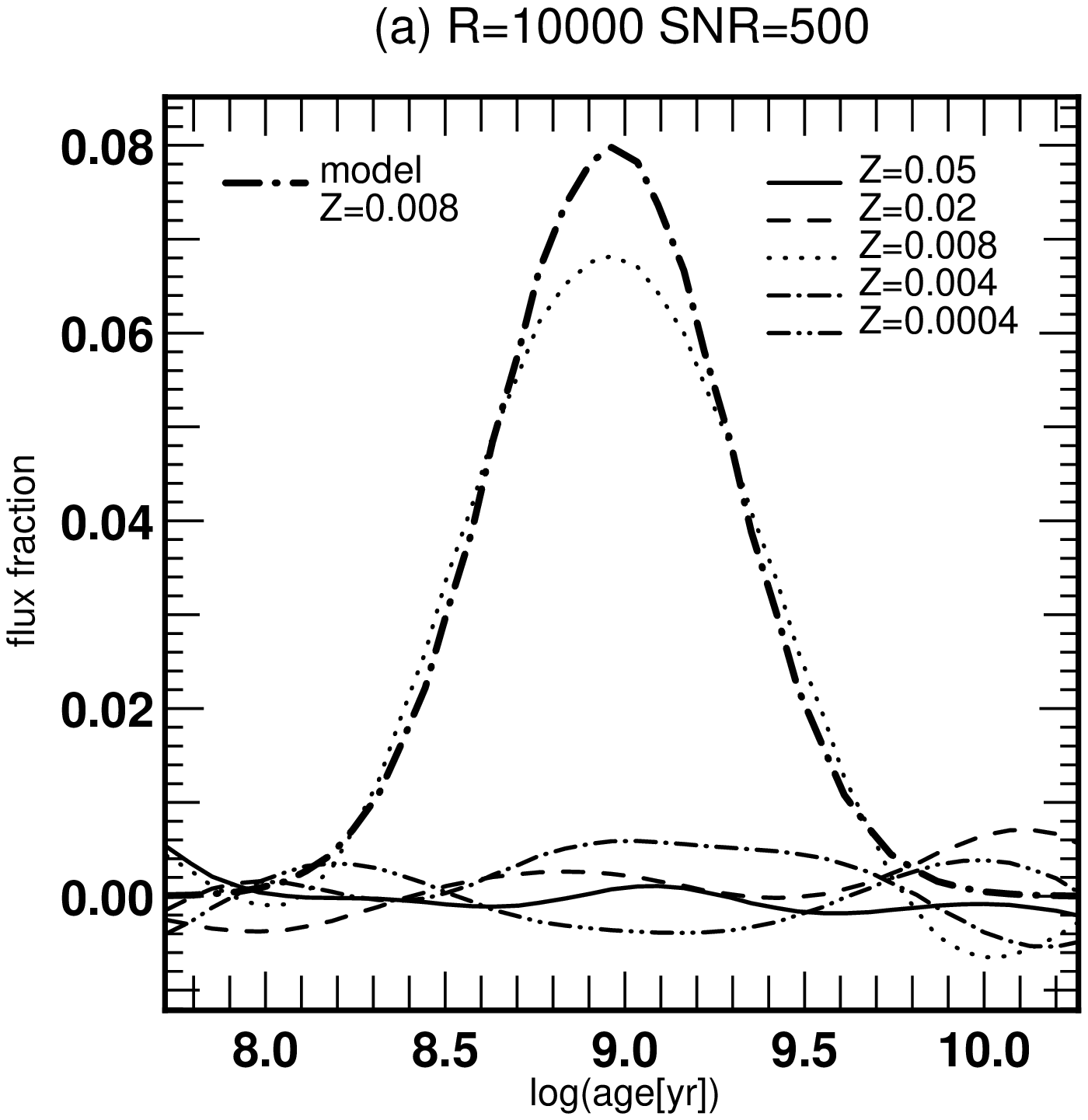}}&
\resizebox{6cm}{6cm}{\includegraphics{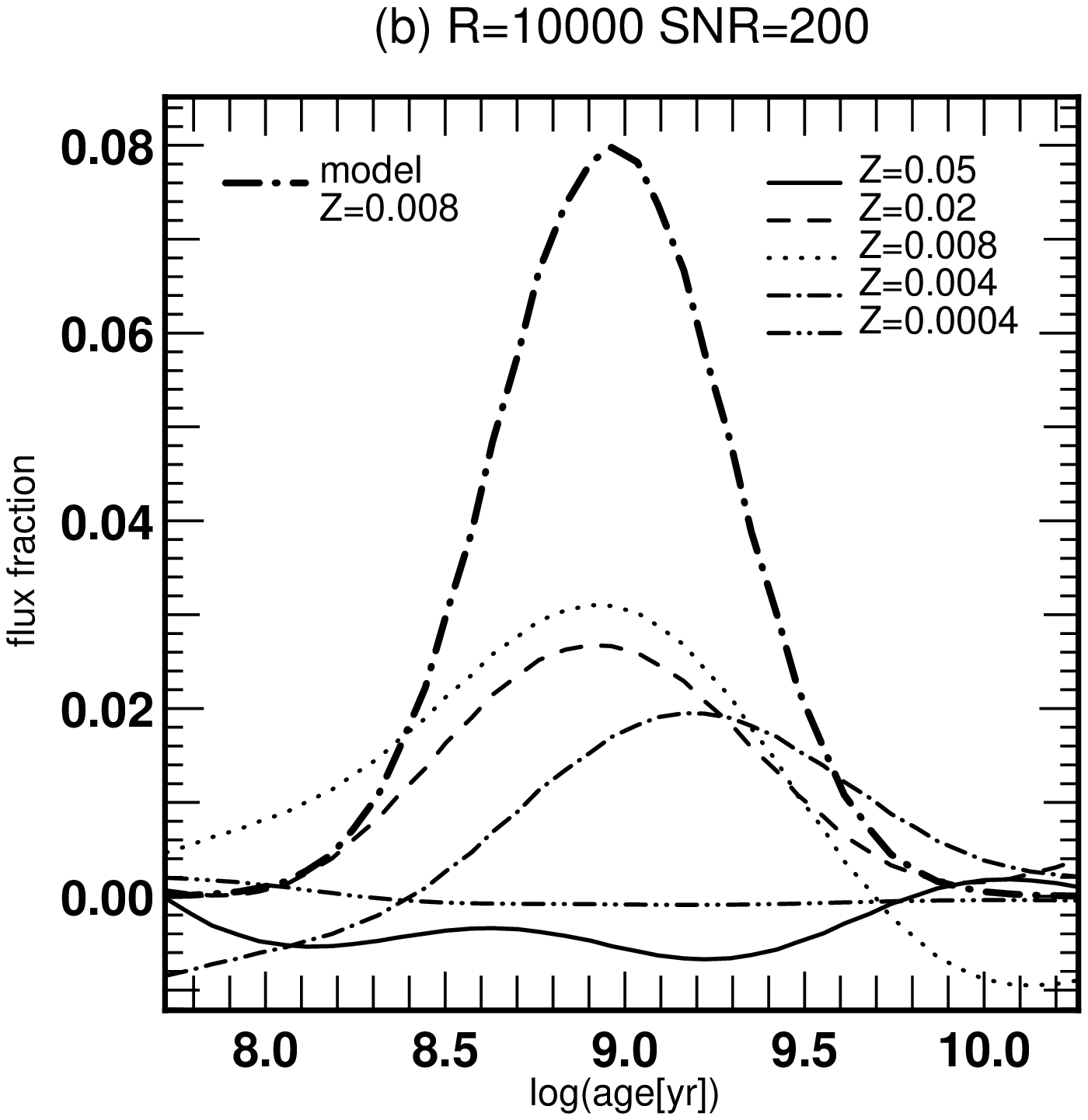}}\\
\resizebox{6cm}{6cm}{\includegraphics{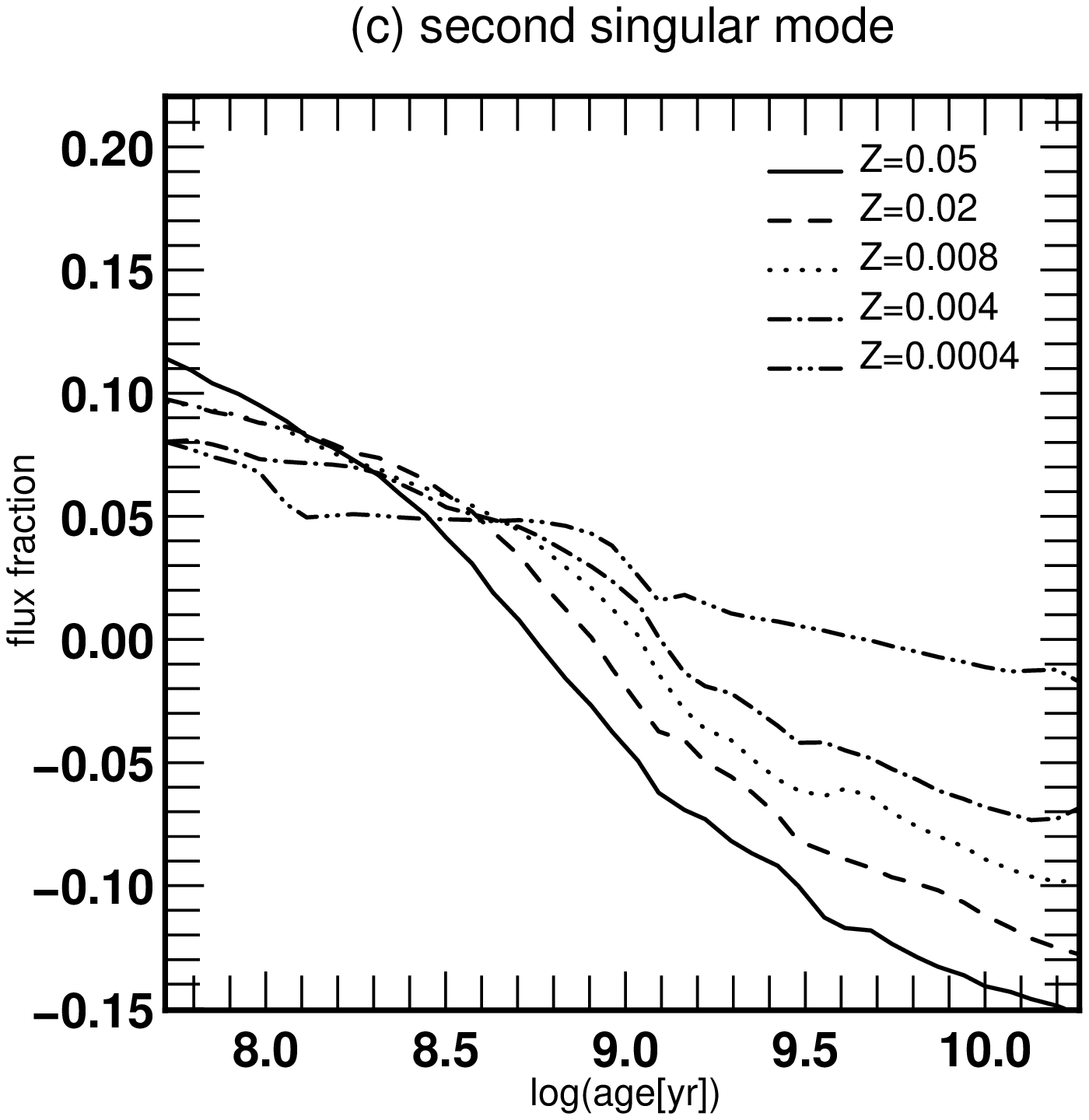}}&
\resizebox{6cm}{6cm}{\includegraphics{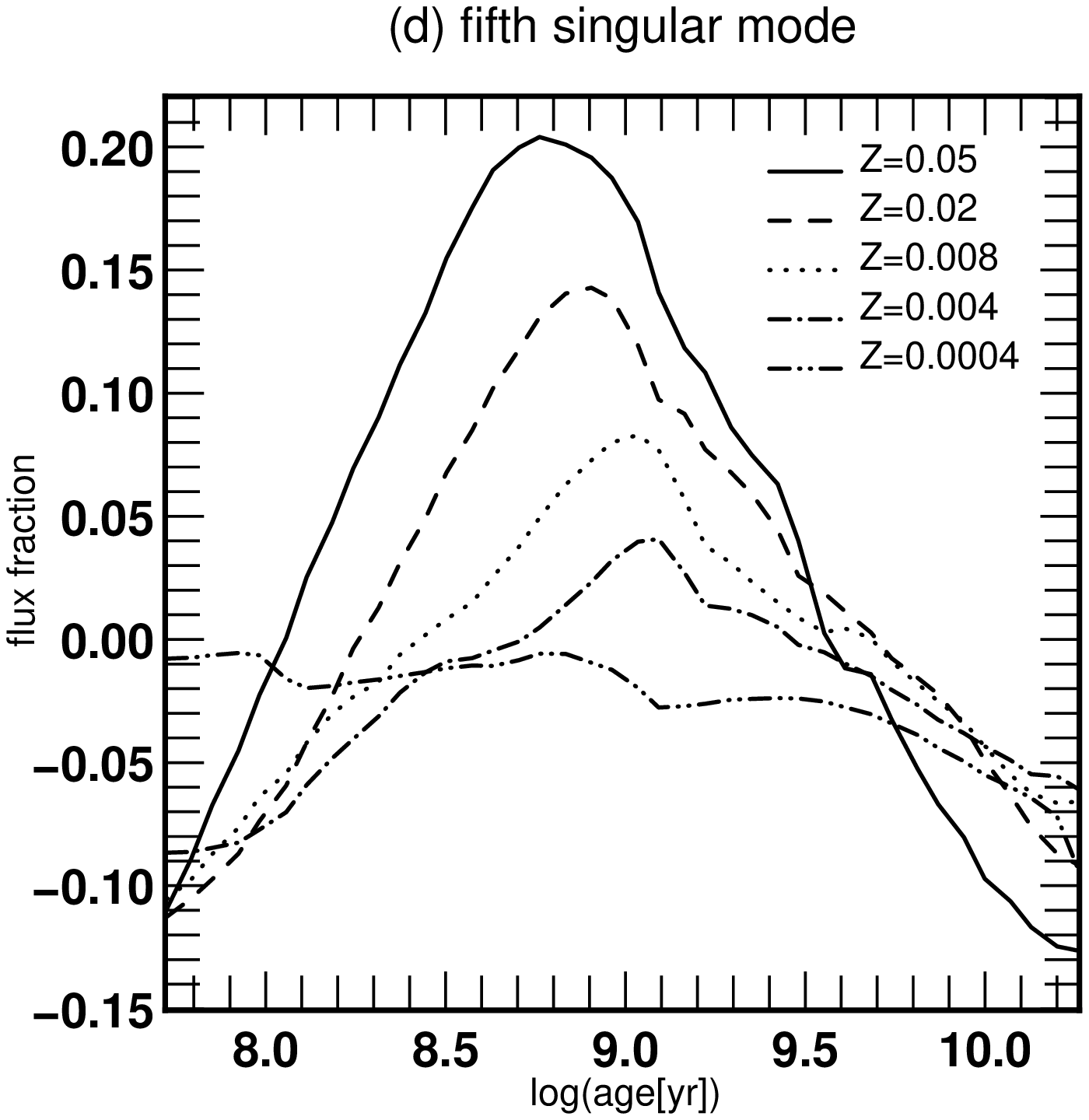}}
\end{tabular}
\caption{(a) and (b): Free metallicity reconstructions of a mono-metallic
  population for $\mathrm{SNR}=500$ and $\mathrm{SNR}=200$ per pixel. For
  high $\mathrm{SNR}$ a mono-metallic population is unambiguously
  recovered, while at lower SNR, a multi-metallic solution appears,
  indicating the degeneracy of the problem. (c) and (d): solution singular
  modes of the $2$D age-metallicity reconstruction problem. The difficulty
  involved in such a reconstruction arises from the very bad conditioning
  number, and the lack of features of the first singular modes in the
  metallicity direction.}
\label{f:2dsingular modes}
\end{figure*}

We carried out the following experiment illustrated in figure \ref{f:2dsingular modes}. We produced
mock data corresponding to a $2$D stellar age and metallicity distribution map
$\M{x}$ and investigated how well we could reconstruct it for a given SNR. In
the example of figure \ref{f:2dsingular modes} (top panels), the model is a
mono-metallic bump centered on $1$ Gyr and $Z=0.008$. The corresponding mock
data is noised and then inverted as in \Eq{xtild} except that
$\M{B}$ is now the multi-metallic {\SSP} basis defined above. The penalization
is Laplacian. In this experiment, we focus on the broadening of the bump in
the metallicity direction as a signature of the age-metallicity degeneracy.

The inspection of the first non attenuated solution singular modes tells us
about the properties of the regularized problem. The panels c and d of
\Fig{2dsingular modes} show the second and the fifth solution singular
modes of the model matrix $\M{B}$. Each of them is an age-metallicity map.
The shapes of the {\SAD} for each metallicity in the second singular mode
are very similar, indicating bad separability between metallicities. Thus,
if only the first singular modes are recovered, the solutions will have a
strong tendancy to be flat in the metallicity direction.

The fifth singular mode is the first one to show a well-defined structure:
a bump in age, elongated in the metallicity direction, with a slight shift
to larger ages with decreasing metallicities. This traces the
age-metallicity degeneracy: a pure mono-metallic population will be
reconstructed in regularized regimes as a composite, mixing younger metal
rich {\SSP s} with older metal poor \SSPs.  The a and b panels of
\Fig{2dsingular modes} show reconstructions of such age-metallicity maps
for $R=10\,000$, $\mathrm{SNR}=500$ and $200$ per pixel. The model consists
of a single bump centered on $1$ Gyr and $Z=0.008$, and the penalization is
Laplacian. For $\mathrm{SNR}=500$ per pixel we see that the population is
effectively reconstructed as a single bump in age and metallicity. The
age-metallicity degeneracy is in this example explicitly broken. The same
experiment with $\mathrm{SNR}=200$ per pixel gives a solution degenerate in
metallicity: the mono-metallic population is seen as the sum of three
mono-metallic sub-populations contributing nearly equally to the total
light. The younger component is more metal-rich, while the older one is
poorer, as is expected for age-metallicity degenerate solutions, and is
similar to the trend seen in the solution singular modes.
In this example, the smoothing parameter was chosen by {\GCV}. More
realizations of this experiment gave similar degenerate solutions. From the
shape of the fifth solution singular mode, we can measure the slope of the
age-metallicity degeneracy, that is the slope defined by the maxima of the
bumps of the singular mode in the age-metallicity plane. We find it to be
equal to $0.3$, which is much smaller than the $3/2$ given in
\citet{worthey94}. Smaller slopes indicate a better definition of age. This
is expected because here we consider the whole optical range and the
continuum as reliable.

As a conclusion, we found $2$D age-metallicity map reconstructions to be
feasible for only very high ${\mathrm{SNR}}\geq 500$. Since this is
comparable or larger than $\mathrm{SNR_{b}}$, we consider it strongly
unphysical.  Moreover, from an observational point of view, such a high
$(\mathrm{SNR},R)$ combination for an outer galaxy is generally unreachable in reasonable time with the present generation
of instruments. Thus, inversions with this complexity and SNR are doubly
challenging. We now address a simplified version of this problem by
reducing the metallicity parameters to a $1$D \AMR.

\subsection{Non-linear age-metallicity recovery}
\label{s:age-metal}

In the rest of the paper we assume that the chemical properties of the
population are represented by an {\AMR} $Z(t)$ of unknown shape. In
contrast to \Sec{linaz}, the sub-population of age $t_j$ is
therefore assigned one and only one metallicity $Z_j$ rather than a
metallicity distribution. In addition, we now allow the {\SED} to be
affected by an extinction $\fext(E,\lambda)$ parameterized by the color
excess $E$.  Finally, accounting for the age distribution $\Lambda(t)$, the
observed {\SED} at rest then writes:
\begin{equation}
  \Frest(\lambda) =
  \fext(E,\lambda)\,\int_\Tmin^\Tmax\Lambda(t)\,
  \Bflux\bigl(\lambda,t,Z(t)\bigr)\,\mathd t  \, .
  \label{e:v3}
\end{equation}
This model is linear in age distribution $\Lambda$, and non linear in
metallicity $Z$ and extinction $E$.  Recall that $\fext$ may be replaced by
other parametric functions of wavelength that could for instance describe
flux calibration corrections.

\subsubsection{Discretization and parameters}

Following the same prescription as in \Sec{age}, but accounting for
extinction, we can derive the discretized version of \Eq{v3}.  Provided
the extinction law is very smooth compared to the size of the wavelength
bins, the model of the sampled {\SED} of the reddened composite
stellar population in the $i$-th spectral bin writes:
\begin{eqnarray}
  s_i &=& \int \Frest(\lambda)\,g_i(\lambda)\,\mathd\lambda \nonumber\\
  &\approxeq& \fext(E,\lambda_i)
  \!\int\! g_i(\lambda)\!\int_\Tmin^\Tmax\!\!\Lambda(t)\,
  \Bflux\bigl(\lambda,t,Z(t)\bigr)\,\mathd t
  \,\mathd\lambda ,
\end{eqnarray}
which simplifies to:
\begin{equation}
  s_i = \fext(E,\lambda_i)\,\sum_{j=1}^{n} B_{i,j}\,x_j
  \,,\quad i \in \{1,..,m\} \, ,
\end{equation}
or in matrix form:
\begin{equation}
  \label{e:model-aze}
  \M{s} = \diag\bigl(\Fext(E)\bigr)\cdot\M{B}\cdot\M{x}\, ,
\end{equation}
where the kernel matrix $\M{B}$ and the vector $\M{x}$ of the age distribution
$\Lambda(t)$ sampled upon time are defined as in \Sec{age}, and
$\diag(\Fext)$ is the diagonal matrix formed from the extinction vector:
\begin{equation}
  \Fext(E)=\T{\bigl(\fext(E,\lambda_1),\ldots,\fext(E,\lambda_m)\bigr)}\,.
\end{equation}
which contains the extinction law seen by the population and depends
non-linearly on the color excess ${E}$.  Note that $\M{B}$ contains the
{\SSP} basis for the {\AMR} vector $\M{Z}$ (the {\AMR} $Z(t)$ sampled in
time).

From a computational point of view, any matrix product involving 
$\diag\bigl(\Fext(E)\bigr)$ is very expensive and can be
profitably implemented using term to term product. However, in order to
save the introduction of confusing operators, we will carry on with the current notation.

\subsubsection{Smoothness a priori with MAP}

The model defined by \Eq{model-aze} is non-linear because of the
dependancies of $\M{T}$ and $\M{B}$ on respectively $E$ and $\M{Z}$. We can
therefore not refer to the classical definition of ill-conditioning.
However, since the simpler problem solved in \Sec{age} is ill-conditioned,
it is expected that the more complex problem treated here will be even more ill-conditioned,
all the more since we now seek two fields plus one extinction parameter. We
will thus add a priori information by implementing smoothness constraints,
and allow the unknowns to have different smoothing parameters. We define
the penalizing function $P_{\mathrm{smooth}}$ by
\begin{equation}
P_{\mathrm{smooth}}(\M{x,Z}) \equiv
\mu_{\M{x}}
 P(\M{x}) +  \mu_{\M{Z}} P(\M{Z}) \, ,
\end{equation}
where $P$ is the standard quadratic function defined by (\ref{e:defP}).
%
%
\subsubsection{Metallicity bounds}
\label{s:bz}
The metallicity range $[\Zmin,\Zmax]$ for which models are available is
bounded.  We must therefore find a way to ensure that the solution lies in
the desired metallicity range by making unwanted values of $Z$
unattractive. To do this we use a binding function $c$ ($c$ stands for
constraint) which is another kind of penalizing function. This technique
was proposed by R. Lane (private
communication\footnote{http://www.elec.canterbury.ac.nz/staff/Academic/rgl/rgl.htm}).
The function $c$ must be flat inside $[\Zmin,\Zmax]$ in order not to
influence the metallicity search and increase gradually outside. We define
$c$ piecewise by
\begin{equation}
  \displaystyle
  c(Z)=
  \quad
  \left\{\begin{array}{ll}
    (Z-\Zmin)^2  & \mbox{if } Z \leq \Zmin \, ,\\
    (Z-\Zmax)^2  & \mbox{if } Z \geq \Zmax \, ,\\
    0 & \mbox{else} \, .
  \end{array}\right.
\end{equation}
The binding function $C$ used in practice is defined by
\begin{equation}
  C(\M{Z})=\sum_j c(Z_j) \, .
  \label{e:defC}
\end{equation}
The penalization function we finally adopt  is
\begin{equation}
  P_{\mu}(\M{x},\M{Z}) \equiv P_{\mathrm{smooth}}(\M{x},\M{Z})
  + \mu_{C} \, C(\M{Z})\, ,
\end{equation}
where a binding parameter $\mu_{C}$ allows to set the repulsiveness of the
exterior of $[\Zmin,\Zmax]$. The objective function,
\begin{displaymath}
  Q_{\mu} = \chi^2(\M{s}(\M{x},\M{Z},E)) + P_\mu(\M{x},\M{Z})\,,
\end{displaymath}
is now fully characterized. Its derivatives are given in the
appendix~\ref{s:agemin}.
\subsection{Simulations of metal dependent LWSAD}
\label{s:sims-aze}
We applied the proposed inversion method to mock data for various stellar
age distributions, {\AMR}s, extinctions and SNR. In this case, choosing an
input model involves choosing the functions $\Lambda(t)$, $Z(t)$, and a
color excess $E$. The corresponding model spectrum is then computed
following (\ref{e:model-aze}). Gaussian noise is added to obtain the
pseudo-data.

\begin{figure*}
\begin{tabular}{ll}
{\includegraphics[width=0.35\linewidth,clip]{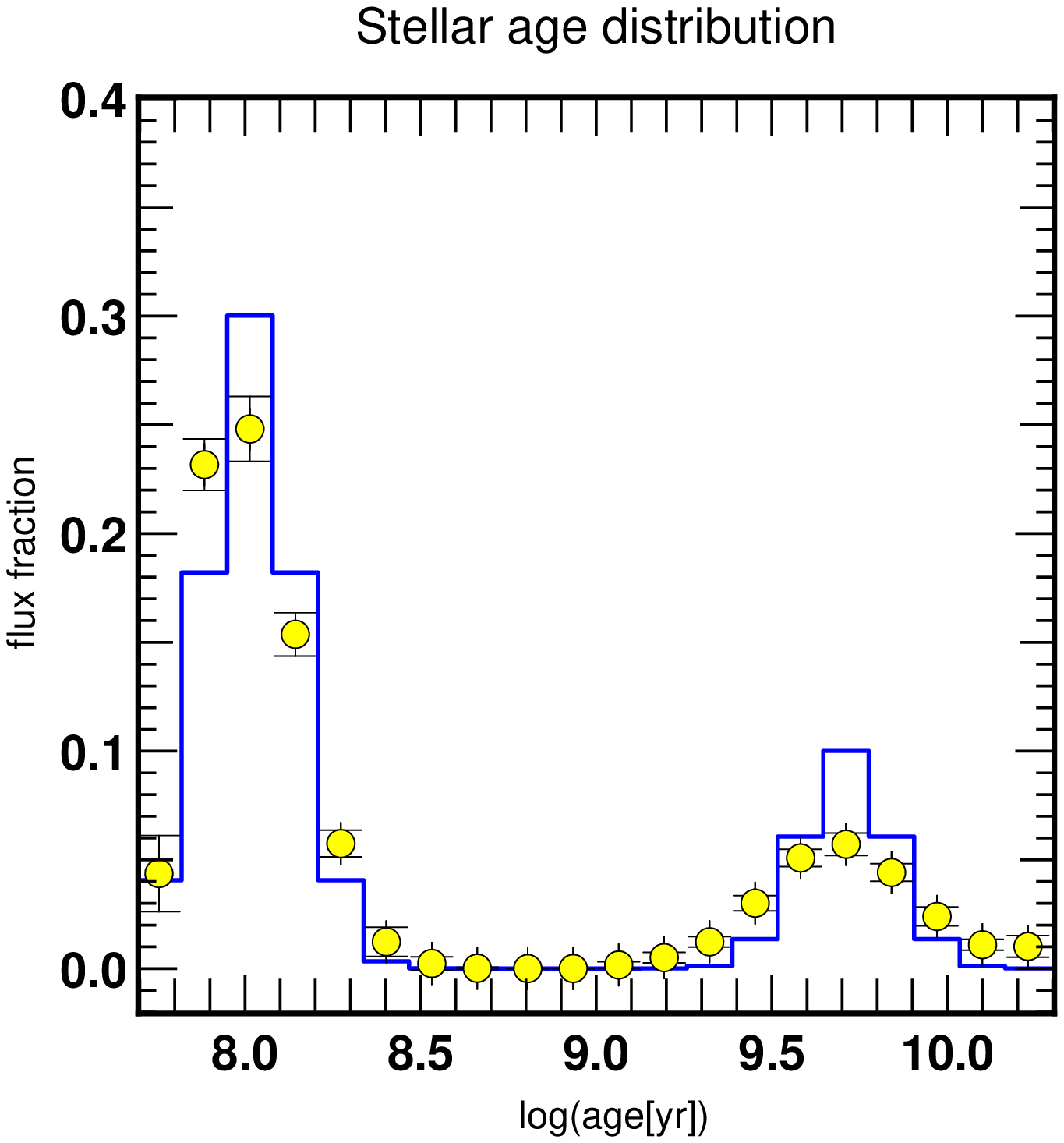}} &
{\includegraphics[width=0.35\linewidth,clip]{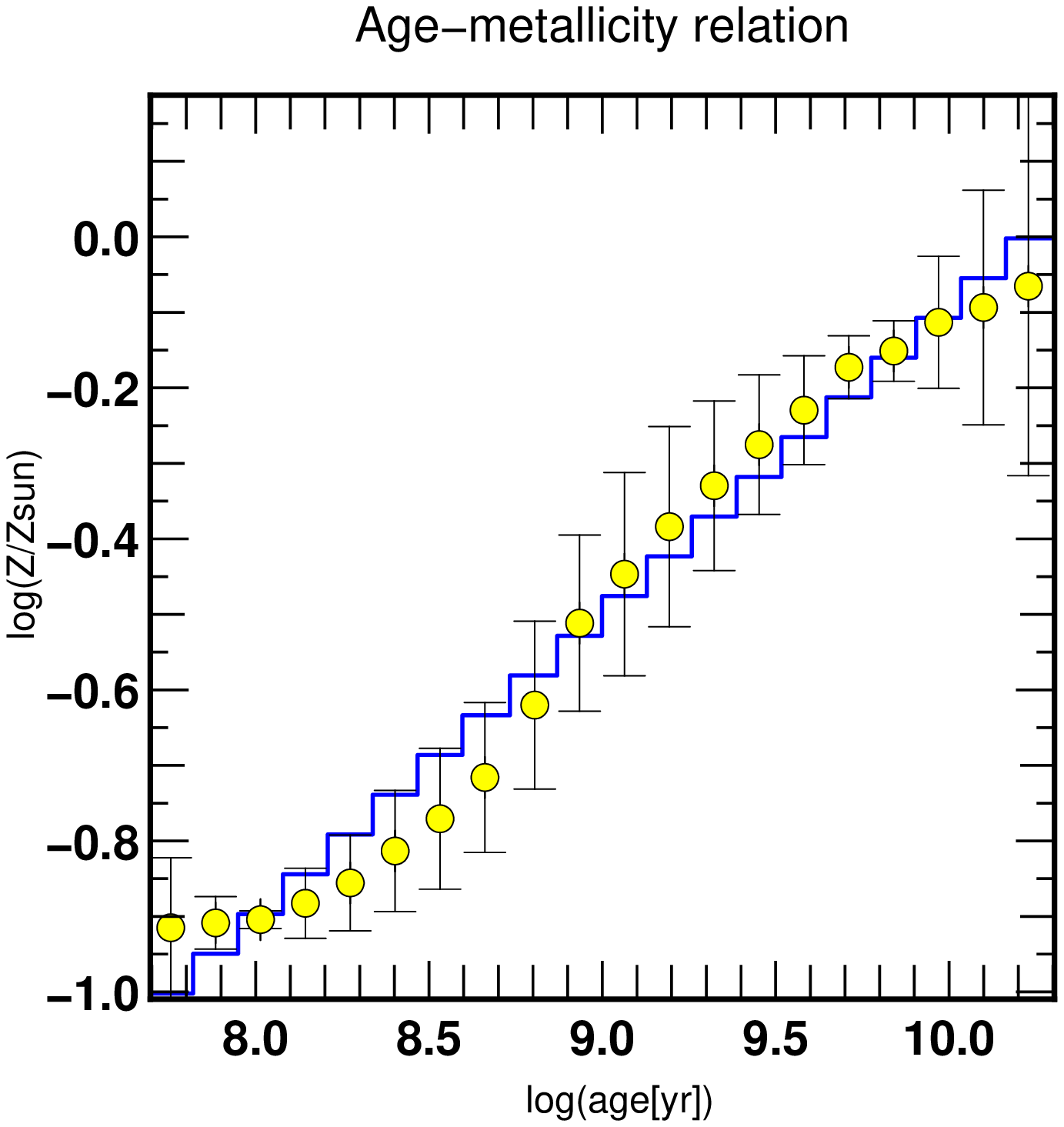}}
\\
{\includegraphics[width=0.35\linewidth,clip]{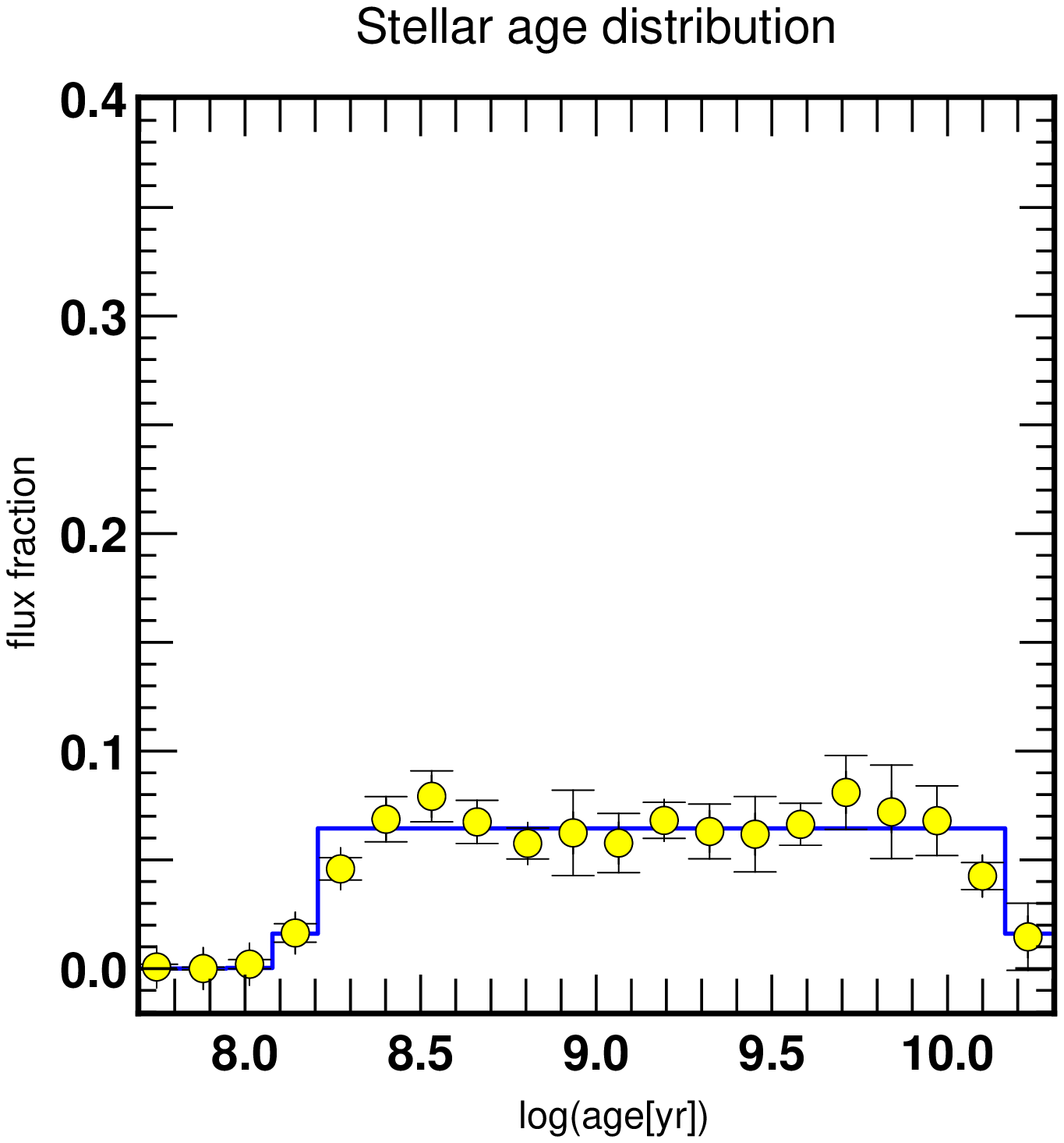}} &
{\includegraphics[width=0.35\linewidth,clip]{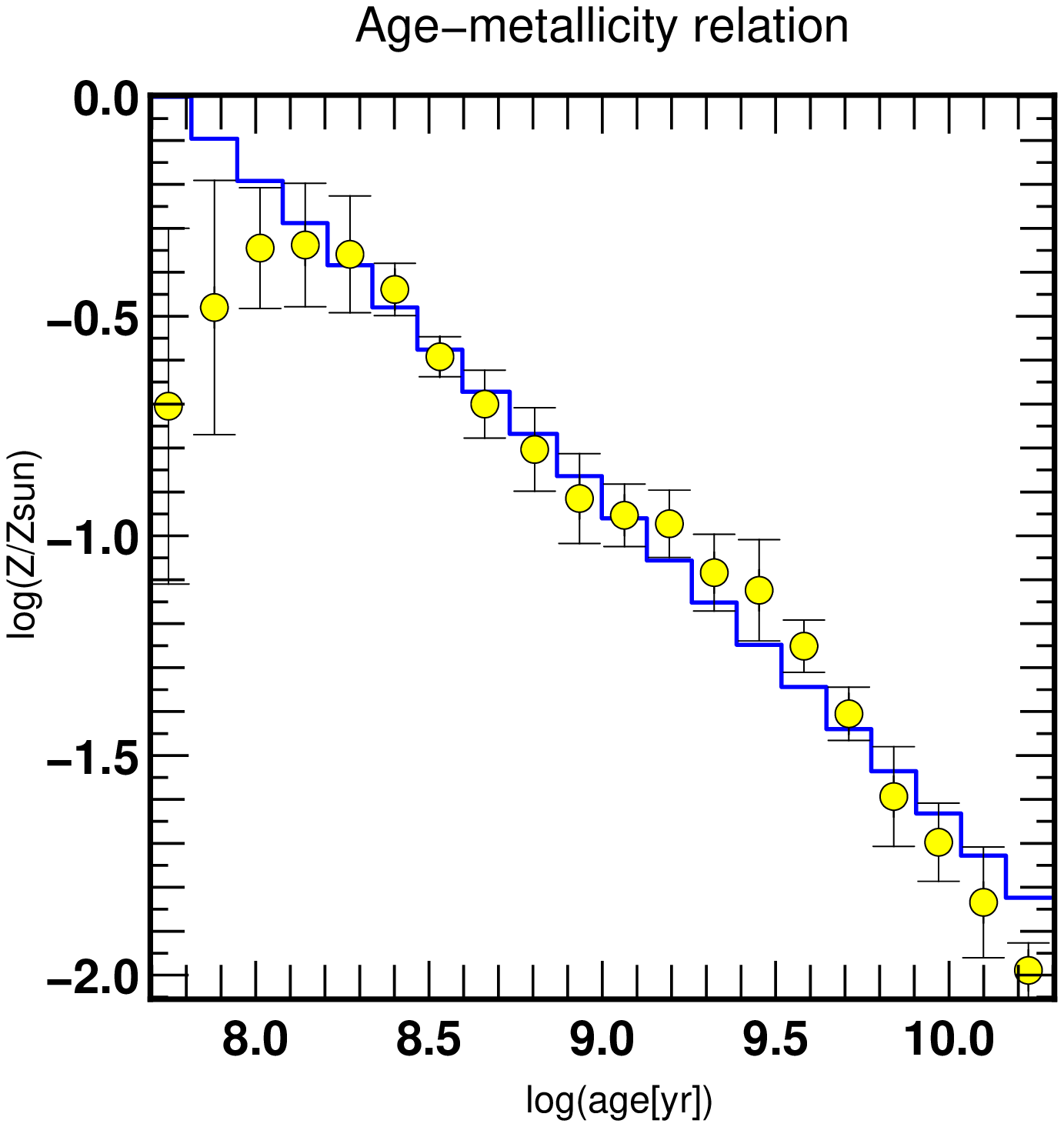}}
\end{tabular}
\caption{Reconstruction of the stellar age distribution (left) and {\AMR} (right)
for $R=10\,000$
and $\mathrm{SNR}=100$ per pixel. The thick
line is the input model. The circles and the bars show respectively the median
and the interquartiles of the recovered solutions for 100 realizations. The metallicities
and flux fractions of the populations with significant contributions are
adequately recovered. In each experiment, the extinction parameter of the
model was
chosen randomly and recovered with good accuracy.
}
\label{f:simaz1e}
\end{figure*}

Figure~\ref{f:simaz1e} shows simulations of reconstructions in the case of
high quality pseudo-data: $R=10\,000$ at $4000-6800$ {\AA} with
$\mathrm{SNR}=100$ per pixel for $100$ realizations. The left panels show
the {\SAD} while the right panels show the \AMR s.  The top row shows
reconstructions of a double-burst population where the two bursts have
different luminous contributions. The young component accounts for $75 \%$
of the light, and its metallicity is a tenth of the old component's, which
contributes only to $25 \%$ of the total light. 
The unbalance between the young and old luminous contributions should make
it more difficult to constrain the old component. Still, the
reconstructions are good in the sense that the bumps are properly centered
and scaled. Metallicities are also adequately recovered.  The reconstructed
{\SAD s} are smoothed versions of the model, as expected.

The bottom line plots illustrate the case of a continuous rather than bumpy
{\SAD}. All ages contribute equally to the light except the youngest and
oldest ones. The model {\AMR} yields a metallicity $Z(t)$ that increases
with time. The rise and decay of the recovered age distribution are
adequately located, and the metallicities have the correct trend. The
metallicities of the youngest component are unconstrained simply because they
do not contribute to the total light.

For each
realization of these simulations, the color excess was a random number
between $0$ and $1$. In each case, it was recovered with an accuracy better
than $10^{-2}$.

\subsection{Age separation of metal dependent LWSAD}
\label{s:sep-aze}

In a realistic observationnal setting, we would like to age-date
superimposed populations. For such investigations, it is essential
to have a good understanding of the limitations of the non-parametric
method.  We therefore investigated again how well we could reconstruct two
superimposed bursts of unknown metallicities and extinction. We
proceeded as in \Sec{agesep}, and the grid of double burst ages is the
same. Both bursts contribute equally to the total light. In a first
set of experiments, the model {\AMR} is arbitrarily chosen as
$\log(Z)=-9.95+0.85 \, $ log(age[yr]), where the age ranges from $50$ Myr to
$15$ Gyr. It is not supposed to be a physically motivated choice, but allows us to
explore about $2$ decades in metallicity. The allowed range for the
solution {\AMR} is $[\Zmin=0.0004, \Zmax=0.05]$.  The extinction
parameter was chosen randomly between $0$ and $0.5$.  The reconstructions
were performed without any a priori for the {\AMR}, {\SAD} or extinction
parameter, apart from the requirement of smoothness. For each pseudo-observational context,
the smoothing parameter was set using the GCV value for the mono-metallic
case and fine-tuned for a small separation between 2
bursts. The smoothing parameter for the {\AMR} was set to a large value
(around $10^3$) because we just wish to recover a global trend of the
metallicity evolution in the reconstruction.  A flat guess for all
variables was the starting point. In every case we converged to a stable
solution in less than $1500$ iterations, corresponding to $\approx 1$ minute on a $1$ Ghz pc
for a $R=10000$ basis (i.e. $14000$ pixels of $0.2$ \AA) with $60$ age bins. The distribution of the reduced $\chi^2$ of the
solutions were found to follow a Gaussian distribution law with unit mean,
showing that each experiment had properly converged.

\begin{figure*}
\begin{tabular}{ll}
\resizebox{6cm}{6cm}{\includegraphics{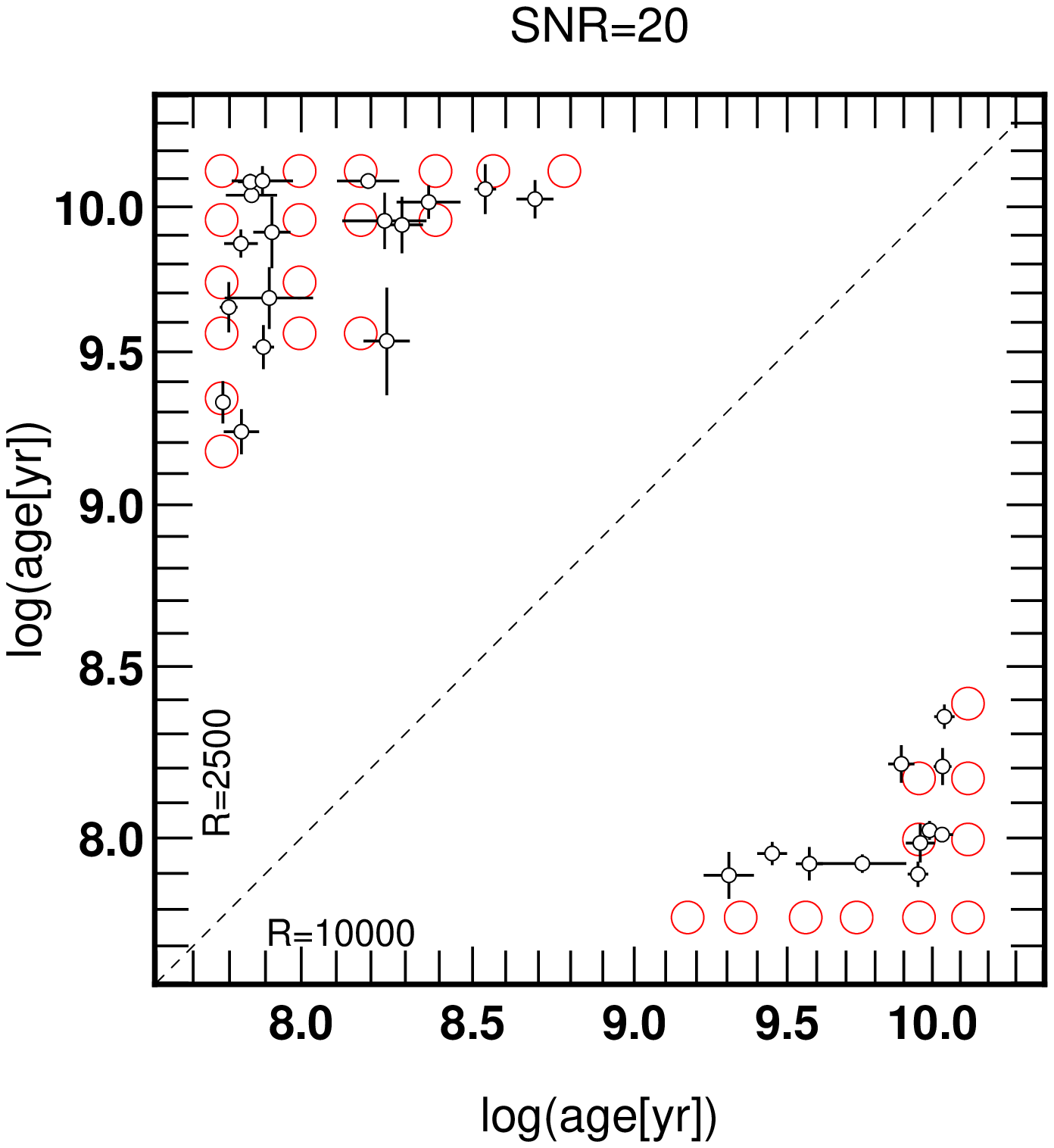}} &
\resizebox{6cm}{6cm}{\includegraphics{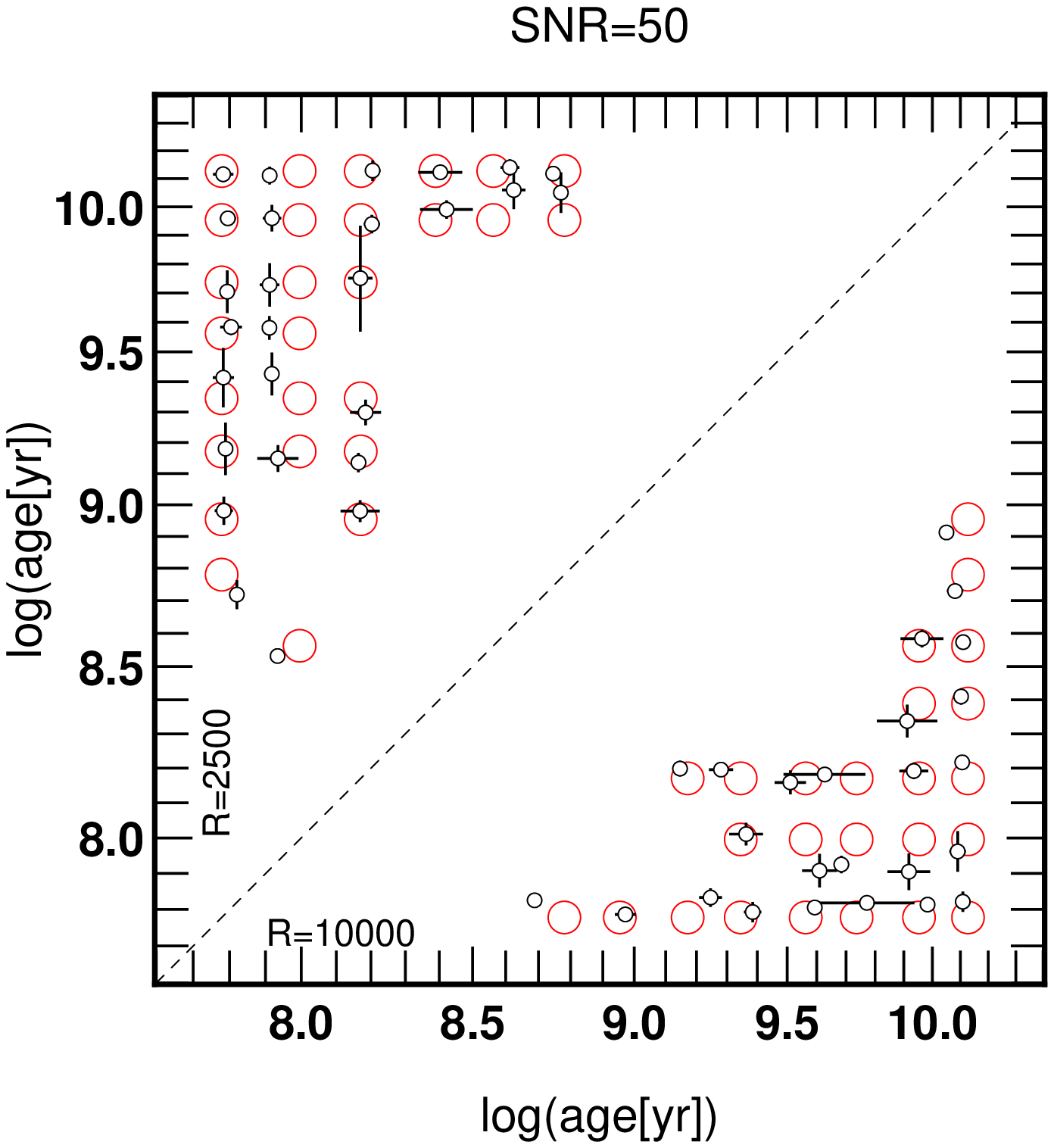}} \\
\resizebox{6cm}{6cm}{\includegraphics{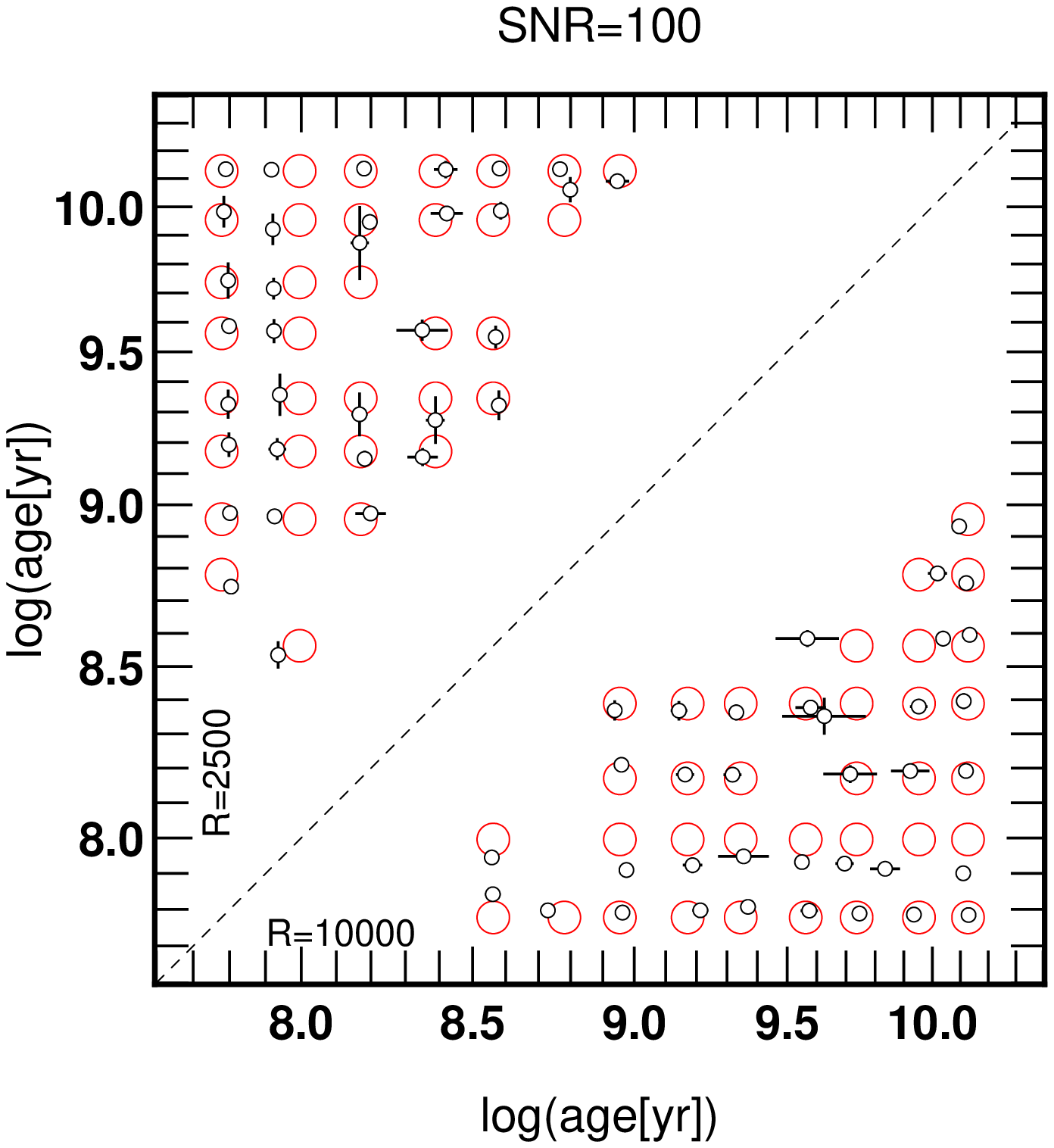}} &
\resizebox{6cm}{6cm}{\includegraphics{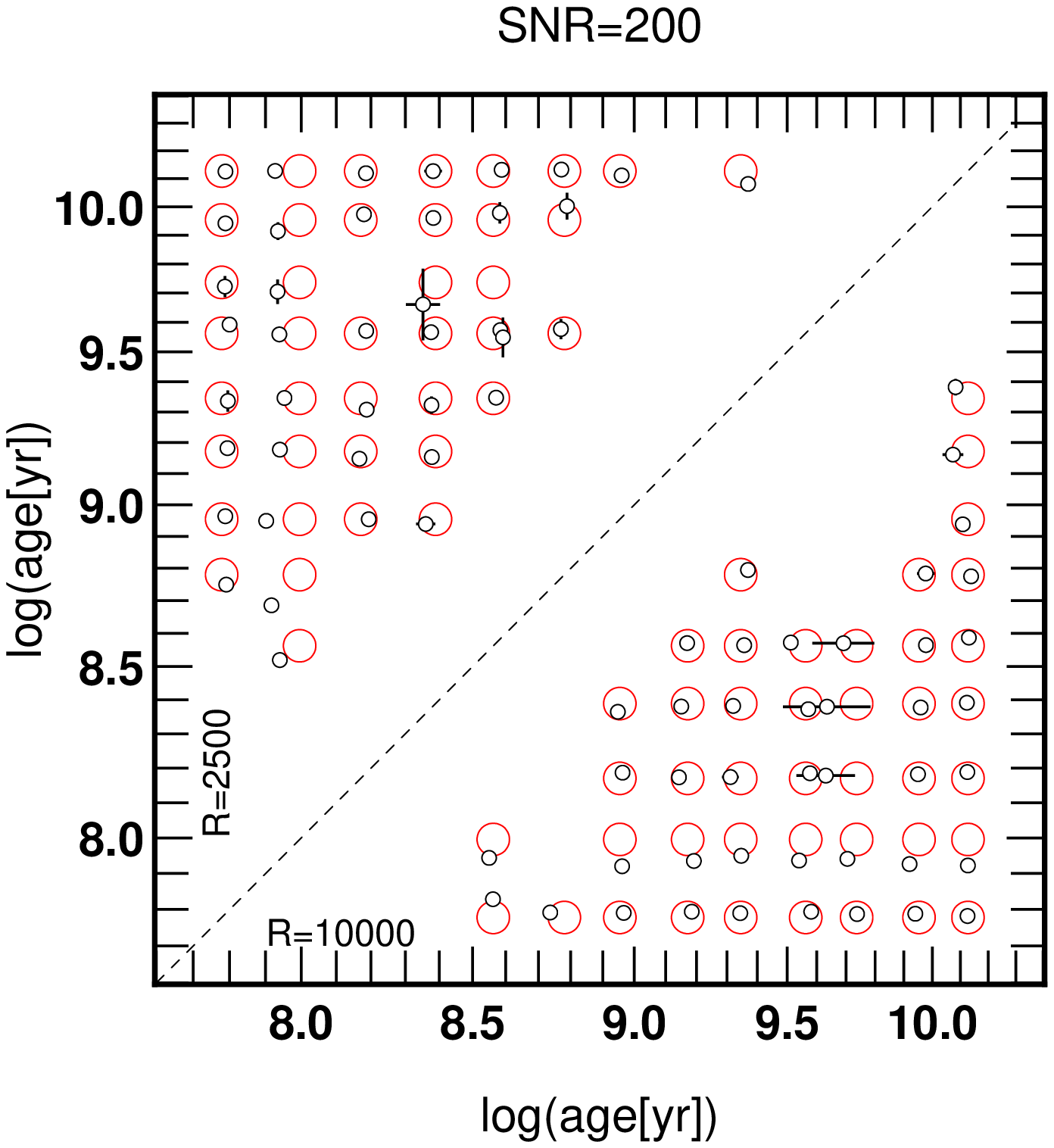}}
\end{tabular}
\caption{Same as \Fig{r-snr-sim} but the metallicities and the
extinction are free parameters. The SNR is given per \AA. The ability to separate close sub-populations
improves with SNR, as does the accuracy of the age estimates.
}
\label{f:simaze}
\end{figure*}

\begin{figure*}
\begin{tabular}{lll}
{\includegraphics[width=0.32\linewidth,height=5.5cm]{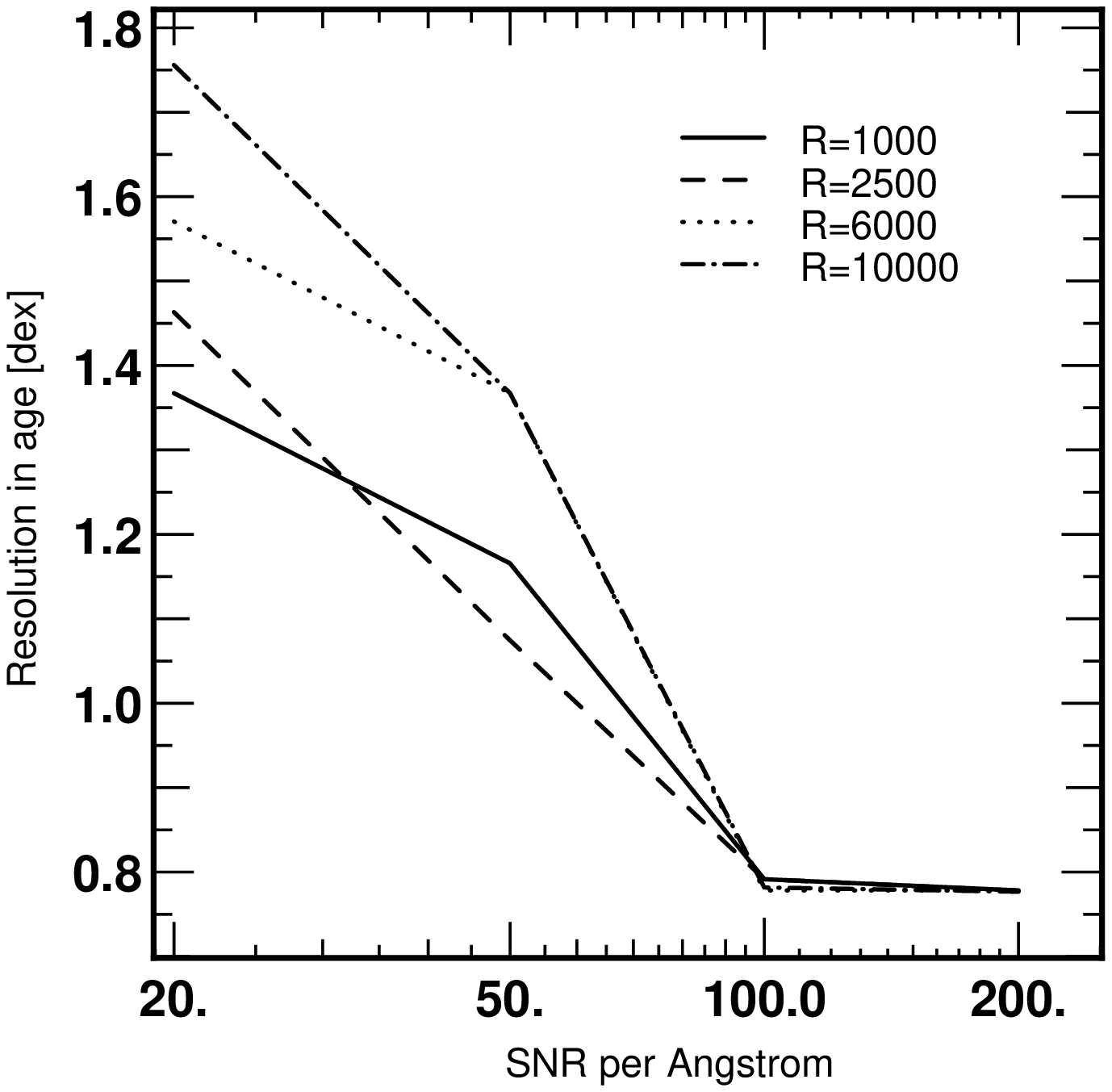}}&
{\includegraphics[width=0.32\linewidth,height=5.5cm]{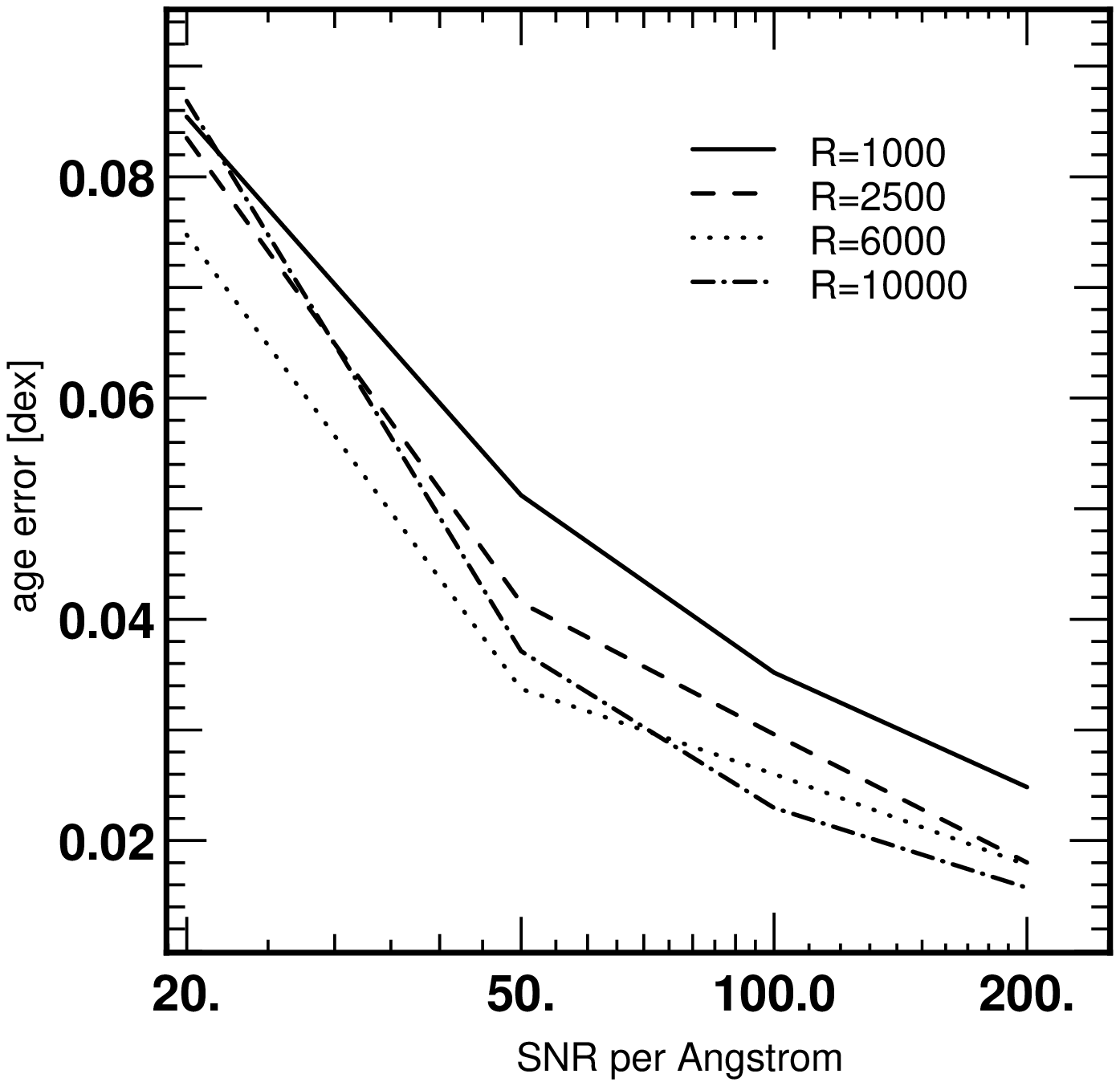}}&
{\includegraphics[width=0.32\linewidth,height=5.5cm]{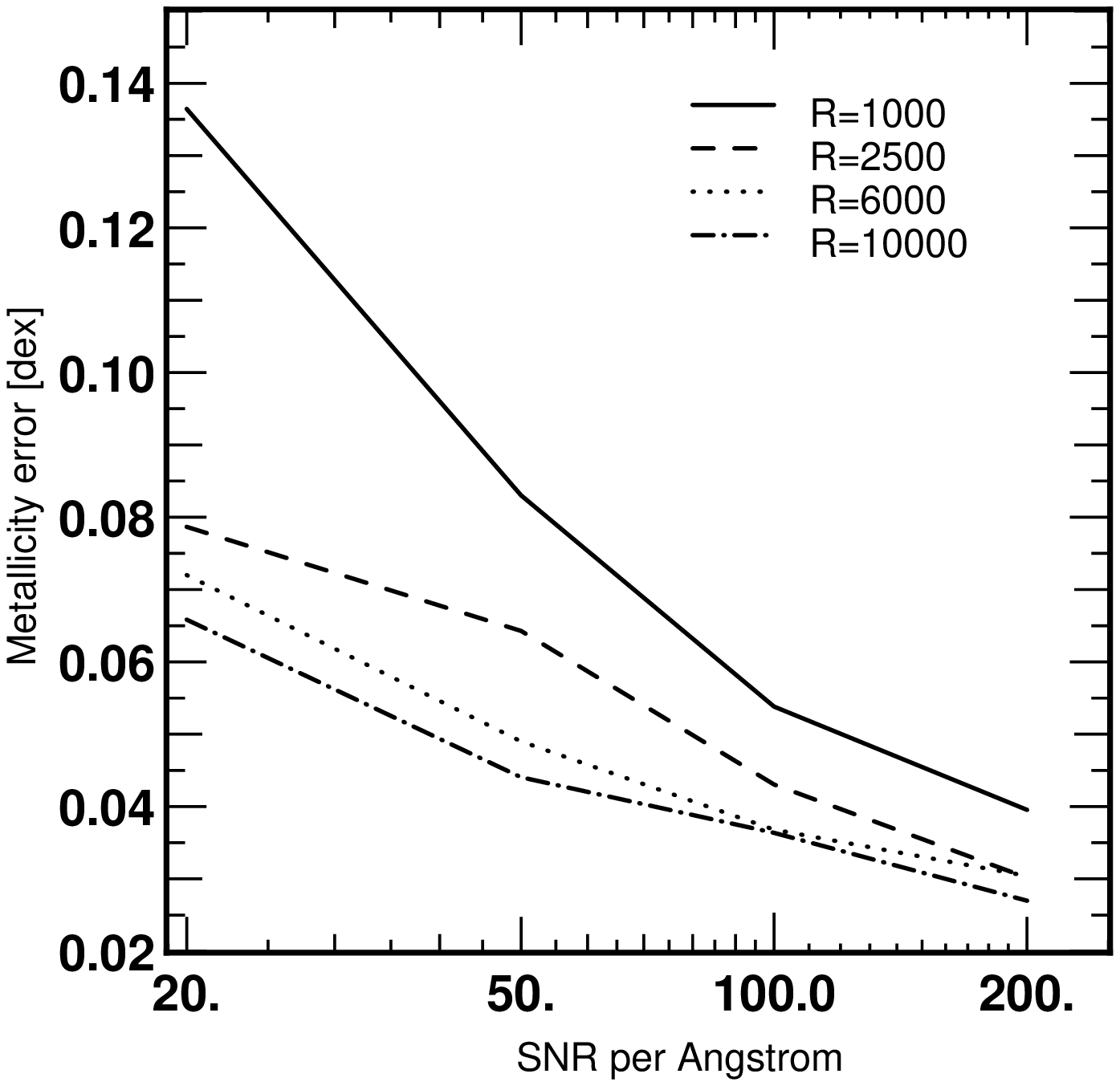}}
\end{tabular}
\caption{\emph{Left}:Resolution in age [dex] versus SNR per {\AA} for various spectral resolutions.
As expected, the resolution in age improves with increasing SNR. It settles around $0.8\;\mathrm{dex}$ for the highest SNR. No significant trend is seen with
spectral resolution. \emph{Middle}: Median error on the age of the bursts [dex] in the successful
separations vs SNR for several resolutions. High resolution experiments give
the smallest errors. \emph{Right}: Same as middle panel but for metallicity
estimates. Again, the best accurracy is obtained at high spectral resolution,
given the same total number of photons.
}
\label{f:resvssnr}
\end{figure*}

We are thus able to give an estimate of the resolution in age versus SNR
and spectral resolution.  Figure~\ref{f:simaze} shows some of the results
of our simulation campaign. On each panel we plotted the results obtained
at $R=2\,500$ (upper octan) and $R=10\,000$ (lower octan). The results for
$R=1\,000$ and R=$6000$ are very similar and are not shown.  The number of
successful inversions rises with increasing SNR, and the unseparable zone
in the diagram shrinks. In the same way, the error bars and bias reduce
with increasing SNR. We give the resolution in age for several SNR per
{\AA} and spectral resolutions in \Fig{resvssnr}. It improves with
increasing SNR, but settles around $0.8\;\mathrm{dex}$ for very high
quality data. The variation of the resolution in age with spectral
resolution is not significant compared to the statistical error
($\approx0.25\;\mathrm{dex}$), so that no trend with spectral resolution
can clearly be deduced. The middle panel shows the median error on the
luminous weighted ages of the two bursts for the successful separations.
The error decreases with increasing SNR down to $0.02\;\mathrm{dex}$ for
$\mathrm{SNR}=200$ per {\AA}, and is significantly lower for the high
resolution experiments (the relative statistical error for this measure is
smaller than $5\%$). We see the same trend in the metallicities median
errors of the double bursts, in the right panel. The smallest error is
obtained for the highest spectral resolution. The general smallness of
these errors is partly explained by the severity of the selection, that
rejects as non separable any ambiguous solution.


Somewhat unexpectedly, the results do not depend on the slope of the age-metallicity relation adopted
for the double-burst models. With a negative slope, a young metal-rich
population is added to an old metal-poor one. In view of the age-metallicity
degeneracy, this should be the least favorable situation for a proper
separation. We performed simulations with positive and negative slopes and
obtained identical results considering the statistical errors given
above. Thus, the age-metallicity degeneracy  is not a limiting factor in our experiment.

\section{Conclusions and prospects}

{Let us sum up our findings relative to the diagnosis of the linear (mono-metallic) problem
(\Sec{age} and \ref{s:valid})
and the  more
realistic non linear problem of recovering simultaneously the \LWSAD, the extinction and the  \AMR\, (\Sec{aze}) in turn,
and close on the observational and methodological prospects of \STECMAP.}

\label{s:1stc}
\label{s:2ndc}

\subsection{Probing the linear problem: the tricks of the trade}
The idealized problem of recovering the non-parametric {\SAD} of a
mono-metallic population seen without extinction is linear. The {\CN} of the kernel is very large and accounts for the ill
conditioning of the problem, i.e.\ pathological sensitivity to noise in the
data.

The noise in the {\SSP} models also limits the number of free parameters that may be
recovered robustly to describe the star formation
history. In textbook inversion problems,
 this number can be estimated quantitatively from the sequence of
singular values of the single stellar population basis.
Here however, this theoretical value is misleading because
the expected signature of the model noise in the singular value spectrum is
not apparent.
We explained this by the correlations between the noise patterns in subsequent
basis spectra.
To obtain the number of free parameters, the singular values are used together
with an independent estimate of the SNR of the basis.
For the optical spectral range covered with Pegase-HR and ages ranging from
$50$ Myr to $15$ Gyr,
the corresponding number is 6. This  makes high frequency variations
of the stellar age distribution unrecoverable, no matter the
data quality, $\mathrm{SNR}_\mathrm{d}$, and the inversion method.

When the dominant error source is the data, the problem may be regularized
by truncating the \SVD\, or reducing the number of age bins so that
$\sigma_1/ \sigma_n \leq \mathrm{SNR_d} \sqrt{m}$. This crude rule can be used
to obtain a quick estimate of the performance  expected for a given
dataset.

The problem can be more profitably regularized without reducing arbitrarily
the number of age bins by imposing the smoothness of the solution, to
obtain a penalized likelihood estimate. This constraint reduces the risk of
overinterpreting the data. The smoothing parameter is set automatically by
\GCV\ for each ${\mathrm{SNR_d}}$, or/and by performing simulations in a
suited pseudo-observational context.

For an adequately regularized problem, we defined the inverse model matrix
and inspected it in order to find the wavelength ranges which are most
discriminative for age determinations. We found that the information is
widely distributed along the optical range (cf.\ Figs.~\ref{f:invmodel} and
\ref{f:invmodelaz}).

The behaviour of the inversion can be predicted by inspecting the {\SVD} or
{\GSVD} of the kernel. The first non-attenuated solution vectors are
responsible for the detailed shape of the regularized reconstructions, and
thus for the generation of artifacts. The general shape of the solution
vectors, and especially the presence/absence and location of their
oscillations, gives an indication in which age ranges the inversion behaves
worst.

{In particular, the inspection of the SVD components revealed that the problem
of recovering flux distributions was less pathological than the problem of
recovering mass fractions. More specifically, the transition rank $i_0$
between signal and noise dominated regimes is independent
from the fiducial model in the recovery of flux fractions.}

Second or third order penalizations gave similarly good results, showing
that the quality of the inversion does not rely strongly on the details of the
regularization.

Requiring the solutions to be positive improves the results
even further, and in particular reduces Gibbs ringing, as can be seen by
comparing \Fig{1bsim} and \Fig{1bqsim}.

One should be aware that the
efficiency of the inverse method cannot be assessed on the basis
of a small set of simulations. Indeed, it is easy to produce good looking
results down to ${\mathrm{SNR_d}}=0.1$ per pixel by carefully choosing the model
age distribution.

We performed an extensive simulation campaign by inverting a
grid of double burst models in several pseudo-observational regimes. If the age
difference between the bursts is larger than $0.4\;\mathrm{dex}$, we were able to separate
the $2$ components and recover their ages with a very small error from high
quality data ($\mathrm{SNR_d}=200$ per {\AA}).

However, the high $\mathrm{SNR_d}$ regime for which we obtained the best
results are questionable.
Indeed, when ${\mathrm{SNR_d}}$ and ${\mathrm{SNR_b}}$ are comparable,
the number of degrees of freedom is imposed by the noise in the basis rather than
in the data.
We therefore consider the extreme regimes with ${\mathrm{SNR_d}}
\geq 200$ per {\AA} unphysical: small oddities (of uncertain nature) in the basis are seen as physically
discriminative information.
Only an improvement of $\mathrm{SNR_b}$ could in principle increase the number of
degrees of freedom. Assuming that the singular values spectrum of the initial
kernel shown in \Fig{SVstuff} is representative of the basis even at higher $\mathrm{SNR_b}$,
we can set the following rules of thumb.
\begin{enumerate}
\item If for example $\mathrm{SNR_d}=100$ per pixel, the maximum number of
  freedom degrees one may consider is of order $8$ ($n=8$ from criterion
  (\ref{e:trunc}) or \Fig{picard}).
\item To ensure that no serious contamination of the singular values by
  noise in the basis happens for $i<8$, one would need $\mathrm{SNR_b}\geq
  1\,000$ per pixel (estimated from \Fig{SVstuff}) ($2\,500$ per
  {\AA}). We caution that this is an extrapolation, and that the actual
  behaviour of {\SSP} spectra at this kind of SNR is not known.
\end{enumerate}

{By comparing the solutions given by {\SVD} and the {\GSOd} kernel we showed
that ill-conditioning remains an issue when working with compressed data.}
%

Finally, the mismatch observed when a mono-metallic population is fitted by a basis of different metallicity allowed us to
constrain this additional metallicity parameter with a $\mathrm{SNR_d}$ as small as $10$ per pixel,
well enough to motivate a feasibility study of the recovery of the age
distribution, the metallicities and the reddening of a composite
stellar population.

\subsection{Beyond the mono-metallic inversion?}

The ill-conditioned problem of recovering a $2$D age-metallicity
distribution of a composite unreddened population can also be recast into a linear
problem. A penalized likelihood estimate can be obtained by means of additionnal smoothness
constraints. The inspection of the regularized inverse model matrix reveals
that a large number of age and metallicity sensitive lines carrying
discriminative information are located all along the optical range.
The shape of the first solution singular modes shows that age-metallicity
degenerate solutions are expected even for $\mathrm{SNR_d}$ as large as
$200$ per pixel. Notwithstanding the above caveat about high SNR, the inversions with such a complexity are thus unfeasible in realistic
regimes from optical integrated light only.

A natural simplification involves assuming that the metallicity of the population can be
described by a one to one non parametric {\AMR}. The problem of recovering the
{\SAD}, the {\AMR} and an extiction parameter then becomes tractable provided
that adequate regularization (smoothness, bound and positivity) is
implemented, and yields a penalized likelihood estimate.

A detailed simulation campaign allowed us to estimate the resolution in
age that can be achieved from optical data in several pseudo-observational
regimes.  If the time elapsed between $2$ instantaneous bursts is larger
than $0.8\;\mathrm{dex}$, they can be separated unambiguously by \STECMAP\ from high
quality data ($\mathrm{SNR_d}=100$ per {\AA}), and their ages and metallicities
can be constrained with an accuracy of respectively $0.02\;\mathrm{dex}$
and $0.04\;\mathrm{dex}$. In such regimes, the age-metallicity degeneracy
is effectively broken. For smaller separation, there is always a mono-burst
or smoother solution that fits the data equally well.  Our experiments
reveal no clear dependency of the resolution in age on the spectral
resolution $R$ ($\geq1\,000$) as long as the SNR per {\AA} (or integration
time) is conserved in the comparative experiments.  As in the preliminary
conclusion for the idealized mono-metallic unreddened problem, it is not
clear whether the extreme $\mathrm{SNR}_\mathrm{d}$ are physical or not, since in
these regimes the noise in the basis is not negligible any more compared to
the noise in the data. In any case, $0.8\;\mathrm{dex}$ should be
considered as a lower resolution limit, for any separation attempt in the
range $\lambda\lambda=[4\,000\;\mbox{\AA}, 6\,800\;\mbox{\AA}]$.

The fact that free extinction does not hinder the inversions indicates that
the continuum is not a critical constraint. Simulations with more complex
corrections on the continuum (not described in this paper) confirm this
point. The information on age and metallicities is carried in the line spectrum.

\subsection{Discussion and prospective}

Perhaps the most intriguing conclusions of this paper are the small number of
degrees of freedom found in an optical {\SSP} basis even with $\mathrm{SNR_b}$ as
large as {\PHR}, and the very anti-intuitive hint that
significantly larger SNR is needed in the basis than in the data to be
analyzed. It highlights the need to study and quantify the influence of
the models noise in linear and non-linear inversions, and to continue and
improve the various steps involved in the construction of the model.

Several directions can be followed, on the basis of
\Sec{problemsSSPs}.  Empirical libraries should improve with the
combination of large collecting areas, and high resolution, large coverage
instruments with massive multi object capacities, which should boost up the construction of
libraries by a significant factor.  The library UVES POP
\citep{uvespop} is an example. With telescopes of the 10m class or
larger, stars in clusters and in Local Group galaxies can be observed to
remedy in part the issue of completeness and some of the biases of solar
neighbourhood libraries (e.g.\ more luminous metal-poor stars, or stars with
modified $\alpha$-element abundances).

On the theoretical side, one should investigate accurately and
systematically what drives the shape of the singular value spectrum of the
SSP basis.  In this paper we concentrated on a \emph{given} SSP model, without
tuning the basis to study the effect of e.g.\ sampling strategies on the
conditioning.  Since the behaviour of an inverse problem depends on the
shape of the solution singular vectors as well, it is a key issue to
understand what drives their shape. Making them smoother and more regular
is a step towards reducing the generation of artifacts.  Clearly, one would
want to question the sampling strategy in $(T,g,Z)$ space in terms of both the conditioning number of $\M{B}$
and the roughness of its singular vectors. In particular, one would like for
instance to
apply an error weighted regularized tomographic interpolation in $(T,g,Z)$
space, in order to construct a noise free spectral basis, which would by
construction prevent from interpolating the noise from one spectrum to
another. Even though the interpolation of the noise patterns of individual stars in
the library may explain the vanishing of the saturation
of the singular values, we still miss a quantitative relation between the
density of library stars in $(T,g,Z)$ space, their SNR, and the slope of
the singular value spectrum.

Ultimately one should aim at designing inverse methods where the errors in
the models are explicitly taken into account (for instance using TLS) in
order to draw a consistent error budget.

The generally very limited separability of successive star formation
episodes in most pseudo-observational settings is in strong contrast with
the results of a number of more optimistic authors.  In particular, if one
is bound to draw cosmological constraints from the stacking of a large set
of noisy \SFHs, it is still essential to check that individual \SFHs\ are
well recovered, since otherwise the median solution is likely to be
dominated by artifacts. Exhaustive testing of the method as we propose is
in this case a mandatory step.

The {\SED} matching procedures and parameter recovery presented here are absolutely not model-dependent and could be used in association
with any other stellar population model as is\footnote{We are preparing a
  public release of the inversion codes}.  It will thus be interesting and
informative to perform the same kind of study (resolution in age,
conditioning) with other existing evolutionary synthesis models, in order
to quantify the amount of information and the constraints to be expected
from observations in other wavelength domains, as the UV, NIR or FIR. It is
expected that increasing the wavelength coverage should improve
significantly the resolution in age and the behaviour of the problem in
general. The possible discrepancies between the models are also a major matter of concern. For instance, are the metallicity constraints using a
given set of {\SSPs} robust to a change of the evolutionary synthesis code?
It will be interesting to test this by producing mock data with one available code
\citep{BC03,GD05} and interpreting them with another one. We expect misfits to arise from
wavelength calibration error, small scale flux calibration errors, and
systematic deviations caused by the use of different evolutive tracks,
IMFs, and stellar libraries. This exercise will allow us to investigate the
amount of error introduced by the models themselves.

The methods we described together with the corresponding error and separability
analysis will be very useful for interpreting large sets of data from
large surveys such as SDSS, 2DFGRS, DEEP2, $\cdots$, and also for upcoming
new generation instruments, especially
high resolution instruments with multi-object or field integral capacities,
for instance FALCON \citep{falcon04}, or MUSE \citep{muse03}.
In this context, astronomers will want to extract kinematical information
as well, and question the relationship between the kinematics and the
nature of the stellar populations. The simultaneous recovery of the
kinematical distribution and the corresponding stellar population via the
non-parametric interpretation of spectra is described in a companion paper.

\section*{Acknowledgments}
We thank the referee, Dr R. Jimenez, for criticisms.
We are grateful to the {\PHR} team for providing us an early version of the
models.  We thank M.~Fioc, R.~Foy, S.~Prunet, A.~Siebert, and J.~Blaizot for useful
comments and helpful suggestions.  We would like to thank D.~Munro for
freely distributing his Yorick programming language (available at
\texttt{http://www.maumae.net/yorick/doc/index.html}), together with its MPI
interface, which we used to implement our inversion algorithm in parallel.  We thank
UKAFF and the MPA-Garching for their hospitality. This work was partly
supported by an EARA studentship.



\bibliographystyle{mn2e}
\bibliography{mybib}
\appendix
\section{Dependence of the signal-noise transition on the fiducial model}
\label{s:fid1}
%
\begin{figure*}
{\includegraphics[width=0.32\linewidth,clip]{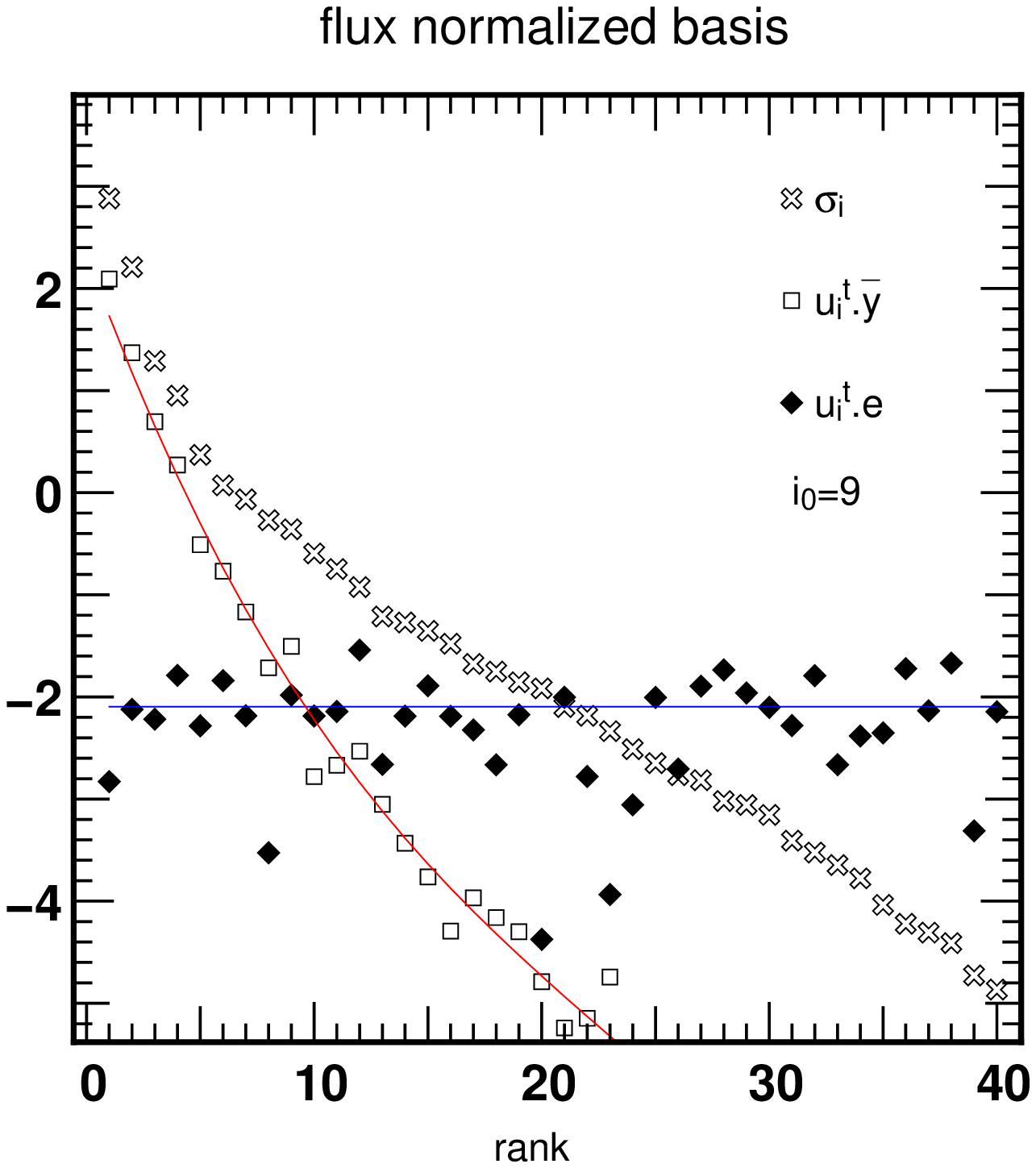}} 
{\includegraphics[width=0.32\linewidth,clip]{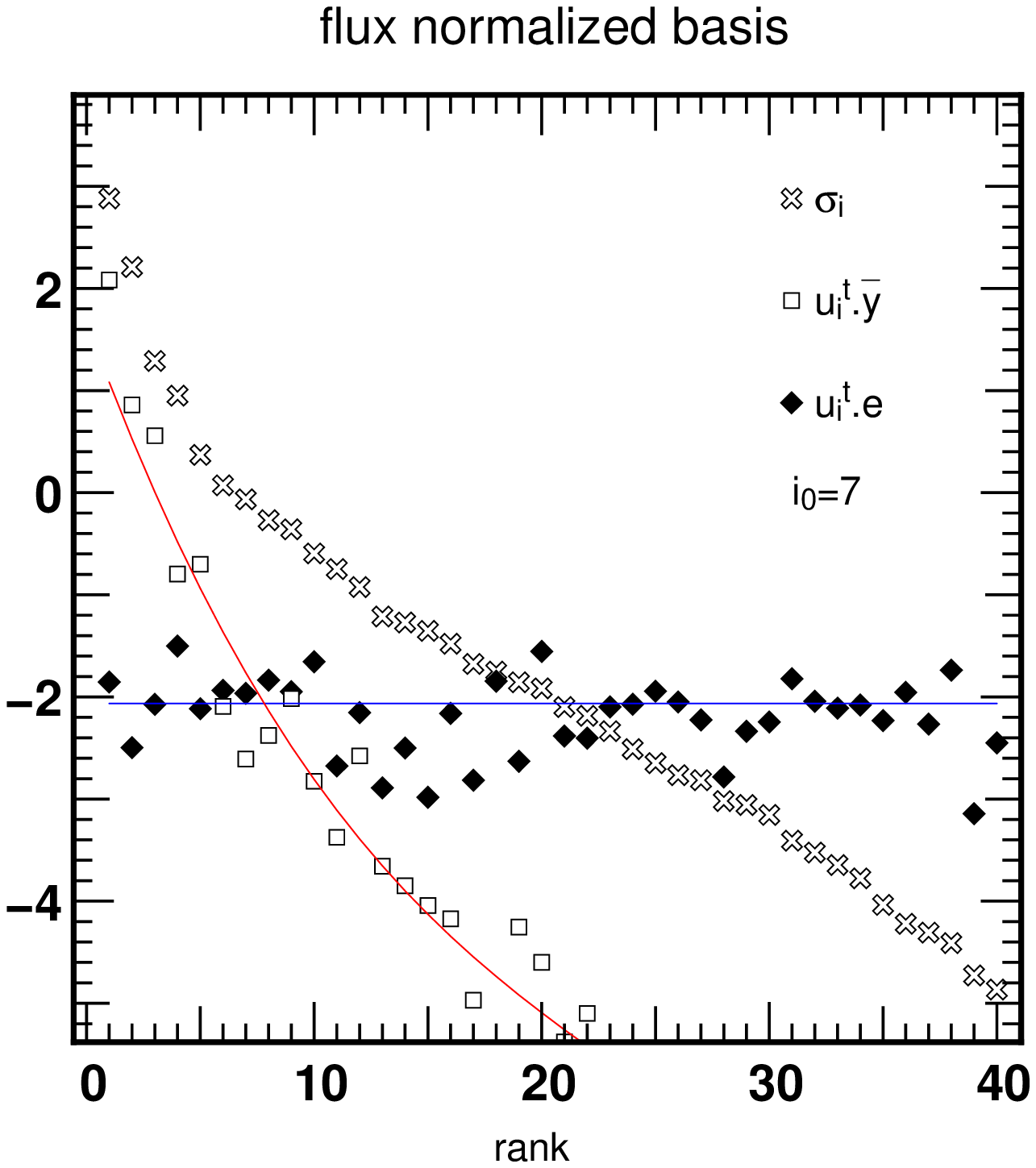}} 
{\includegraphics[width=0.32\linewidth,clip]{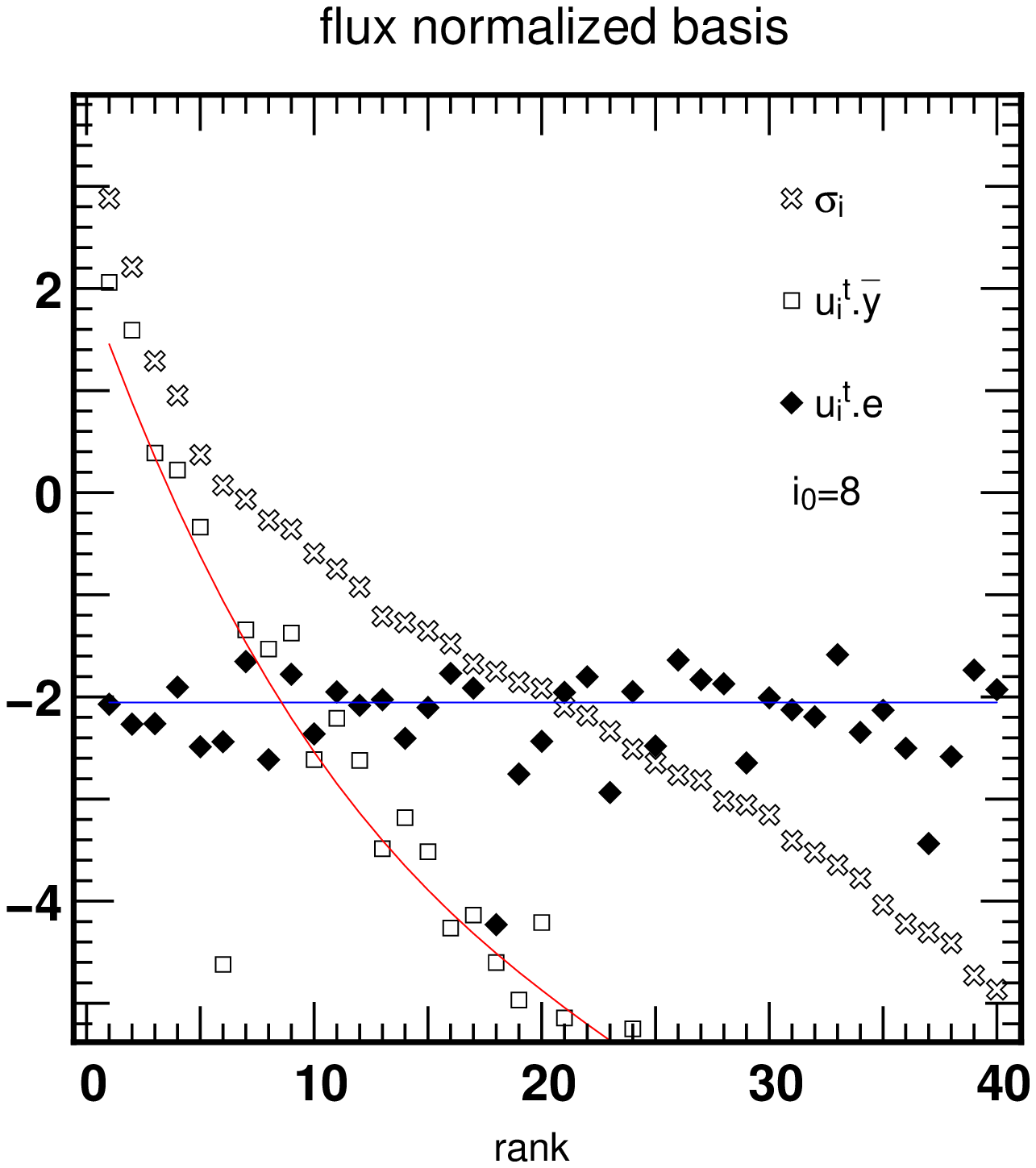}} 
{\includegraphics[width=0.32\linewidth,clip]{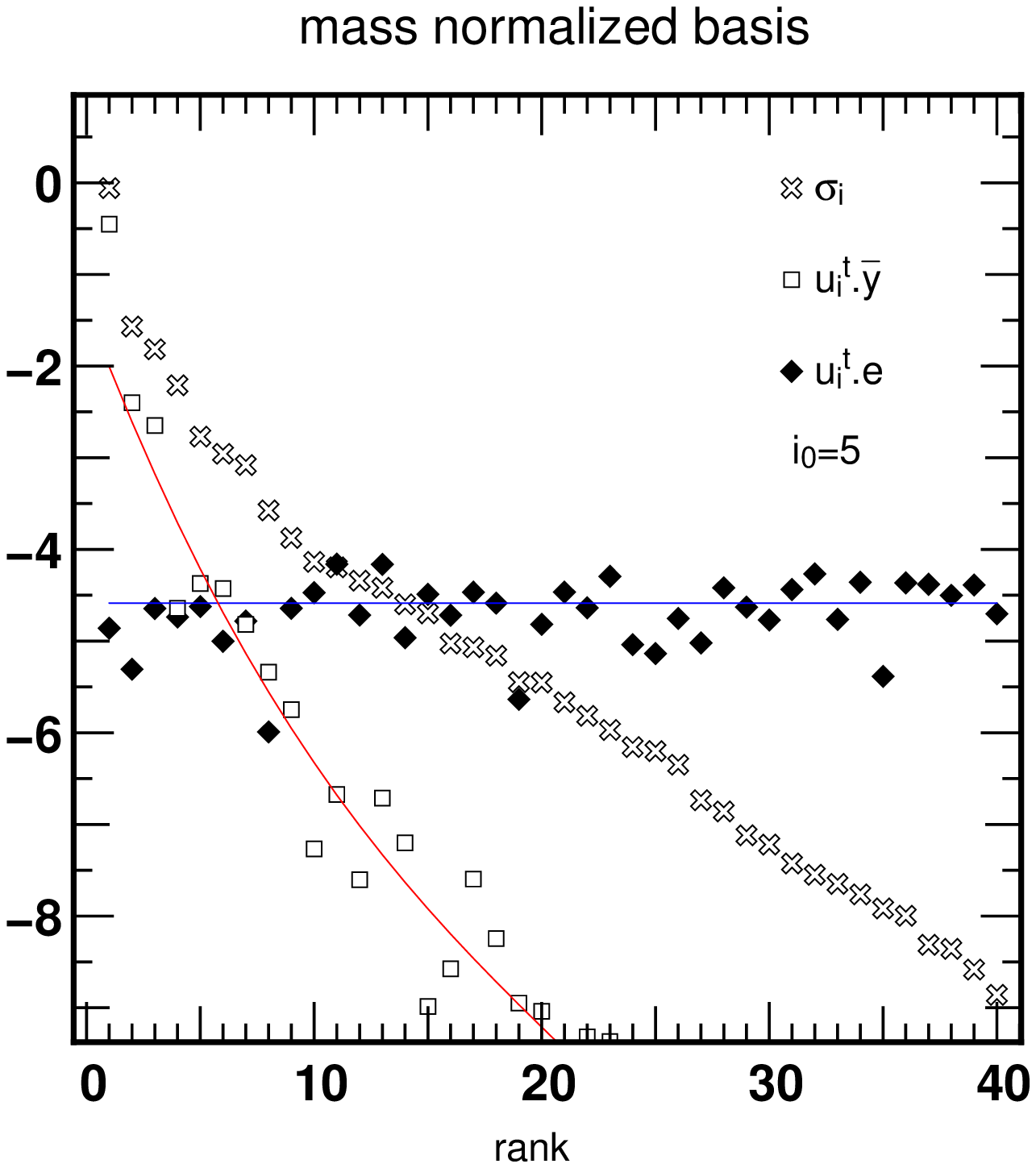}} 
{\includegraphics[width=0.32\linewidth,clip]{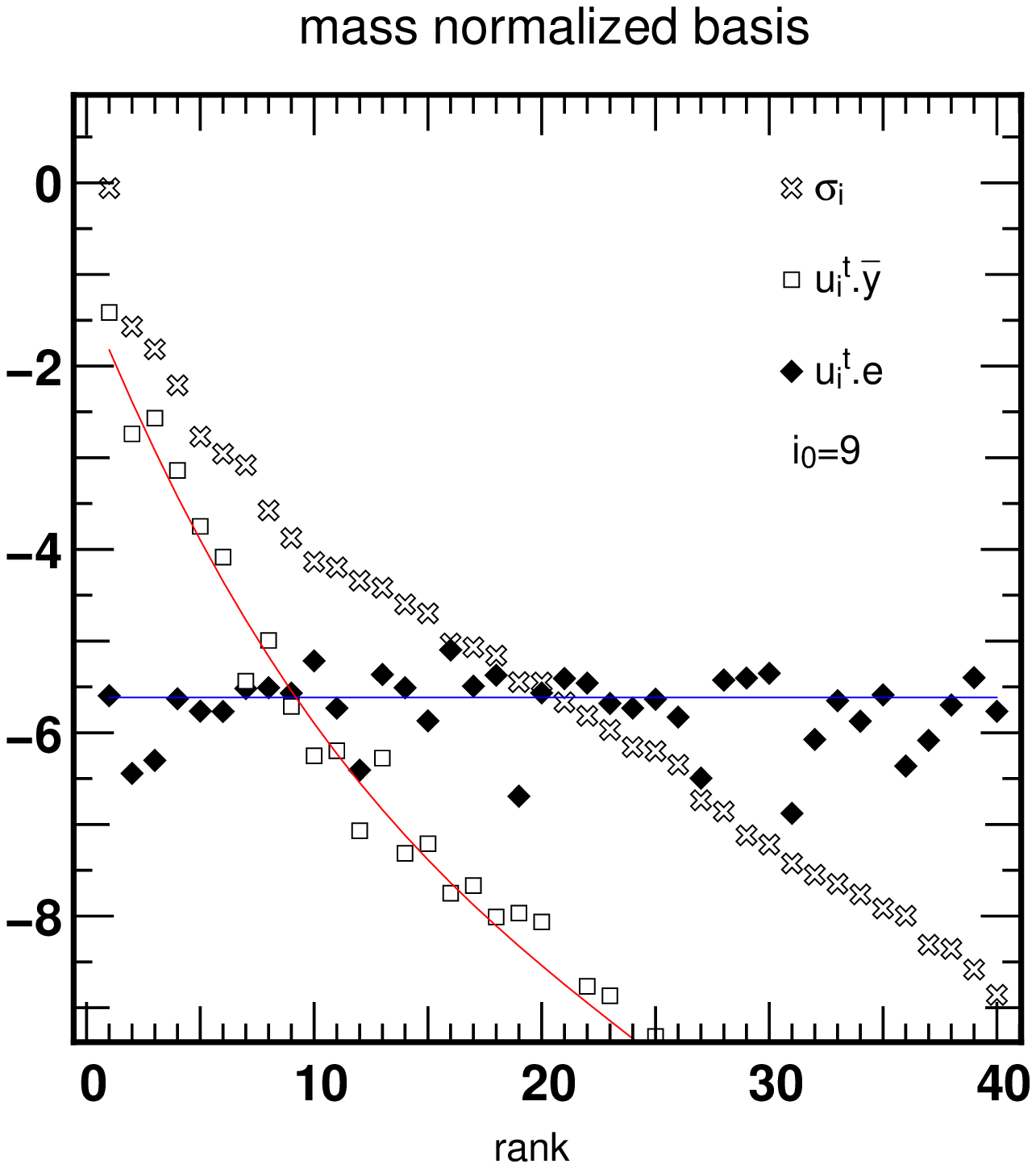}} 
{\includegraphics[width=0.32\linewidth,clip]{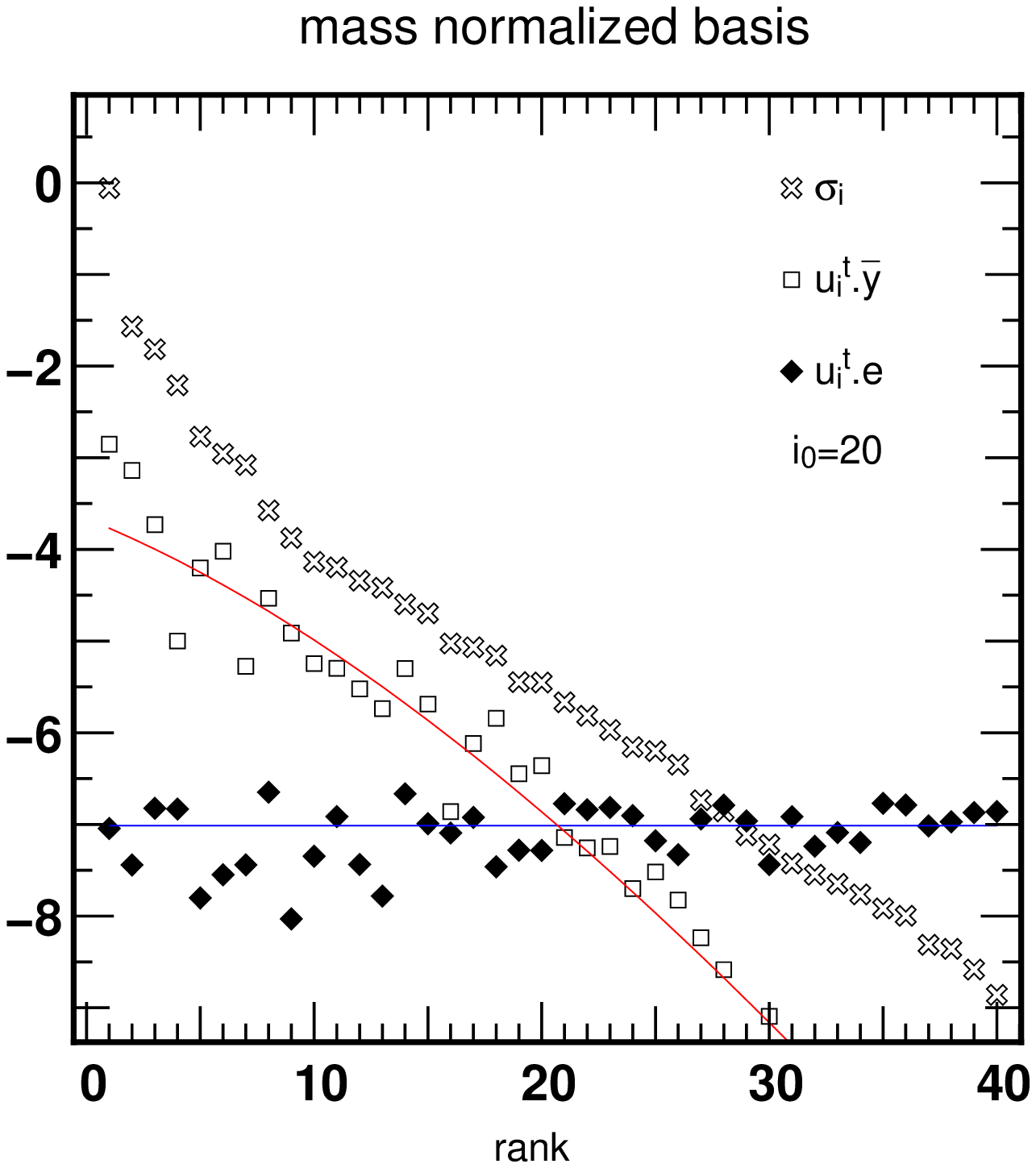}} 
{\includegraphics[width=0.32\linewidth,clip]{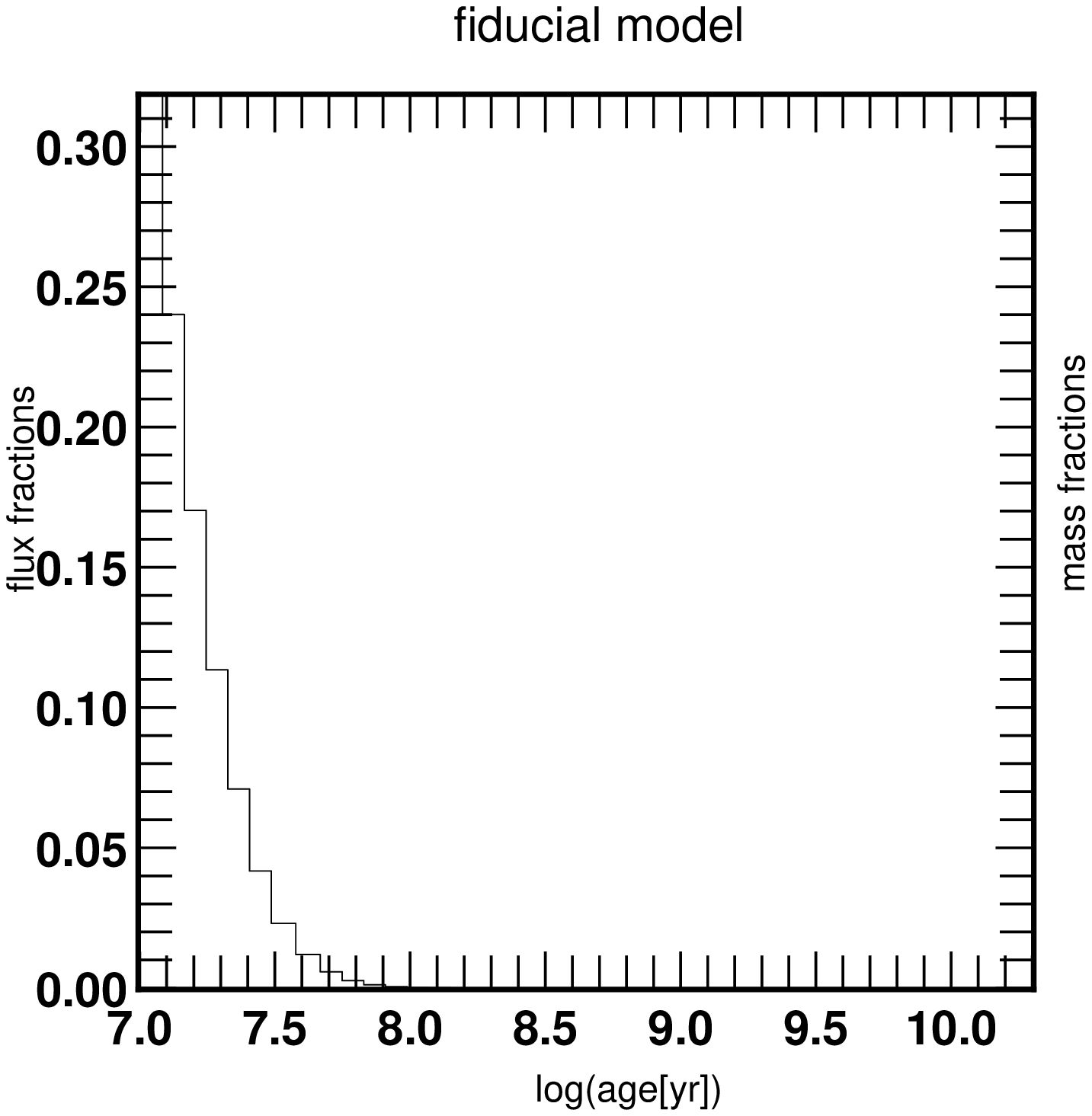}} 
{\includegraphics[width=0.32\linewidth,clip]{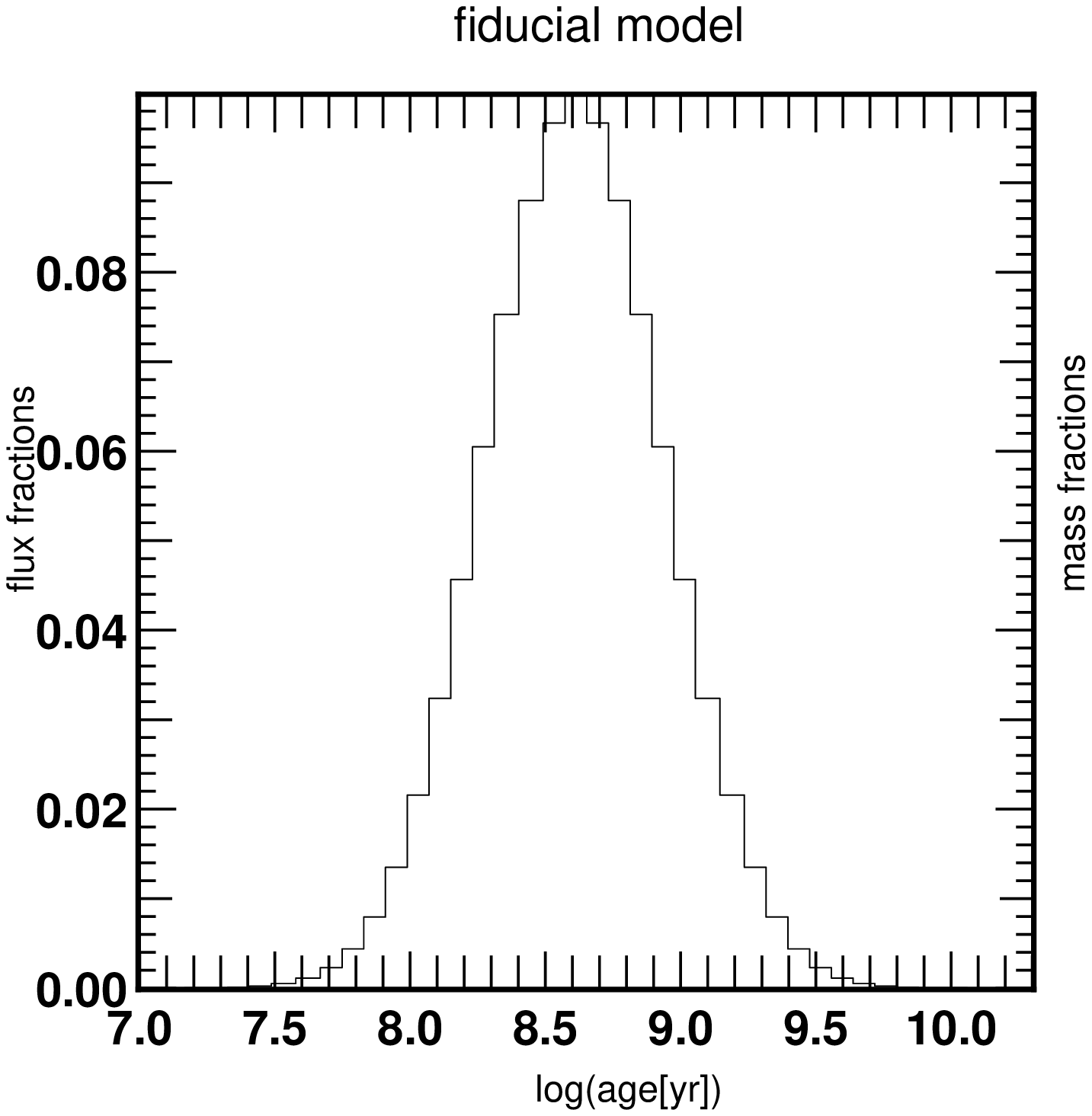}} 
{\includegraphics[width=0.32\linewidth,clip]{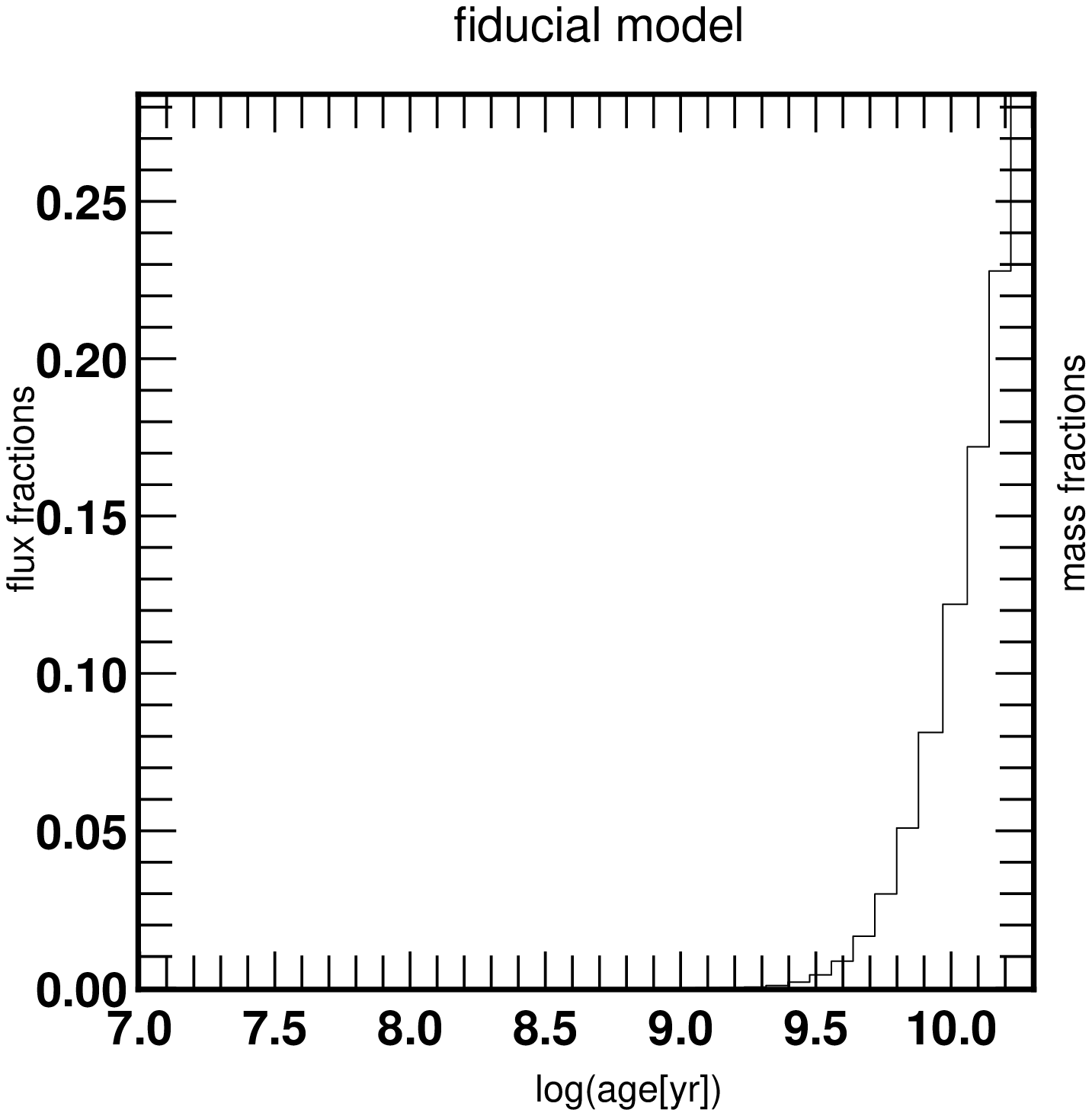}}
\caption{{Study of the location of the signal-noise transition rank as a function of 
the fiducial model. The figures are the same as \Fig{picard},
with the same pseudo-observational setting (SNR=100 per pixel),
for a flux-normalized (top) and a mass-normalized basis (middle) respectively, for 3
different 
fiducial models $\overline{\M{x}}$, given at the bottom of each column. 
Polynomial fits are given for the signal and noise singular coefficients.
The transition rank $i_0$ is given in each figure as the intersection of
these fits.
For the
mass-normalized basis, the rank of
the transition between signal and noise dominated regimes spans a wide range of values depending on the fiducial
model $\overline{\M{x}}$. On the contrary,
for the flux-normalized basis, the transition rank is rather constant
 with regard to modifications of the age of the fiducial model.}}
\label{f:fid2}
\end{figure*}
{In this section we clarify the relation between the transition rank $i_0$ between the noise and signal dominated regimes (the
intersection of the $\T{\M{u}_i} \cdot \overline{\M{y}}$ with the $\T{\M{u}_i}
\cdot \M{e}$) and the
fiducial model, as defined in \Sec{ill} and \Fig{picard}. More specifically, we explore the
shift of the transition by varying the age of the fiducial model, for a flux-normalized and a
mass-normalized basis. The results are shown in \Fig{fid2}. The fiducial
models are given in the bottom of each column. Note that the $y$ axis is labeled
``flux fractions'' on the left and ``mass fractions'' on the right. 
This is to recall that the interpretation of the model curve differs, depending
on the adopted normalization of the basis.
Compared to \Fig{picard}, we added a 3rd order polynomial fit to the signal singular
coefficients and a constant fit to the noise coefficients. This allows to
detect automatically and objectively the transition rank $i_0$, as the
intersection of the two fits. 

For the mass-normalized basis, the transition moves from the 5-th
rank (for the youngest fiducial model) up to the 20-th (for the oldest
fiducial model). 
On the other hand, the location of the transition for the flux-normalized
basis is rather unaffected by changes of the fiducial model and remains around
rank 7-9.}
\section{Gradients of  $Q_\mu$}

\label{s:agemin}

The direct linear solution which minimizes the objective function $Q_\mu$
can only be used in the case of a linear model (with respect to the
parameters) and without constraints (such as positivity).  For all other
cases, the objective function $Q_\mu$ can only be minimized by means of an
iterative method.  The most efficient and yet simple to use of these
methods require the computation of the objective function and of its
gradient. These optimization methods are: the conjugate gradients and
variable metric methods (e.g.\ BFGS).  In practice for non-linear problems,
variable metric methods have been found to require less iterations and less
function evaluations than conjugate gradient ones \citep{thiebautp02}.  For
that reason, we used the limited memory variable metric method VMLM-B
implemented in the OptimPack package written by E.\,Thi\'ebaut for Yorick
(http://www.maumae.net/yorick/doc/index.html).

Since the efficiency of these iterative optimization algorithm rely on the
correctness of the gradient of $Q_\mu$ (i.e.\ partial derivatives of
$Q_\mu$ with respect to the free parameters), we devote this appendix to
the derivation of such partial derivatives for the different cases
considered in this paper.  Whenever it was possible (i.e.\ in the linear
case), the iterative solutions were tested against the analytical
solutions, and were found to be identical down to machine precision.

\subsection{Simple Linear Model}

In the linear problem, the gradients of $Q_\mu$ have simple expressions:
\begin{eqnarray}
  \frac{\partial\chi^2}{\partial\M{x}} &=& -2\,\T{\M{B}}\cdot\M{W}\cdot
  \left(\M{y} - \M{B}\cdot\M{x} \right)\,, \\
  \frac{\partial P}{\partial\M{x}} &=& 2\,\T{\M{L}} \cdot \M{L} \cdot \M{x}\,.
  \label{e:dQdx1}
\end{eqnarray}

\subsection{Age-Metallicity-Extinction Gradients}

For the resolution of the age-metallicity-extinction problem (\Sec{aze}),
the objective function $Q_\mu$ is a $\chi^2$ penalized by regularization
terms and a binding function.  The regularization terms being the same as
in the linear case, their gradients are given by \Eq{dQdx1}.  The
gradient of the binding function $C$ for a metallicity vector $\M{Z}$
reads:
\begin{equation}
  \left(\pdrv{C}{\M{Z}}\right)_j =
  \left\{\begin{array}{ll}
      2\,(Z_j - \Zmin) & \mbox{for } Z_j < \Zmin  \, ,\\
      2\,(Z_j - \Zmax) & \mbox{for } Z_j > \Zmax\, , \\
      0                & \mbox{else} \, .
    \end{array} \right. 
\end{equation}
In order to derive the gradients of the $\chi^2$ term for more complex
(non-linear) models, it is useful to rewrite this term as:
\begin{equation}
  \chi^2 = \T{\M{r}}\cdot\M{W}\cdot\M{r} \, ,
\end{equation}
where, for sake of simplicity, we introduced the vector of residuals
$\M{r}$ defined, in this case, by:
\begin{equation}
  \M{r} \triangleq \M{y} - \diag(\Fext)\mdot\M{B}\mdot\M{x} \, .
\end{equation}
Then the derivative of the $\chi^2$ term with respect to any free
parameter, say $\alpha$, writes:
\begin{equation}
  \pdrv{\chi^2}{\alpha}
  = 2\,\T{\pdrv{\M{r}}{\alpha}}\cdot\M{W}\cdot\M{r} \, .
\end{equation}
Considering the different type of free parameters, we obtain:
\begin{eqnarray}
  \pdrv{\chi^2}{\M{x}}
  &=& -2\,\T{\M{B}}\cdot\diag(\Fext)
      \cdot\M{W}\cdot\M{r} \, , \label{e:dQdx} \\
  \pdrv{\chi^2}{\M{Z}}
  &=& -2\,\T{\M{x}}\cdot\T{\pdrv{\M{B}}{\M{Z}}}\cdot\diag(\Fext)
      \cdot\M{W}\cdot\M{r} \, , \label{e:dQdz} \\
  \pdrv{\chi^2}{E}
  &=& -2\,\T{\M{x}}\cdot\T{\M{B}}\cdot\diag\!\left(\pdrv{\Fext}{E}\right)
      \cdot\M{W}\cdot\M{r} \, . \label{e:dQde}
\end{eqnarray}
In the above expressions, $\partial\M{B}/\partial\M{Z}$ is derived
directly from the {\SSP} basis $\Bflux(\lambda,t,Z)$:
\begin{equation}
  \left(\pdrv{\M{B}}{\M{Z}}\right)_{i,j} \triangleq
  \left(\pdrv{\Bflux(\lambda,t,Z)}{Z}\right)_{t=t_{j},Z=Z_{j},\lambda=\lambda_i} \, .
\end{equation}
Similarly, the term $\partial\Fext/\partial E$ derives
directly from the chosen extinction law $\fext(E,\lambda)$:
\begin{equation}
  \left(\pdrv{\Fext}{E}\right)_{i} \triangleq
  \left(\pdrv{\fext(E,\lambda)}{E}\right)_{E,\lambda=\lambda_{i}} \, .
\end{equation}

\section{Generalized singular value decomposition}
\label{s:GSVD}

This section introduces briefly the Generalized singular value
decomposition which is used in the main text to understand how
regularization damps smoothly the singular vectors according to the SNR.
In short, the {\GSVD} of $(\M{B},\M{L})$ is defined by:
\begin{equation}
  \M{B} = \M{U}\cdot\M{\Sigma}\cdot\T{\M{V}}\,,
  \quad\mbox{and}\quad
  \M{L} = \M{Q}\cdot \M{\Theta}\cdot\T{\M{V}} \, ,
\end{equation}
where $\M{U}$ and $\M{Q}$ are both orthogonal. The matrix $\M{V}$ is
non-singular and its columns $\M{v}_i$ are $\T{\M{B}}{\cdot}\M{B}$ and
$\T{\M{L}}{\cdot}\M{L}$ orthonormal i.e.\
$\T{\M{V}}{\cdot}\T{\M{B}}{\cdot}\M{B}{\cdot}\M{V} = \M{\Sigma}^2$ and
$\T{\M{V}}{\cdot}\T{\M{L}}{\cdot}\M{L}{\cdot}\M{V} = \M{\Theta}^2$.  The
matrices $\M{\Sigma}$ and $\M{\Theta}$ are diagonal:
$\M{\Sigma}=\diag(\sigma_1,\sigma_2,\dots,\sigma_n)$ and
$\M{\Theta}=\diag(\theta_1,\theta_2,\dots,\theta_n)$, with the $\sigma_i$ in
increasing order and the $\theta_i$ decreasing.  See \citet{hansen} for a
more detailed description of {\GSVD} and its properties.

\section{GSO versus SVD}
\label{s:CDvsSVD}

\begin{figure*}
\begin{tabular}{ccc}
\includegraphics[width=0.4\linewidth,height=0.4\linewidth]{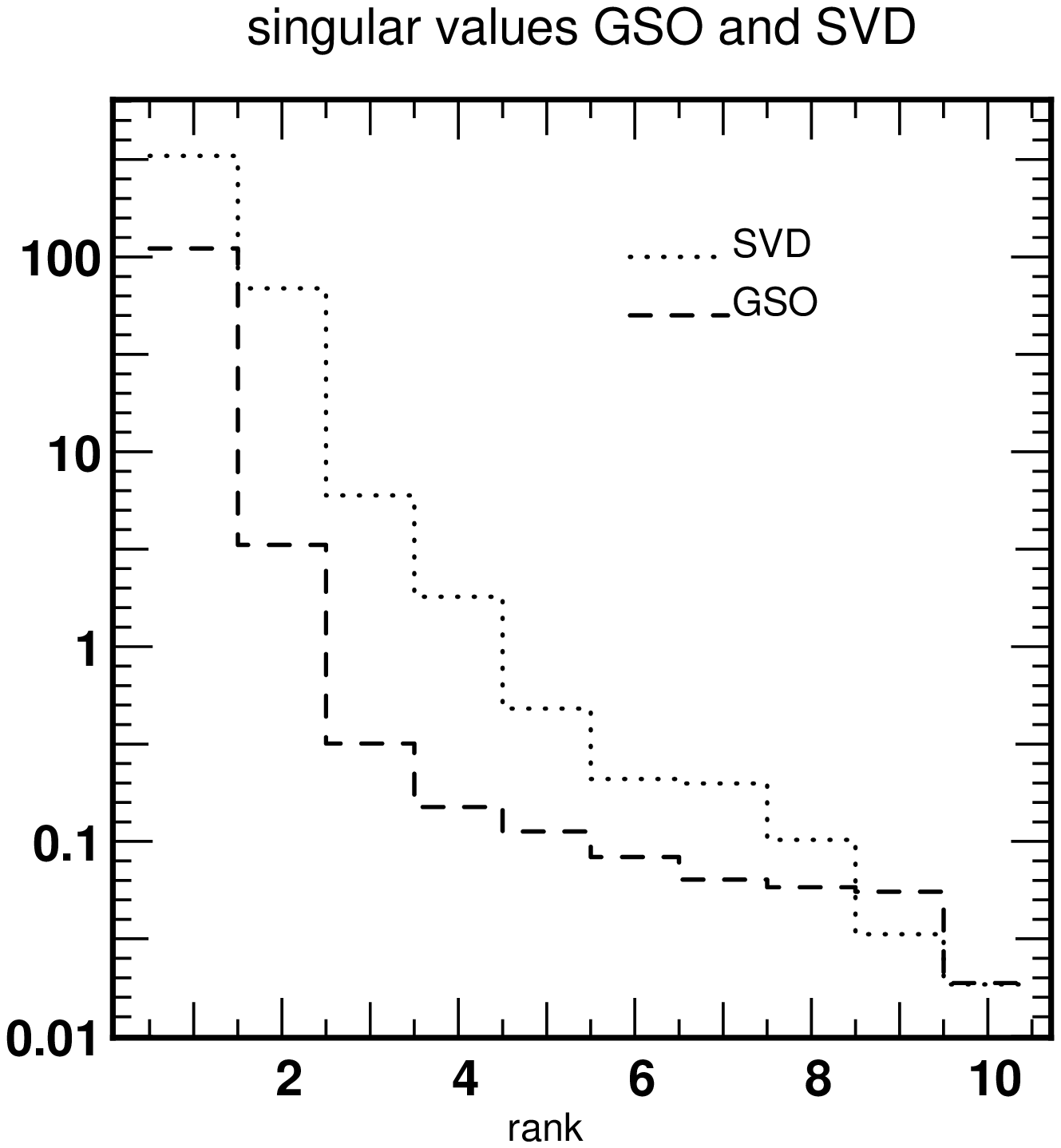} &
\includegraphics[width=0.4\linewidth,height=0.4\linewidth]{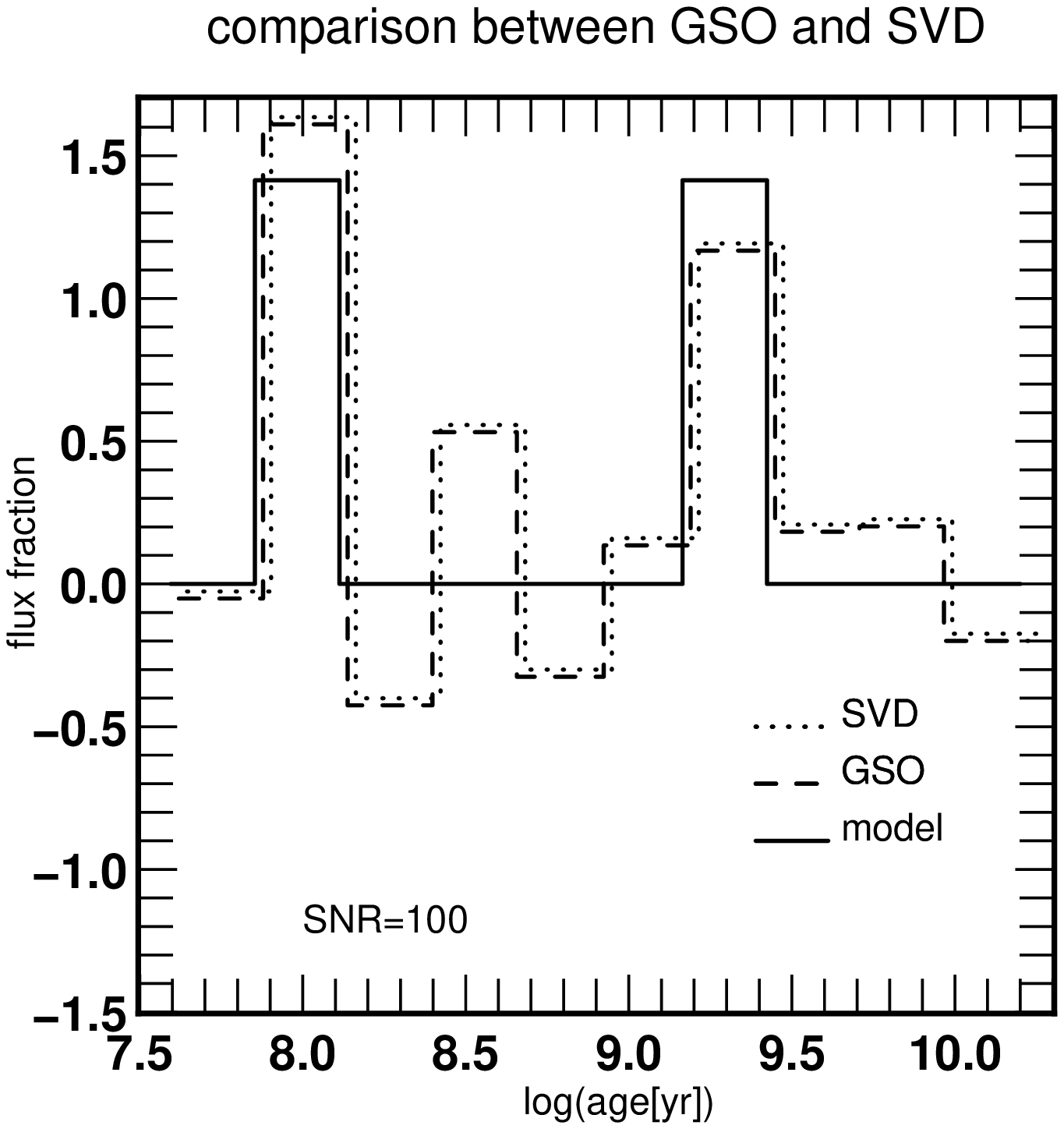}
\end{tabular}
\caption{\emph{Left}: Singular values of the {\GSO} and the {\SVD} of the kernel. Both
  decays are characteristic of an ill-conditioned problem. \emph{Right}:
  Solutions found using the {\GSO} and the {\SVD} (slightly offset for clarity)
  for simulated data with $\mathrm{SNR}=100$ per pixel, $R=10\,000$. They
  are identical down to machine precision, showing the similarity between
  both formulations.}
\label{f:MOPED}
\end{figure*}
In the main text, we claim that {\GSO } amounts to {\SVD} in the linear regime
(mono-metallic and extinction-less populations) in the absence of truncation. Let us demonstrate
and discuss  this briefly.

In the mono-metallic extinctionless case, we can expand the kernel $\M{B}$ as:
\begin{equation}
  \M{B} = \M{O} \cdot \M{\Sigma} \cdot \M{V} \, ,
  \label{e:GSO}
\end{equation}
where $\M{O}$ is the Gram-Schmidt orthonormalized kernel obtained from
$\M{B}$, and $\M{\Sigma}=\diag(\sigma_1, \dots, \sigma_n)$ is a
diagonal matrix such that $\M{\Sigma} \cdot \M{V}= \T{\M{O}} \M{B}$ is the
passage matrix from the initial coordinates of the kernel $\M{B}$ to the
orthonormalized basis. In this sense, the $\sigma_i$ are the norms of the
vectors of the passage matrix. It is interesting to compare this expansion
with the SVD: the kernel $\M{O}$ is orthonormal and the matrix $\M{\Sigma}$
is diagonal, but the matrix $\M{V}$ is not orthogonal.

Thus, the expansion of \Eq{GSO} is not exactly identical to that
corresponding to \SVD.  Still, as long as none of the $\sigma_i$ is zero,
the matrix $\M{V}$ is inversible, and as for the \SVD, we can
write the solution $\M{x}$ as
\begin{equation}
  \M{x} = \M{V}^{-1} \cdot \M{\Sigma}^{-1} \cdot \T{\M{O}} \cdot \M{y}
  = \sum_{i=1}^n \frac{\T{\M{O}}_i\cdot\M{y}}{\sigma_i}
  \left(\M{v}^{-1}\right)_i \, ,
\label{e:GSO-sol}
\end{equation}
where $\M{y}=\M{B}\cdot\M{x}$ is the data, and the $(\M{v}^{-1})_i$ are the
columns of ${\M{V}}^{-1}$. We will in this section abusively call the
$\sigma_i$ the singular values, the $\M{O}_i$ and $(\M{v}^{-1})_i$
respectively the data singular vectors and the solution singular vectors.
The solution $\M{x}$ is the sum of the singular coefficients $\T{\M{O}_i}
\cdot \M{b}$ (the ``compressed datum'' proposed by MOPED's authors) divided
by the singular values $\sigma_i$ times the solution singular vector
$(\M{v}^{-1})_i$.  The left panel of \Fig{MOPED} shows the singular values
of the SVD and the GSO expansion of the kernel. Their very similar decay
indicates similar behaviour of the inverse problem. The right panel
of \Fig{MOPED} shows for a moderately ill-conditioned
example ($R=10\,000$, $\mathrm{SNR_d}=100$, 10 age bins, solar metallicity,
$\sigma_{1}/\sigma_{10}=2\,\sqrt{m}\,\mathrm{SNR}_{\mathrm{d}}$) the
solutions found by applying \Eq{GSO-sol} and \Eq{svdsol} corresponding to the
two expansions. As expected from the
{\CN} and $\mathrm{SNR_d}$, both are fairly noisy, but the important point
is that they are actually equal down to machine precision. Thus, even though
there is a slight formulation difference between these two expansions, they
practically give the same solutions.




\label{lastpage}
\end{document}